\documentclass[useAMS,usenatbib]{mn2e}
\usepackage{graphicx}
\usepackage{natbib,amssymb,color,amsmath}
\usepackage{times} 

\def\farcm{\hbox{$.\mkern-4mu^\prime$}}

\newcommand{\bi}{\begin{itemize}}
\newcommand{\ei}{\end{itemize}}

\PassOptionsToPackage{hyphens}{url}\usepackage{hyperref}
\usepackage[hyphens]{url}
\usepackage{breakurl}

\newcommand{\replya}[1]{{#1}}

\defcitealias{mhb06}{M06}
\defcitealias{jsb13}{J13a}
\defcitealias{jsh13}{J13b}
\title[CFHTLenS: Halo ellipticity constraints]{CFHTLenS: 
Weak lensing constraints on 
the ellipticity of galaxy-scale 
matter haloes
and the galaxy-halo misalignment}

\author[Schrabback  et al.]{Tim Schrabback$^{1,2,3}$\thanks{E-mail:
schrabba@astro.uni-bonn.de},
Stefan Hilbert$^{4,5}$,
Henk Hoekstra$^{3}$, 
Patrick Simon$^{1}$, 
\newauthor
 Edo van Uitert$^{1,6}$, 
Thomas Erben$^{1}$, 
Catherine Heymans$^7$, 
Hendrik Hildebrandt$^{1}$,  
\newauthor
Thomas D.~Kitching$^8$, 
Yannick Mellier$^{9,10}$, 
Lance Miller$^{11}$, 
Ludovic Van Waerbeke$^{12}$, 
\newauthor
Philip Bett$^{1}$, 
Jean Coupon$^{13}$, 
Liping Fu$^{14}$, 
Michael J. Hudson$^{15,16}$, 
\newauthor
Benjamin Joachimi$^{6}$, 
Martin Kilbinger$^{9,10}$, 
Konrad Kuijken$^{3}$.
\\ 
        \vspace*{3pt}
\\
$^{1}$ Argelander Institute for Astronomy, University of Bonn, Auf dem H{\"u}gel 71, 53121 Bonn, Germany\\
$^{2}$ Kavli Institute for Particle Astrophysics and Cosmology, Stanford University, 382 Via Pueblo Mall, Stanford, CA 94305-4060, USA\\
$^{3}$ Leiden Observatory, Leiden University, Niels Bohrweg 2, NL-2333 CA Leiden, The Netherlands\\
$^{4}$ Excellence Cluster Universe, Boltzmannstr. 2, 85748 Garching, Germany\\
$^{5}$ Ludwig-Maximilians-Universit{\"a}t, Universit{\"a}ts-Sternwarte, Scheinerstr. 1, 81679 M{\"u}nchen, Germany\\
$^{6}$ Department of Physics and Astronomy,  University College London, Gower Street, London, WC1E 6BT, UK\\
$^7$ The Scottish Universities Physics Alliance, Institute for Astronomy, University of Edinburgh, Blackford Hill, Edinburgh EH9 3HJ, UK \\ 
$^8$ Mullard Space Science Laboratory, University College London, Holmbury St Mary, Dorking, Surrey RH5 6NT, UK \\
$^9$ Institut d'Astrophysique de Paris, UMR7095 CNRS, Universit\'e Pierre \& Marie Curie, 98 bis boulevard Arago, F-75014 Paris, France \\
$^{10}$ CEA/Irfu/SAp Saclay, Laboratoire AIM, F-91191 Gif-sur-Yvette, France \\
$^{11}$ Department of Physics, Oxford University, Keble Road, Oxford OX1 3RH, UK \\
$^{12}$ Department of Physics and Astronomy, University of British Columbia, 6224 Agricultural Rd, Vancouver BC, V6T 1Z1, Canada \\
$^{13}$ Astronomical Observatory of the  University of
Geneva, ch. d'Ecogia 16, 1290 Versoix, Switzerland\\
$^{14}$ Shanghai Key Lab for Astrophysics, Shanghai Normal University, 100 Guilin Road, 200234, Shanghai, China \\
$^{15}$ Department of Physics \& Astronomy, University of Waterloo, Waterloo, ON, N2L 3G1, Canada \\
$^{16}$ Perimeter Institute for Theoretical Physics, 31 Caroline St. N., Waterloo, ON, N2L 2Y5, Canada 
}

\begin{document}

\date{Accepted 2015 August 30. Received 2015 August 28; in original form 2015 July 15}

\pagerange{\pageref{firstpage}--\pageref{lastpage}} \pubyear{2015}

\maketitle

\label{firstpage}

\begin{abstract}
We present weak lensing constraints on the ellipticity of galaxy-scale 
matter haloes and the galaxy-halo misalignment.
Using data from the Canada-France-Hawaii Telescope Lensing Survey
(CFHTLenS), we measure the weighted-average ratio of the aligned projected ellipticity components of galaxy matter haloes and their embedded galaxies, $f_\mathrm{h}$, split by galaxy type.  We then compare our observations to measurements taken from the Millennium Simulation, assuming different models of galaxy-halo misalignment.  
Using the Millennium Simulation we verify that the statistical estimator used removes contamination from cosmic shear.
We also detect an additional signal in the simulation,  which we interpret as the impact of intrinsic shape-shear alignments between the lenses and their large-scale structure environment.
These alignments are likely to have caused some of the previous observational constraints on  $f_\mathrm{h}$ to be biased high.
From CFHTLenS  we find \mbox{$f_\mathrm{h}=-0.04\pm 0.25$} for early-type galaxies,
which is consistent with current  models for  the galaxy-halo misalignment
predicting \mbox{$f_\mathrm{h}\simeq 0.20$}.
For late-type galaxies we measure \mbox{$f_\mathrm{h}=0.69_{-0.36}^{+0.37}$} from CFHTLenS.
This can be compared to the simulated results which yield \mbox{$f_\mathrm{h}\simeq 0.02$} for misaligned late-type models.
\end{abstract}


\section{Introduction}
In the standard cosmological paradigm,
 galaxies, groups, and clusters are embedded in large haloes of -- mostly dark -- matter.
Numerical simulations of cosmic structure formation predict that these haloes are roughly triaxial \citep[e.g.][]{jis02}, and that
their average density profiles closely follow the Navarro-Frenk-White profile \citep[NFW,][]{nfw96,nfw97}.
In projection, these should approximately appear elliptical.
This prediction can be tested observationally on the scales of galaxies and
clusters.
One approach is to constrain halo shapes via the use of baryonic tracers such as satellite galaxies \citep[e.g.][]{hol69,bra05,lfc05,ybm06,app07,mkj07,flm07,flw09,bpn08,ojl09,agb10,nat11}, the distribution of stellar velocity \citep{olm00}, satellite tidal streams \citep{ili01,lrl12,veh13}, HI gas \citep{baj08,bfk10}, or planetary nebulae
\citep{nrc11}.
Such constraints can be compared to the output of hydrodynamical simulations that aim at modelling galaxy formation, linking the baryonic and dark matter components, and which can provide predictions on the relative (mis-) alignment of galaxies, their dark matter hosts, and the surrounding large-scale structure, as well as  on the impact of the baryons on  halo shapes \citep[e.g.][]{bkg05,klk10,vsh11,tmm14,lpc15,dbr15,vcs15}.

As an observational alternative, gravitational lensing
 probes
the total projected mass distribution directly without
relying on  visible tracers.
Strong gravitational lensing can provide information on the inner shapes of the
mass distribution in massive clusters \citep[e.g.][]{lms13} and galaxies \citep{wfd10,dbm11,shk12}, but at
these scales baryons have a non-negligible influence.
At larger scales,  constraints on the projected mass distribution can be
obtained with weak gravitational lensing, which probes the 
coherent distortions 
 imprinted onto the observed shapes of background galaxies  by the tidal
 gravitational field \citep[e.g.][]{bas01,sch06}.
Even at these scales baryons contribute to the total mass distribution probed by weak lensing (e.g.~in the form of satellite galaxies),
but
the dark matter is expected to dominate. 

Weak gravitational lensing 
has been successfully used in a wide range of applications, including
for example cosmic shear studies which are sensitive to the growth of
large-scale structure \citep[e.g.][]{shj10,kfh13,hgh13,heh14,kha14,ssw15}, the mass
calibration of galaxy clusters \citep[e.g.][]{laa14,kgu15,fwm15,hhm15}, and constraints on the
{\it azimuthally-averaged} mass profiles of galaxy matter haloes
\citep[e.g. ][]{msk06,uhv11,ltb12,vuh14,hgc15}.

In principle, weak lensing is also sensitive to halo ellipticity as the
gravitational shear at a given radius is larger along the direction of the
major axis of the projected halo compared to the projected minor axis \citep[e.g.][]{nar00,brw00}.
For very massive clusters, the weak-lensing signal is strong enough to provide individual halo ellipticity
constraints \citep{ckc09,oto10}.  Constraints were also obtained in 
\replya{stacked analyses of larger cluster samples \citep{evb09,obd12}}.
For less massive, galaxy-scale haloes, the signal can still be detected statistically by stacking very large samples.

A measurement of this effect through stacking requires one to align
the shear field around all lenses prior to stacking, 
such that the major axes  of all projected haloes are aligned.
The orientations of the (mostly dark) matter haloes, however, are
not directly observable. One
 approach which has  been used instead for weak lensing halo shape studies,
is to approximate the orientations of the projected haloes by the orientations
of their
 galaxy images \citep{hyg04,mhb06,phh07,uhs12}.
Also, it has typically been assumed that, 
 on average and in projection, more elliptical lenses are
hosted by more elliptical haloes, 
 as supported by simulations for early type galaxies \citep{ncp06}.
A key parameter which is then extracted is the average aligned ellipticity ratio
between the ellipticities of
the projected halo and the observed lens light distribution
\mbox{$f_\mathrm{h}  = \langle
  \cos(2\Delta\phi_\mathrm{h,g})|e_\mathrm{h}|/|e_\mathrm{g}| \rangle$},
 where the averaging typically  includes a weighting scheme 
that depends on $|e_\mathrm{g}|$.
Here, $\Delta\phi_\mathrm{h,g}$ is the angle between the major axis of the projected
galaxy light distribution and the major axis of its projected 
matter halo.
Thus, only in the case of perfect alignment (\mbox{$\Delta\phi_\mathrm{h,g}=0$}), \mbox{$f_\mathrm{h}$} reduces to
the actual ellipticity ratio. 
However, in practice one expects a considerable 
random misalignment  between the observed shapes of galaxies and 
 matter
haloes, as suggested both by
numerical  simulations of galaxy formation \citep[e.g.][]{oef05,ctd09,bef10,dmf11,dlk14,wlk14,tmm14} and observations
that approximate 
matter haloes via the distribution of satellites
 \citep[e.g.][]{ojl09}.
This should substantially  reduce
\mbox{$f_\mathrm{h}$} and wash out the halo shape signature,  making it
 difficult to detect observationally
\citep[][]{bet12}.
 It is important to test this prediction observationally, both to
  improve our understanding of galaxy formation, but also to inform models
  of intrinsic galaxy alignments. Such alignments of galaxies with their
  surrounding mass distribution are an important physical contaminant for
  cosmic shear studies \citep[e.g.][]{hgh13,jsh13}.
Of particular concern are  shape-shear intrinsic
  alignments \citep{his04}.
Here, the ellipticities of foreground galaxies
  are aligned with their surrounding large-scale structure, which lenses the
  background sources.
So far, most constraints on shape-shear  alignments come from studies
investigating the alignment of galaxies with their surrounding galaxy
distribution \citep[e.g.][]{mbb11,jma11,ljf13,zyw13,smm15}.
Interestingly, the aligned halo shape signature we want to
  extract directly contributes to the shape-shear intrinsic
  alignment signal  at small scales \citep{bra07}.

In addition to the expected small signal, there are 
further
observational challenges:
 While a potential additional alignment of lenses and sources 
 does not affect
the  
azimuthally-averaged 
 galaxy-galaxy lensing signal, this is no longer the case for the most
 simple  estimator of the anisotropic halo ellipticity signal.
Such a source-lens alignment can, for example, be introduced  by incomplete
removal of instrumental signatures such as the point-spread function (PSF).
In addition, gravitational lensing by structures in front of the lens causes extra alignment.
The latter is commonly referred to as cosmic shear, 
or sometimes
as multiple deflections \citep{bra10}.
For example, \citet{hob10} study the impact of cosmic shear 
on
halo ellipticity constraints and conclude that observational estimates of
the ratio of the shears along the lens major and minor axes would need to be
compared to Monte Carlo simulations for interpretation.
However, they do not consider the  modified estimator
introduced by \citet[][hereafter \citetalias{mhb06}]{mhb06}, which leads to a cancellation of such spurious signal at the relevant
(small) scales, and which we employ in the current study.

 Previous observational constraints on halo ellipticity from weak
  lensing are somewhat inconclusive: On the one hand, \citet{hyg04} and \citet{phh07}
  find indications for positive $f_\mathrm{h}$ using magnitude-selected lens
  samples and simple estimators that do not correct for systematic shear. On
the other hand, \citetalias{mhb06} and \citet{uhs12} separate lenses by colour
and correct for systematic shear, but do not detect significantly non-zero
$f_\mathrm{h}$. For example, \citetalias{mhb06} find \mbox{$f_\mathrm{h}=0.60 \pm 0.38$}
for red and  \mbox{$f_\mathrm{h}=-1.4^{+1.7}_{-2.0}$} for blue lenses using
data from the Sloan Digital Sky Survey (SDSS).
Very recently, \citet{clj15} reported a  significant detection of the
signature of halo ellipticity around luminous red galaxies, also employing SDSS data.

In this study we use weak lensing data from the 
Canada-France-Hawaii
Telescope
  Lensing Survey  \citep[CFHTLenS,][]{hwm12,hek12,ehm13,mhk13a} 
to derive 
updated constraints on the ellipticity of galaxy 
matter haloes from weak lensing.
These data allow us to constrain the signal for galaxies subdivided 
into bins of photometric type and stellar mass.
The former division allows us to approximately 
separate the lens sample into early and late type galaxies, which have
different predictions for the expected weak lensing halo shape signal.
The latter  division optimises the  total measurement signal-to-noise as
stellar mass acts as a proxy for halo mass and therefore the signal strength.

In addition, we study a simulated weak lensing halo shape signal based on
the Millennium Simulation \citep{swj05}, employing the ray-tracing analysis
from \citet{hhw09} and lens shapes computed in \citet{jsb13,jsh13}.
On the one hand, this provides a signal prediction given current galaxy-halo
(mis-) alignment models \citep{bet12} that we can compare the CFHTLenS results to. On the other hand, it allows us to study the impact of cosmic shear on the
measurement and check for deviations from the simple model prediction of
isolated elliptical NFW haloes.

This paper is organised as follows:
Sect.\thinspace\ref{se:methodandtests} describes the formalism of the 
halo ellipticity measurements using weak lensing, and the verification tests which we have
conducted using a simple simulation.
In Sect.\thinspace\ref{se:data} we summarise properties of the CFHTLenS
data, discuss the
selection of lens and source galaxies, 
and present the measured shear signal and
 constraints on the aligned ellipticity ratio $f_\mathrm{h}$.
We present our analysis of the simulated data based on the Millennium
Simulation in Sect.\thinspace\ref{sec:millennium}.
We then discuss our results and conclude in Sect.\thinspace\ref{se:discussion}.
 In addition, we present a consistency check for shape measurements of background sources in the vicinity of bright foreground lenses using image simulations in Appendix \ref{app:testlf}.

For the computation of angular diameter distances we by default assume a flat
$\Lambda$CDM cosmology with \mbox{$\Omega_\mathrm{m}=0.3$},
\mbox{$\Omega_\Lambda=0.7$}, \mbox{$H_0=70 h_{70}$ km/s/Mpc}, 
 and \mbox{$h_{70}=1$}, which is consistent with the 
best-fitting cosmological parameters from both WMAP9 \citep{hlk13}
and Planck \citep{planck15cosmo} at the \mbox{$\sim 1-2\sigma$} level. 
 The only exception is our analysis of the simulated data from the Millennium Simulation, for which we use the input cosmological parameters of the simulation \mbox{$\Omega_\mathrm{m}=0.25$},
\mbox{$\Omega_\Lambda=0.75$}, \mbox{$H_0=73 h_{73}$ km/s/Mpc}, 
 and \mbox{$h_{73}=1$} \citep{swj05}.
All magnitudes are in the AB system. 
Stellar masses $M_*$ are given in units
of solar
masses  $M_\odot$.

\section{Method}
\label{se:methodandtests}

\subsection{Formalism}
\label{su:method}

The methodology of our analysis largely follows the
approach and notation that was
introduced by 
\citetalias{mhb06} and additionally applied in \citet{uhs12}.
It allows for
the correction of spurious signal originating from cosmic shear or
instrumental distortions.

\subsubsection{Constraining the isotropic galaxy-galaxy lensing signal}

 In weak lensing studies the shape of a galaxy is typically
described by the complex ellipticity
\begin{equation}
e=e_1+\mathrm{i}e_2=|e|\mathrm{e}^{2\mathrm{i}\phi} \,.
\end{equation}
 For the ellipticity definition employed here
 \citep[see][]{mhk13a}\footnote{This ellipticity definition is often
   referred to as $\epsilon$ in the literature. Here, we denote it
   as $e$ to be consistent with \citet{mhk13a}.}
and the case of an idealised source with elliptical isophotes, 
the absolute value of the ellipticity is given by
\mbox{$|e|=(a-b)/(a+b)$}. 
Here,  $a$ and $b$ are the  major and minor axes of the ellipse, while 
$\phi$ corresponds to  the position angle of the major axis from the
$x$-axis of the coordinate system.
In this definition the ellipticity of each background galaxy provides an
unbiased but very noisy estimate of the reduced gravitational shear $g$ in
the direction of the source
\begin{equation}
\mathrm{E}(e)=g=\frac{\gamma}{1-\kappa}\simeq \gamma \,,
\label{eq:ellipticity_expectation}
\end{equation}
where $\mathrm{E}$ indicates the expectation value, while $\gamma$ and $\kappa$ denote the
shear and convergence.
 In principle, all structures along the line of sight from the source
  to the observer contribute to the net shear and convergence \citep[see
  e.g.][]{sch06}. However, in galaxy-galaxy weak lensing analyses, such as
  the  study presented here, one correlates the
  shear inferred from background sources with the positions of
  foreground lenses. In this case, only structures at the lens redshift
  contribute to the net signal (other structures add noise\footnote{For a
    weak lensing
    halo shape analysis foreground structures in front of the lens cause an
    extra alignment of sources and lenses, which does introduce spurious
    signal for simple estimators. However, this signal is accounted for via the
    formalism explained further below in this section.}),
allowing us to express the  correlated convergence
\mbox{$\kappa=\Sigma/\Sigma_\mathrm{c}$} as the ratio between the surface mass density $\Sigma$ and the critical surface mass density 
\begin{equation}
  \label{eqn:sigmacrit}
  \Sigma_{\mathrm{c}} = \frac{c^2}{4\pi G}\frac{1}{D_{\mathrm{l}} \beta }\,.
\end{equation}
Here,  $c$ denotes the speed of light in vacuum and $G$ the gravitational constant.
The geometric lensing efficiency $\beta$ is defined as
\begin{equation}
\beta=\mathrm{max}\left[0,\frac{D_\mathrm{ls}}{D_\mathrm{s}}\right] \,. 
\end{equation}
 $D_\mathrm{s}$, $D_\mathrm{l}$, and $D_\mathrm{ls}$ indicate the physical angular diameter distances to the source, to the lens, and between lens and source, respectively (note that the equations in \citetalias{mhb06} are expressed for comoving distances instead). 

Galaxy-galaxy weak lensing analyses study the stacked shear field around foreground lens galaxies. 
For this it is useful to decompose the shear and correspondingly the ellipticities of background galaxies into the tangential component and the 45 degrees-rotated cross component
\begin{eqnarray}
e_\mathrm{t} & = & - e_1 \cos{2 \theta} - e_2 \sin{2\theta}\,,\\
\label{eq:et}
e_\times & = & + e_1 \sin{2\theta} - e_2 \cos{2 \theta}  \,,
\label{eq:ex}
\end{eqnarray}
where $\theta$ is the azimuthal angle with respect to the  lens position as
measured from the $x$-axis  of the coordinate system used.

The majority of the previous galaxy-galaxy weak lensing analyses have only
studied the profile of the azimuthally-averaged tangential shear
$\gamma_\mathrm{t}$, which relates to the differential surface mass density
profile \mbox{$\Delta \Sigma (r)\equiv \bar{\Sigma}(<r)-\Sigma(r)$},
where \mbox{$\bar{\Sigma}(<r)$} is the mean convergence within radius $r$,
as \mbox{$\Delta \Sigma =\gamma_\mathrm{t} \Sigma_{\mathrm{c}}$}.
We estimate the differential surface density from the source galaxy ellipticities as
\begin{equation}
\widehat{\Delta \Sigma}(r)=\frac{\sum_{i} w_i \Sigma_{\mathrm{c},i}^{-2}\left(e_{\mathrm{t},i}\Sigma_{\mathrm{c},i}\right)}{\sum_i w_i \Sigma_{\mathrm{c},i}^{-2}} =
\frac{\sum_{i} w_i \Sigma_{\mathrm{c},i}^{-1}e_{\mathrm{t},i}}{\sum_i w_i \Sigma_{\mathrm{c},i}^{-2}}
\, ,
\label{eq:isoestsimple}
\end{equation}
where we sum over lens--source pairs in 
 an annulus
 around $r$.
 In this section we indicate estimators with a hat 
for clarity,
  but we drop it in the subsequent sections when presenting results.  
In our analysis we 
employ inverse-variance weights 
$w_i$
for the source
shape estimates.
These weights
 account for both measurement noise and the intrinsic ellipticity
distribution    \citep[][]{mhk13a}. We use uniform  source shape weights
for simulated data.
The main reason for conducting the analysis in terms of $\Delta\Sigma$,
which is a re-scaled version of the shear,
instead of the shear directly, is to adequately account for the redshift dependence of
the weak lensing signal. 

 We note that the tangential ellipticity components $e_\mathrm{t}$ of
sources provide estimates for the tangential component of the reduced shear
$g_\mathrm{t}$, while $\Delta\Sigma$ is defined in terms of the tangential component of shear
$\gamma_\mathrm{t}$.
In galaxy-galaxy weak lensing, typically \mbox{$|\gamma|\ll 1$} and \mbox{$|\kappa|\ll
  1$}.
Hence, 
many studies have typically approximated the reduced shear with the shear as
indicated in (\ref{eq:ellipticity_expectation}).
Here we implicitly account for the difference when fitting the  azimuthally-averaged tangential shear profiles, as we find  
a small but non-negligible
impact for our most massive lenses.
When studying the anisotropy in the shear field, as detailed below, we
however  ignore reduced shear corrections as they cancel out to leading order.

In our analysis we fit the isotropic part of the measured shear profile with an NFW
shear profile prediction according to \citet{wrb00} in order to constrain
\mbox{$r_{{200}\mathrm{c}}$}, the radius corresponding to a mean over-density
that is $200$ times the critical density at the lens redshift, from the data
itself. For this, we employ the mass-concentration relation of NFW haloes
from \citet{dsk08}.

\subsubsection{Constraining the anisotropic galaxy-galaxy lensing signal}

The formalism to study the anisotropic weak lensing shear field around
elliptical lenses was introduced by \citet{nar00} for the case of an
elliptical isothermal sphere, and further developed and generalised for
other density profiles in \citetalias{mhb06}.
Here we largely follow the notation from \citetalias{mhb06}, and introduce a few additional
quantities.

Similarly to \citetalias{mhb06}, we model the 
 stacked and scaled tangential
 shear field
as a combination of an isotropic 
profile
 $\Delta \Sigma_\mathrm{iso}(r)$ and some azimuthal variation as
\begin{equation}
\Delta \Sigma_\mathrm{model}(r,\Delta\theta)=\Delta \Sigma_\mathrm{iso}(r)\left[1+4 f_\mathrm{rel}(r) |e_\mathrm{h,a}| \cos(2\Delta\theta)\right] \,.
\label{eq:sigmamodel_eh}
\end{equation}
Here,  $\Delta\theta$ denotes the position angle with respect to the
major axis of the lens galaxy. 
We do not know the orientations of the matter halo ellipticities
$e_\mathrm{h}$ on the sky.
Thus, 
we have to approximate them with the orientations of their corresponding
galaxy ellipticities $e_\mathrm{g}$ when stacking the anisotropic shear field.
Accordingly, our analysis is only sensitive to the average component
\begin{equation}
|e_\mathrm{h,a}| = \langle \cos(2\Delta\phi_\mathrm{h,g}) |e_\mathrm{h}|
\rangle \simeq  \langle \cos(2\Delta\phi_\mathrm{h,g})
\rangle |e_\mathrm{h}|
\label{eq:eha}
\end{equation}
of the halo ellipticity that is aligned with the galaxy ellipticity, where
$\Delta\phi_\mathrm{h,g}$ indicates the misalignment angle.
Here we average over the misalignment distribution, 
  which we assume does not depend on $|e_\mathrm{h}|$.
Following \citetalias{mhb06}, we make the assumption that the absolute value of the halo
ellipticity is proportional to the absolute value of the galaxy ellipticity
\begin{equation}
|e_\mathrm{h}|=\tilde{f_\mathrm{h}}|e_\mathrm{g}| \,.
\label{eq:eheg}
\end{equation}
While there will be deviations from this assumed linear scaling in reality,
it provides a reasonable approximate weighting scheme \citep[see also][who
explore additional schemes]{uhs12}. 

In equation (\ref{eq:sigmamodel_eh}), $f_\mathrm{rel}(r)$ 
describes the relative
asymmetry in the shear field for an elliptical halo of ellipticity
$|e_\mathrm{h,a}|$.
It depends on the assumed density profile and needs to be computed
numerically for non-power-law profiles (see \citetalias{mhb06})\footnote{Our tests conducted in Sect.\thinspace\ref{su:testsims} indicate that
there is only a weak dependence of
$f_\mathrm{rel}(r)$ on the halo ellipticity  itself, which can be ignored for
the expected halo ellipticities.}.
To recover  the notation of \citetalias{mhb06}, we define
\begin{eqnarray}
f_\mathrm{h}&=&\tilde{f_\mathrm{h}}\langle \cos(2\Delta\phi_\mathrm{h,g}) \rangle=
\frac{|e_\mathrm{h}|}{|e_\mathrm{g}|} \langle\cos(2\Delta\phi_\mathrm{h,g})\rangle\,, \label{eq:fhfromeg}\\
f(r)&=&f_\mathrm{rel}(r) f_\mathrm{h}\,. \label{eq:frfsfh}
\end{eqnarray}
Then, equation (\ref{eq:sigmamodel_eh}) reduces to
\begin{equation}
\Delta \Sigma_\mathrm{model}(r,\Delta\theta)=\Delta \Sigma_\mathrm{iso}(r)\left[1+4 f(r) |e_\mathrm{g}| \cos(2\Delta\theta)\right] \,.
\label{eq:sigmamodel}
\end{equation}
\citetalias{mhb06} show that the joint solution for the estimators of  the
isotropic and anisotropic shear field components is given by
\begin{eqnarray}
\widehat{\Delta \Sigma_\mathrm{iso}}(r) & = & \frac{\sum_{i} w_i \Sigma_{\mathrm{c},i}^{-1}e_{\mathrm{t},i}}{\sum_i w_i \Sigma_{\mathrm{c},i}^{-2}}\,,\label{eq:estimate_iso}\\
\widehat{f(r) \Delta \Sigma_\mathrm{iso}(r)} &= & \frac{\sum_{i} w_i \Sigma_{\mathrm{c},i}^{-1}e_{\mathrm{t},i} |e_{\mathrm{g},i}|\cos(2\Delta\theta_i)}{4 \sum_{i} w_i \Sigma_{\mathrm{c},i}^{-2} |e_{\mathrm{g},i}|^2\cos^2(2\Delta\theta_i)}\, ,
\label{eq:fdeltasigma_estimate}
\end{eqnarray}
 where the summation is again over lens--source pairs in a separation interval around $r$.
Note that the factor 2 difference in equations (\ref{eq:sigmamodel}) and
(\ref{eq:fdeltasigma_estimate}) compared to equations (4) and (6) in \citetalias{mhb06}
originates from the different ellipticity definition used by \citetalias{mhb06}.
To ease the comparison to \citetalias{mhb06} we decided to not rescale $f(r)$, but rather to
write out the factor 2 difference explicitly.

In practice, (\ref{eq:fdeltasigma_estimate}) is not a useful estimator for
constraining halo ellipticity as it is susceptible to a systematic signal
if the ellipticities of lenses and sources are aligned because of 
an additional effect.
This could arise from instrumental systematics such as imperfectly corrected PSF
anisotropy, but also from cosmic shear by structures in front of the lens.
This can easily be understood: For example, an intrinsically round lens
(with an isotropic halo)
would appear elliptical because of this extra shear or systematic.
Sources would also have an extra shear component parallel to the lens ellipticity. 
In the coordinates defined by the observed lens ellipticity this appears as
an increased shear along the lens minor axis and a decreased shear along the lens
major axis. Accordingly, this would be interpreted as an ``anti-aligned'' halo
with \mbox{$f_\mathrm{h}<0$}.

To cancel this systematic contribution, \citetalias{mhb06} suggest to 
 include an additional term in the
estimator 
that 
 is based on
 the ellipticity cross component $e_\times$ (\ref{eq:ex}),
and which 
is given by
\begin{equation}
\widehat{f_{45}(r) \Delta \Sigma_\mathrm{iso}(r)} =  - \frac{\sum_{i} w_i \Sigma_{\mathrm{c},i}^{-1}e_{\times,i} |e_{\mathrm{g},i}|\sin(2\Delta\theta_i)}{4 \sum_{i} w_i \Sigma_{\mathrm{c},i}^{-2} |e_{\mathrm{g},i}|^2\sin^2(2\Delta\theta_i)}\, .
\label{eq:fdeltasigma_estimate45}
\end{equation}
Equations (\ref{eq:sigmamodel}) and (\ref{eq:fdeltasigma_estimate45}) obtain nearly equal
contributions from 
systematic
effects aligning the lens and source
ellipticities, as long as the shear correlation function
\mbox{$\xi_-(r)=\langle \tilde{\gamma}_\mathrm{t} \tilde{\gamma}_\mathrm{t}-  \tilde{\gamma}_\times   \tilde{\gamma}_\times\rangle (r)$} is sufficiently small, where $\tilde{\gamma}$ indicates the additional ``systematic'' shear. This is the case at
the relevant small $r$ (see our test for cosmic shear in
Sect.\thinspace\ref{sec:millennium}, and the discussion in \citetalias{mhb06}).
Hence, the estimator \mbox{$\widehat{\left(f(r)-f_{45}(r)\right)\Delta
  \Sigma_\mathrm{iso}(r)}$}  can be used to probe halo ellipticity free from
the systematic contribution.

 Importantly,  $f_{45}(r)$ also contains signal from the flattened
halo, so that both  $f(r)$ and $f_{45}(r)$ need to be modelled.
Here we scale the model for $f_{45}(r)$ 
in correspondence to equation (\ref{eq:frfsfh}) as
\begin{equation}
f_{45}(r)=f_\mathrm{rel,45}(r) f_\mathrm{h}\,.
\label{eq:f45fs45fh}
\end{equation}
\citetalias{mhb06}  compute predictions for 
$f_\mathrm{rel}(r)$ and $f_\mathrm{rel,45}(r)$
for several elliptical
density profiles numerically.
We restrict our analysis to elliptical NFW 
profiles as they provide
good fits to the isotropic shear signal of the CFHTLenS data within the considered radial range,
see Sect.\thinspace\ref{se:signal_errors_fitrange}.
For NFW density profiles \citetalias{mhb06} compute model predictions in terms of
$r/r_\mathrm{s}$, where $r_\mathrm{s}$ is the NFW scale radius (we linearly
interpolate between the discrete values provided by \citetalias{mhb06}). Accordingly, 
$f_\mathrm{rel}(r)$ and $f_\mathrm{rel,45}(r)$ 
 can be computed for arbitrary
NFW profiles and halo masses.
We verify the numerical predictions from \citetalias{mhb06} 
in
Sect.\thinspace\ref{su:testsims}, where we find very good agreement with \citetalias{mhb06}
for $f_\mathrm{rel}(r)$.
However,
our analysis  indicates an opposite sign for 
$f_\mathrm{rel,45}(r)$ 
compared to the \citetalias{mhb06} models\footnote{We note that different sign
  definitions exist in the literature for equation (\ref{eq:ex}), which might
  have led to an inconsistent model prediction derived in \citetalias{mhb06}.} (see Sect.\thinspace\ref{su:testsims}).
Hence, while we use the model interpolation scheme from \citetalias{mhb06},
we implicitly use an 
opposite sign for $f_\mathrm{rel,45}(r)$.

\subsubsection{Estimating $f_\mathrm{h}$}
\label{se:estimate_fh}

It is a primary goal of our analysis to constrain the aligned
ellipticity ratio $f_\mathrm{h}$.
From the estimators in equations (\ref{eq:estimate_iso}),
(\ref{eq:fdeltasigma_estimate}), and (\ref{eq:fdeltasigma_estimate45}), and
from the model descriptions in equations (\ref{eq:frfsfh}) and
(\ref{eq:f45fs45fh}),
one could be tempted to define an estimator as
\begin{equation}
\widehat{f_\mathrm{h}^\mathrm{biased}}(r)=\frac{\widehat{y}(r)}{\widehat{x}(r)}
\label{eq:fhest_biased}
\end{equation}
with
\begin{eqnarray}
\widehat{y}(r)&=&
\frac{1}{f_\mathrm{rel}(r)-f_\mathrm{rel,45}(r)}\widehat{\left(f-f_{45}\right)\Delta
  \Sigma_\mathrm{iso}(r)}\,,\\
\widehat{x}(r)&=&\widehat{\Delta
  \Sigma_\mathrm{iso}}(r)\,.
\end{eqnarray}
However, the estimator (\ref{eq:fhest_biased}) would be biased due to the
occurrence of the noisy $\widehat{x}(r)$ in the denominator.
In addition, we want to combine the estimates from the different radial
bins, and possibly also different galaxy samples.
One solution could be to divide by the best-fit model for $\Delta
  \Sigma_\mathrm{iso}(r)$, e.g. obtained  from an NFW shear profile fit.
However, here we adopt the alternative approach suggested by \citetalias{mhb06}, which is based on 
\citet{bli35a,bli35b} and \citet{fie54}:
We are interested in the ratio \mbox{$m=y/x$} of two random variables. 
In our case, $m$ corresponds to $f_\mathrm{h}$ assuming
$f_\mathrm{rel}(r)$ and $f_\mathrm{rel,45}(r)$
model the relative asymmetry
  in the shear field well.
We assume that  $y$
and $x$ have a Gaussian distribution, which is a reasonable approximation in
galaxy-galaxy lensing given the dominant shape noise.
We have multiple estimators $\widehat{y_i}$,  $\widehat{x_i}$ from the different
radial bins.
Also, in some cases we want to combine the constraints from multiple galaxy
samples originating from different redshift or stellar mass bins, which
provide additional  $\widehat{y_i}$,  $\widehat{x_i}$.
For each $i$, the quantity \mbox{$\widehat{y_i}-m\widehat{x_i}$} is a Gaussian
random variable drawn from an $N(\mu=0,\sigma^2=\tilde{w}_i^{-1})$ distribution with \mbox{$\tilde{w}_i^{-1}=\sigma_{\widehat{y}_i}^2+m\sigma_{\widehat{x}_i}^2$}.
Accordingly, the summation 
\begin{equation}
\frac{\sum_i\tilde{w}_i(\widehat{y_i}-m\widehat{x_i})}{\sum_i\tilde{w}_i}\sim N\left(0,\frac{1}{\sum_i\tilde{w}_i}\right)
\end{equation}
over all measurements is also a Gaussian random variable, where the
distribution is taken at fixed $m$.
We then determine frequentist confidence intervals at the $Z\sigma$ level 
as
\begin{equation}
\frac{-Z}{\sqrt{\sum_i\tilde{w}_i}}<\frac{\sum_i\tilde{w}_i(\widehat{y_i}-m\widehat{x_i})}{\sum_i\tilde{w}_i} <\frac{Z}{\sqrt{\sum_i\tilde{w}_i}}\,.
\end{equation}
By stepping through a grid in $m$ we identify the best-fitting
value
 that provides the desired estimator 
\begin{equation}
\widehat{f_\mathrm{h}}=m(Z=0)\,,
\end{equation}
as well as 68 per cent confidence limits  \mbox{$m(Z=\pm 1)$}.
At the best-fitting  \mbox{$m(Z=0)$} we also compute a reduced $\chi^2$ of
the fit as 
\begin{equation}
\chi^2/\mathrm{d.o.f.}=\frac{\sum_{i=1}^{i=n}\tilde{w}_i(\widehat{y_i}-m(Z=0)\widehat{x_i})^2}{n-1}\, .
\end{equation}
\replya{This approach assumes that off-diagonal covariance elements can be
  neglected. We find that this is indeed the case 
 (see Sections \ref{se:signal_errors_fitrange} and \ref{se:millen_shearcat}).}

\subsection{Shear-field of an elliptical NFW halo}

\label{su:testsims}

In order to test our analysis pipeline and the model predictions from \citetalias{mhb06},
we first analyse a simplistic simulation containing a single, elliptical NFW halo.
 A more detailed simulation is presented in
  Sect.\thinspace\ref{sec:millennium},
using data from the
Millennium Simulation and including more
realistic galaxy samples, galaxy
misalignment, and cosmic shear.

For the basic test presented here, 
we generate an isotropic NFW convergence $\kappa$ profile \citep{wrb00} on a fine $4096^2$ grid
(\mbox{$10^\prime \times 10^\prime$}) for a 
\mbox{$M_{200c}=10^{12} M_\odot/h_{70}$} halo at \mbox{$z_\mathrm{l}=0.3$} assuming the
\citet{dsk08} mass-concentration relation and sources at
\mbox{$z_\mathrm{s}=1$}. 
We shear the $\kappa$ profile along the $x$-axis of the grid with various strengths to make it
 elliptical ($e_{\mathrm{h},1}>0$). 
 Then the major and minor axes  $a$ and $b$ of
 $\kappa$  iso-density contours fulfil \mbox{$|e_\mathrm{h}|=(a-b)/(a+b)$}.
Next, we compute  the corresponding shear field using the
\citet{kas93} formalism.
To test our fitting procedure  with a regular covariance matrix we generate  50 shape noise realisations with a very
small ellipticity dispersion (arbitrarily chosen 
\mbox{$\sigma_e=0.004$}),
 which were added to the shear
field.
For this test we simply use the halo ellipticity as the lens
ellipticity, hence \mbox{$f_\mathrm{h}=1$}.

  \begin{figure}
   \centering
   \includegraphics[width=8cm]{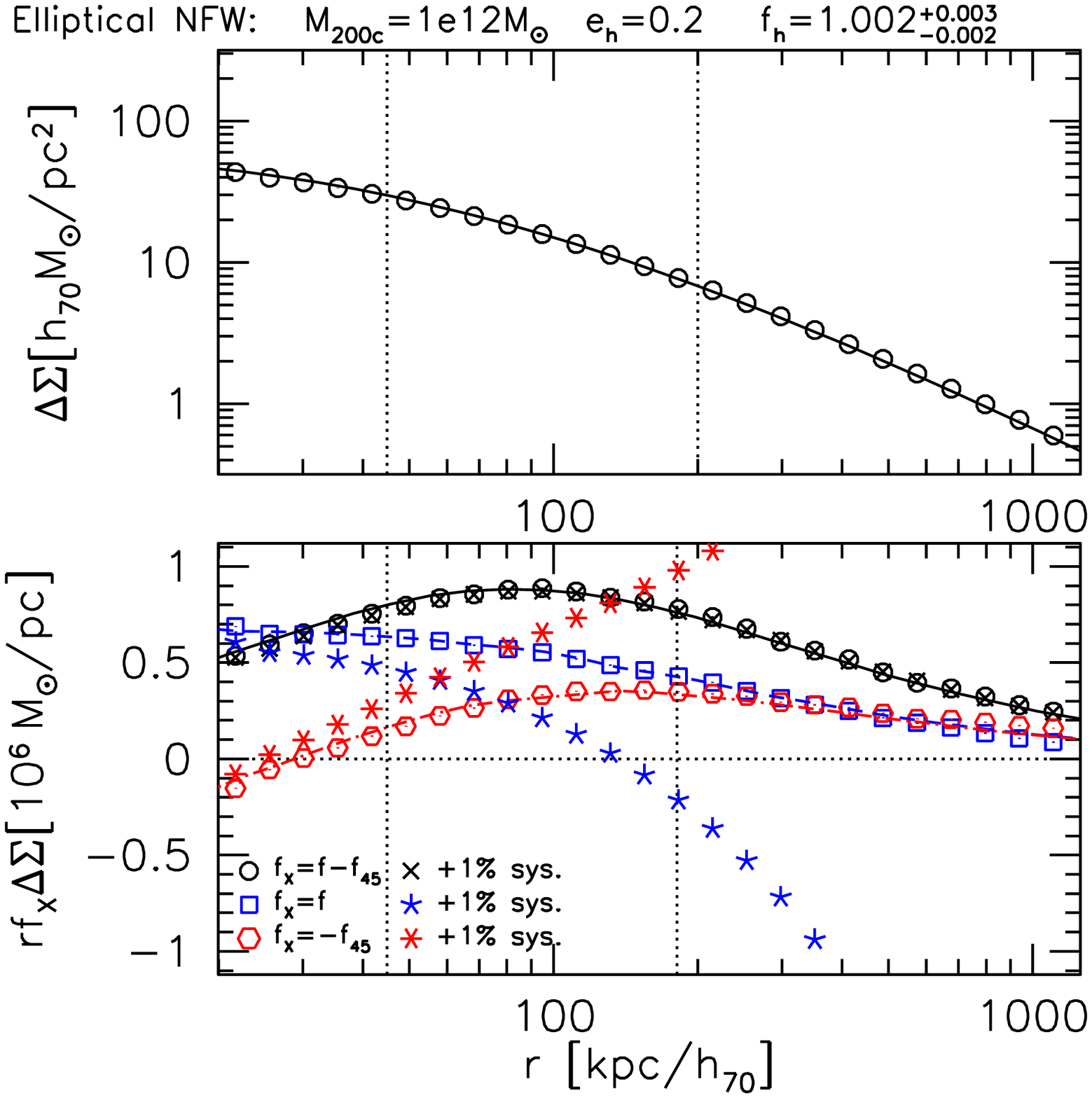}
   \caption{Test of the models and our analysis pipeline using a simple NFW  $\kappa$ profile for a 
\mbox{$M_{200c}=10^{12} M_\odot/h_{70}$} halo at \mbox{$z_\mathrm{l}=0.3$} and \mbox{$z_\mathrm{s}=1$}  sources.
The halo was sheared on a grid to an ellipticity \mbox{$e_\mathrm{h}=0.2$}
and the corresponding
shear
field was computed according to \citet{kas93}. The {\it top} and {\it bottom} panels show the
isotropic and anisotropic shear profiles, respectively. For the anisotropic signal (note the scaling by $r$)
the 
 open blue squares, red hexagons, and black circles show $f\Delta\Sigma$,
$-f_{45}\Delta\Sigma$, and $(f-f_{45})\Delta\Sigma$, respectively. 
 The curves show corresponding model
predictions from \citetalias{mhb06} for $e_\mathrm{h}=0.2$ and $f_\mathrm{h}=1$, where we have swapped the sign for the $-f_{45}$ model (see Sect.\thinspace\ref{su:method}).
 The 
additional symbols show the same measured quantities, but with an
extra constant 1\% shear added to both lens and source ellipticities
 (45$^\circ$
rotated from the lens orientation). 
The vertical dotted lines indicate the fit range.
}
   \label{fi:simnfw}
    \end{figure}

  \begin{figure}
   \centering
   \includegraphics[width=8cm]{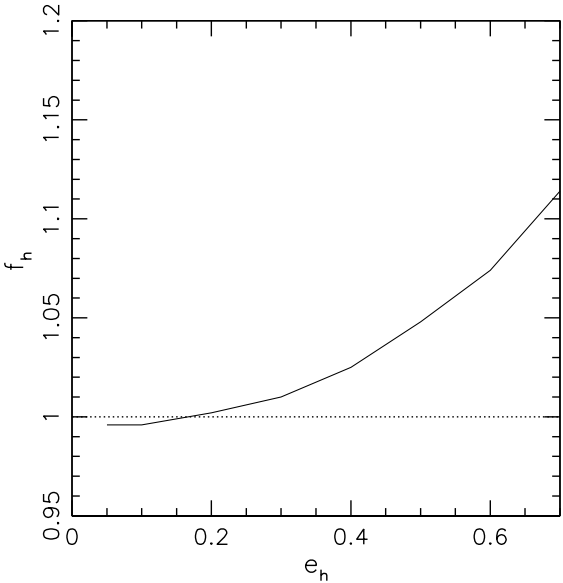}
   \caption{Test of the halo ellipticity recovery using the linearised
     elliptical shear
     field model from \citetalias{mhb06} (corrected for the sign inconsistency for
     $f_{45}$). Here the same type of elliptical NFW models was used as in
     Fig.\thinspace\ref{fi:simnfw}, but for a range of halo ellipticities
     $e_\mathrm{h}$. Perfect recovery corresponds to the dotted
     \mbox{$f_\mathrm{h}=1$} line.
 Deviations for typical halo ellipticities are only a few per cent.
   }
   \label{fi:simnfwrecover}
    \end{figure}

We plot the
 measured 
isotropic and anisotropic shear profiles for an \mbox{$|e_\mathrm{h}|=0.2$} halo
in
Fig.\thinspace\ref{fi:simnfw}.
The anisotropic signal (open symbols in the bottom panel) is well described by the best-fitting model using the
\citetalias{mhb06} prescription once we account for the sign error for
$f_{45}$ (see Sect.\thinspace\ref{su:method}).
Here we accurately recover the input halo ellipticity
(\mbox{$f_\mathrm{h}=1.002^{+0.003}_{-0.002}$}).
For comparison, we also show
the measured signal 
 if we artificially add a constant (scale independent) spurious alignment of
the lens and sources corresponding to a shear of
\mbox{$\gamma_2=1\%$}\footnote{If we always orient the lens for this test along the
  $x$-axis ($e_1$ only), we need to apply a systematic shear along the uncorrelated 45
  degrees-rotated field diagonal ($\gamma_2$). We verified that we obtain
  identical results with a systematic 
shear along a random direction if we randomly rotate the lens and its shear
field prior to applying the systematic shear. This would resemble reality
more closely with random intrinsic lens orientations with respect to the
orientation of the PSF or the cosmic shear field.},
to roughly illustrate the effect of residual PSF anisotropy or cosmic shear
(cross- and star-shaped symbols in the bottom panel of Fig.\thinspace\ref{fi:simnfw}). 
While the $f\Delta\Sigma$ and  $-f_{45}\Delta\Sigma$ components are individually disturbed, their
combination \mbox{$(f-f_{45})\Delta\Sigma$} is unaffected by the systematic shear as expected. 

We repeat this analysis with various halo ellipticities $e_\mathrm{h}$ and plot the
recovered \mbox{$f_\mathrm{h}(e_\mathrm{h})$} in
Fig.\thinspace\ref{fi:simnfwrecover}.
 This shows that the 
linear relation 
assumed in equations (\ref{eq:frfsfh}) and (\ref{eq:f45fs45fh})
is a good approximation, but leads to a slight
overestimation of $f_\mathrm{h}$ for very elliptical haloes.
However, as the deviations from \mbox{$f_\mathrm{h}=1$} are small 
for typical expected halo ellipticities\footnote{For example, in our analysis of haloes from the Millennium Simulation (see Sect.\thinspace\ref{sec:millennium}) that are populated by early type galaxies, we find \mbox{$\langle|e_\mathrm{h}|\rangle=0.163$}, and only 0.5\% of the haloes have \mbox{$|e_\mathrm{h}|>0.5$}.} of \mbox{$|e_\mathrm{h}|\sim 0.2$}, and given
the large statistical uncertainties of CFHTLenS, we ignore this deviation in our analysis.

\section{Observational constraints from CFHTLenS}
\label{se:data}
\subsection{The CFHTLenS data}

To constrain the aligned projected ellipticity ratio $f_\mathrm{h}$ 
observationally, we employ weak lensing data from the
Canada-France-Hawaii Telescope 
Lensing Survey (CFHTLenS).
It has an
effective area of 154 deg$^2$ 
imaged in the
$ugriz$ broad band filters using MegaCam on CFHT
with a
$5\sigma$ limiting magnitude  
in the detection $i$-band and $2^{\prime\prime}$ apertures  of $i_\mathrm{AB} \sim 24.5-24.7$ \citep{ehm13}. 
These data were obtained as part of the 
Wide-component of the Canada-France-Hawaii Telescope (CFHT) Legacy Survey which completed observations in early
2009. 

The CFHTLenS team has reduced the imaging data using the \textsc{THELI} pipeline \citep{ehl09,ehm13},  measured PSF-corrected galaxy shapes in the $i$-band 
using \emph{lens}fit \citep{mkh07,mhk13a,kmh08}, and estimated photometric
redshifts (photo-$z$s) from the $ugriz$ data
employing the \textsc{BPZ}
algorithm \citep{ben00,cbs06} 
as detailed in \citet{hek12},
 with photo-$z$ tests presented in \citet{bwh13}.
Details on the CFHTLenS analysis pipeline and
cosmology-independent 
systematic tests are presented in \citet{hwm12}.

\replya{To subdivide the foreground lens galaxies in our analysis we employ
 stellar mass estimates computed using \textsc{LePhare} \citep{acm99,iam06}. 
They are based on the stellar population
 synthesis (\textsc{SPS}) package of \citet{brc03} and assume a stellar
initial mass function from \citet{cha03}.
 From this we
 built 18 templates with two different metallicities and nine
 exponentially decreasing star formation rates. We allow the dust
 extinction to vary between 0.05 to 0.3 mag using a \citet{cab00} extinction
   law, 
 and 57 starburst ages ranging from 0.01 to 13.5
 Gyr \citep[for more details see Sec. 2.1 of][]{vuh14}.}


\subsection{KSB shapes for bright foreground galaxies}

The \emph{lens}fit shape measurement algorithm  has been
optimised to obtain
accurate  shape estimates  for the typically faint and only moderately   resolved
distant source galaxies \citep{mhk13a},
as needed for an 
unbiased cosmological weak lensing analysis \citep[e.g.][]{kfh13,hgh13,fke14,kha14}.
For our analysis of galaxy halo shapes we additionally require ellipticity
estimates for the  foreground lens
galaxies.
Many of these are fairly bright and extended, which  may result in an exclusion of a galaxy
during the  \emph{lens}fit shape analysis.
This can be caused either because of a size comparable
or exceeding the employed postage stamp size (48 pixels), 
or outer isophotes significantly overlapping with neighbouring galaxies,
or the presence of substantial substructure in the galaxy, which is not well described by
the employed bulge+disk model \citep[see][]{mhk13a}.

In order to not exclude these galaxies from our sample of lens galaxies,
we 
 perform shape measurements using the KSB algorithm
originally proposed
by \citet{ksb95} and \citet{luk97}. 
Based on weighted brightness moments this shape measurement algorithm 
is more robust to the presence of resolved substructure or nearby galaxies.
Here we employ the implementation of the algorithm detailed in 
\citet{hfk98} and \citet{hfk00}, which was tested in the blind challenges 
of the STEP project \citep{hwb06,mhb07} and also employed for 
earlier weak lensing analyses of CFHTLS data \citep{hmw06}.
While there are indications for remaining residual systematics in earlier
cosmological weak lensing studies of CFHTLS data using KSB \citep{kbg09,hwm12},
we note that these are expected to mostly originate from poorly resolved,
low
signal-to-noise galaxies. 
In contrast, the studied lens galaxies 
have high signal-to-noise\footnote{In the highest-redshift redshift slice
  (\mbox{$0.5<z<0.6$}), the median signal-to-noise ratio defined as
 FLUX\_AUTO/FLUXERR\_AUTO from SExtractor is 181 (193) for the blue (red)
 lenses in the lowest stellar mass bins considered.} and are well-resolved.
Thus, they are less sensitive to noise-related biases \citep[e.g.][]{mev12,rka12,kzr12,vkj14} and require smaller PSF corrections.
 Also see \citet{shc15} where the KSB results for bright galaxies are compared to GALFIT shapes \citep{phi02}.
Furthermore, as demonstrated in Sect.\thinspace\ref{su:testsims},
additional ellipticity correlations between lenses and sources due to imperfect
PSF anisotropy correction or cosmic shear cancel out in
the analysis at the relevant scales thanks to the employed estimator.
Thus, the application of KSB shapes for lens galaxies without  \emph{lens}fit shapes
does not compromise the systematic accuracy of our measurement, but only 
increases the statistical precision.
Note that differing from earlier KSB studies of CFHTLS data we conduct
galaxy shape measurements on individual exposures and not stacks, 
similarly to the \emph{lens}fit analysis, and combine the shape estimates on
the catalogue level  as the weighted mean estimate.

\subsection{Lens sample}
\label{sec:cfhtlens_lenssample}
To obtain  a sample of foreground lens galaxies 
we  preselect relatively bright objects (\mbox{$i<23.5$})
that are resolved
(\mbox{$\mathrm{CLASS\_STAR<0.5}$}, \mbox{$\mathrm{star\_flag}=0$}, non-zero
  shape weights from \emph{lens}fit or KSB)
and feature a single-peaked
photometric redshift probability distribution function
\citep[$\mathrm{ODDS}>0.9$, see][]{hek12}. 
We select lenses within the range of best-fitting photometric redshifts
\mbox{$0.2<z_\mathrm{b}<0.6$}, which we subdivide into four lens redshift
slices of width \mbox{$\Delta z_\mathrm{b}=0.1$}.
We split the galaxies into
red (\mbox{$T_\mathrm{BPZ} \le 1.5$}) and blue lenses
(\mbox{$1.5<T_\mathrm{BPZ} < 3.95$}) using the photometric type
$T_\mathrm{BPZ}$ from \textsc{BPZ}.
In order to approximately sort the lenses by halo mass and optimise the
measurement 
signal-to-noise ratio,
 we also subdivide them according to stellar mass
\mbox{$\log_{10}M_*$}
as detailed in Table \ref{tab:lenses}. 
When measuring the anisotropic shear signal, contributions from different lenses are weighted according to the lens ellipticity (see e.g.~Eq.~\ref{eq:fdeltasigma_estimate}). 
Here we restrict the analysis to lenses in the well-constrained ellipticity range \mbox{$0.05<|e_\mathrm{g}|<0.95$}.

An important aspect of our study is the direct comparison of our
measurements from CFHTLenS to results from mock data based on the Millennium
Simulation (see Sect.\thinspace\ref{sec:millennium}). For the mock data we only have measurements for central
haloes but not for satellites (see Sect.\thinspace\ref{se:sims_mr_lenses}). 
To ensure that the results are comparable, we aim to minimise the fraction of
satellites in our CFHTLenS lens samples.
In addition, our model assumes that the anisotropic shear signal 
is caused by isolated elliptical NFW mass distributions (Sect.\thinspace\ref{se:methodandtests}).
This may be a reasonable approximation for centrals, but it is likely
a poor descriptions for  satellites embedded into a larger halo.

\citet{vuh14}  fit the
isotropic galaxy-galaxy weak lensing signal
 around red and blue lenses in CFHTLenS using a halo model approach. 
They find that the satellite fraction is typically low both for blue lenses
and  for red lenses which have a high stellar mass.
However, it increases steeply for
red lenses towards lower stellar mass.
To reduce the fraction of satellites in our sample, we therefore generally
exclude red lenses with stellar mass \mbox{$\log_{10} M_*<10$}.
For the lowest stellar mass bin included in
our analysis for red lenses (\mbox{$10<\log_{10} M_*<10.5$}), 
\citet{vuh14}
estimate a satellite fraction
of
\mbox{$\alpha=0.23\pm 0.02$}.
To further reduce the fraction of satellites, we apply an
additional cut to the galaxies in this bin using the internal flag from
\textsc{SExtractor} \citep{bea96}.
This removes any galaxy
which
is 
either partially blended with another object, or which has a nearby neighbour possibly
affecting the measurement of the \texttt{MAG\_AUTO} magnitude.
Many of these galaxies are located close to a brighter early
type galaxy, as expected for satellites.
We do not filter on this flag for the other stellar mass bins. In
particular, this would exclude many bright early type galaxies, which are
presumably centrals but have  faint nearby neighbours.

\replya{
Given that the remaining fraction of satellites in our lens sample
  is low, we expect that they have a negligible impact on our
  results\footnote{\replya{A net alignment  of satellites  with respect to the tangential shear field of their parent
halo could introduce  spurious signal.
 However, we expect that this has
      a negligible impact on our study for the following reasons: First, the
      satellite fraction is low for our lens samples. Second,  \citet{shc15}
     place tight limits on the net alignment of cluster galaxies, showing that
     it must be very weak. Given our selection of lenses with relatively
      high stellar mass, we expect that many of the satellites present in the
      sample are actually cluster members. Finally, our analysis using
      $(f-f_{45})\Delta\Sigma$ reduces the impact of spurious alignment
      between lenses and sources, which also applies here to first order.}}.
As consistency check for this we investigate the impact of the lens
bin with the highest satellite fraction on our joint constraints in
Sect.\thinspace\ref{se:cfhtles_constraints_fh}.
}

\begin{table}  
\caption{The lens galaxy samples.
\label{tab:lenses}}
\begin{center}
\begin{tabular}{cccc}
\hline
\hline
Colour & Stellar mass  $[M_\odot]$& \mbox{$N(0.2\le z_\mathrm{l}< 0.6)$} & $\sigma_e$\\
\hline
Red &\mbox{$10<\log_{10} M_*<10.5$} & 81763 & 0.35\\
Red &\mbox{$10.5<\log_{10} M_*<11$} & 93032  & 0.30\\
Red &\mbox{$\log_{10} M_*>11$} & 25982 & 0.23\\
\hline
Blue &\mbox{$9.5<\log_{10} M_*<10$} & 166604 & 0.38\\
Blue &\mbox{$10<\log_{10} M_*<10.5$} & 91692 & 0.36\\
Blue &\mbox{$\log_{10} M_*>10.5$} & 33612 & 0.30\\
\hline
\end{tabular}
\end{center}
\vspace{0.2cm}

{\flushleft
Note. --- Overview over the sub-sample of lens galaxies used:
{\it Column 1:} Split between red (\mbox{$T_\mathrm{BPZ} \le 1.5$}) and blue lenses
(\mbox{$1.5<T_\mathrm{BPZ} < 3.95$})  using the photometric type
$T_\mathrm{BPZ}$ from BPZ.
{\it Column 2:} Stellar mass range.
{\it Column 3:} Number of selected lenses in the redshift
interval \mbox{$0.2\le z_\mathrm{l}< 0.6$}.
{\it Column 4:} Ellipticity dispersion of the selected lenses with \mbox{$0.05<|e_\mathrm{g}|<0.95$} combining both ellipticity components.
\\
}
\end{table}

\subsection{Source sample}
From all galaxies with
{\it lens}fit shape measurements and non-vanishing shape weights, 
we select those with a best-fitting photometric redshift in the range
\mbox{$z_\mathrm{l,upper}+0.1<z_\mathrm{b}<1.3$} as our source background sample,
where \mbox{$z_\mathrm{l,upper}$} indicates the upper limit of the corresponding 
lens redshift slice. Above the upper limit at \mbox{$z_\mathrm{b}=1.3$} the
$ugriz$ CFHTLenS photometric redshifts become unreliable \citep{hek12,bwh13} 
  with large redshift uncertainties and
  a partial contamination by low-redshift galaxies.
 Thus,
we exclude these galaxies from our analysis to be conservative.
 As done by \citet{vuh14}, we 
further optimise the separation of lenses and sources by
also  removing source galaxies whose 95\% redshift confidence regions computed by BPZ overlap with the lens redshift slice.
When we compute the galaxy-galaxy lensing signal we weight the contributions
from the individual source galaxies according to their effective geometric lensing
efficiency \mbox{$\beta^\mathrm{eff}_i=\int
  \beta(z_{\mathrm{l,c}},z_\mathrm{s})p_i(z_\mathrm{s})\mathrm{d}z_\mathrm{s}$}.
This is estimated from the full photometric redshift
probability distributions of the sources $p_i(z_\mathrm{s})$
\citep[see][]{hek12} and  uses the  centre\footnote{We use the centre of the thin lens
  redshift slice instead of the individual best-fit lens redshift 
for  computational efficiency. We note that the influence is negligible for
our study as verified by varying the width of the slices.} of
the thin lens redshift slice $z_{\mathrm{l,c}}$.

\citet[][]{mhk13a} study the impact of noise bias for \emph{lens}fit  shape measurements
  and derive an empirical correction for multiplicative
bias as a function of galaxy size and signal-to-noise ratio. 
Following \citet{vuh14} we account for the net effect of this bias on the
estimated shear profiles 
taking the individual shape weights
and lensing efficiencies of the sources into account.  
We note that this bias correction  cancels out for the halo ellipticity constraints except for its impact on the estimate of $r_\mathrm{200c}$, which  is used as upper limit for the fit range and to estimate $r_\mathrm{s}$ for the anisotropic shear field model.

\citet{hwm12} conduct a number of non-cosmological tests to flag CFHTLS fields which are likely affected by residual shape systematics. 
To be conservative, we base our primary analysis on the 129 out of 171 fields which pass these systematics tests\footnote{See
\url{http://www.cadc-ccda.hia-iha.nrc-cnrc.gc.ca/community/CFHTLens/README_catalogs_release.txt}.}
(``pass fields''), but for comparison we also provide constraints obtained from all fields.
As explained in Sect.\thinspace\ref{su:method}, 
the impact of both residual shape measurement systematics
and  foreground cosmic shear 
cancel out in our analysis as long as their 
shear correlation function $\xi_-$  is sufficiently small on the scales of interest. 
As we will see in Sect.\thinspace\ref{se:ms_cs}, this approach leaves
negligible residuals from the expected cosmic shear signal. Accordingly, 
we expect  that the fields  classified as being useful for cosmic shear
measurements can also be used to derive robust constraints on the aligned
halo ellipticity.

%
  \begin{figure*}
   \centering
  \includegraphics[width=5.8cm]{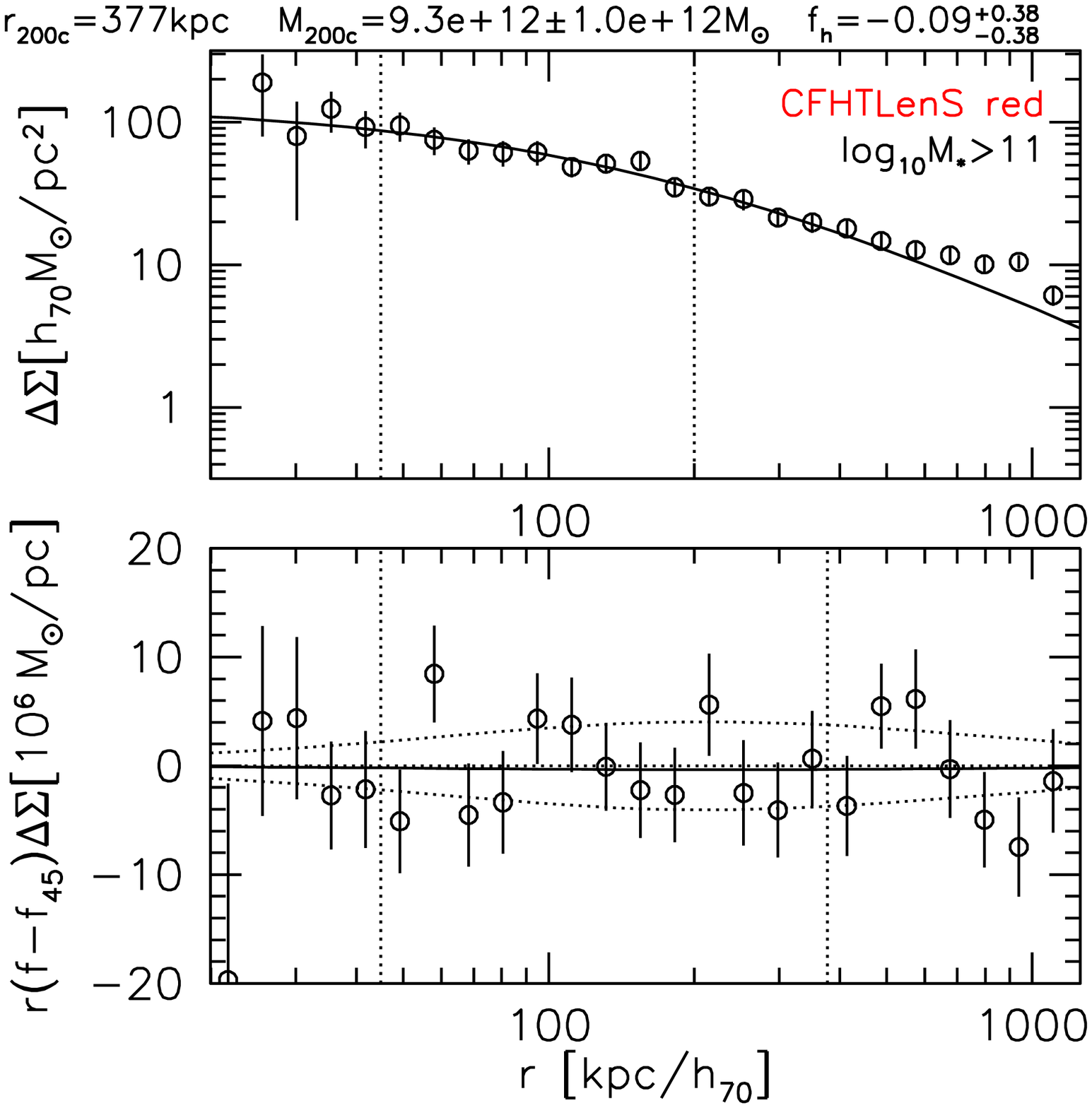}
  \includegraphics[width=5.8cm]{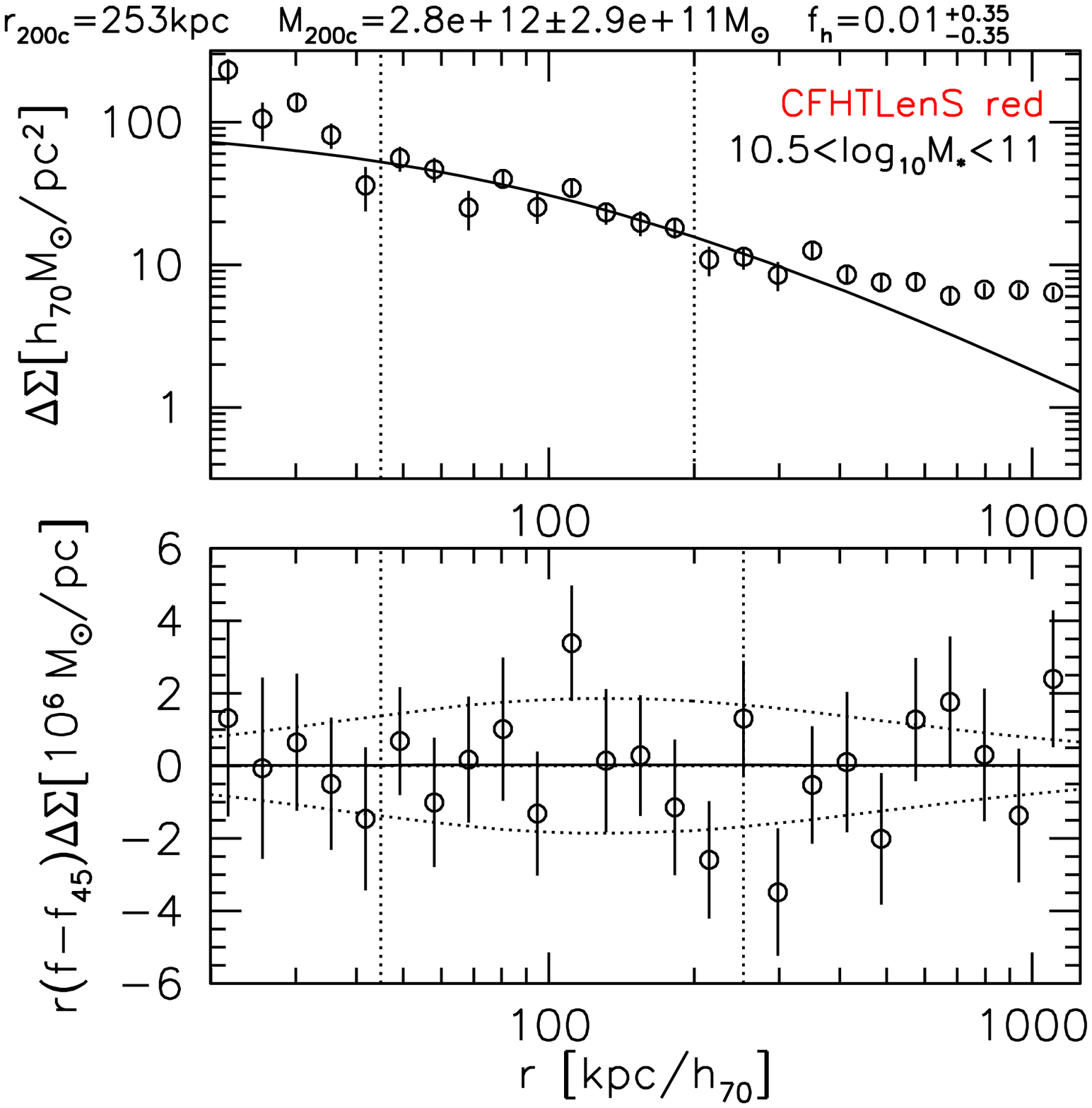}
  \includegraphics[width=5.8cm]{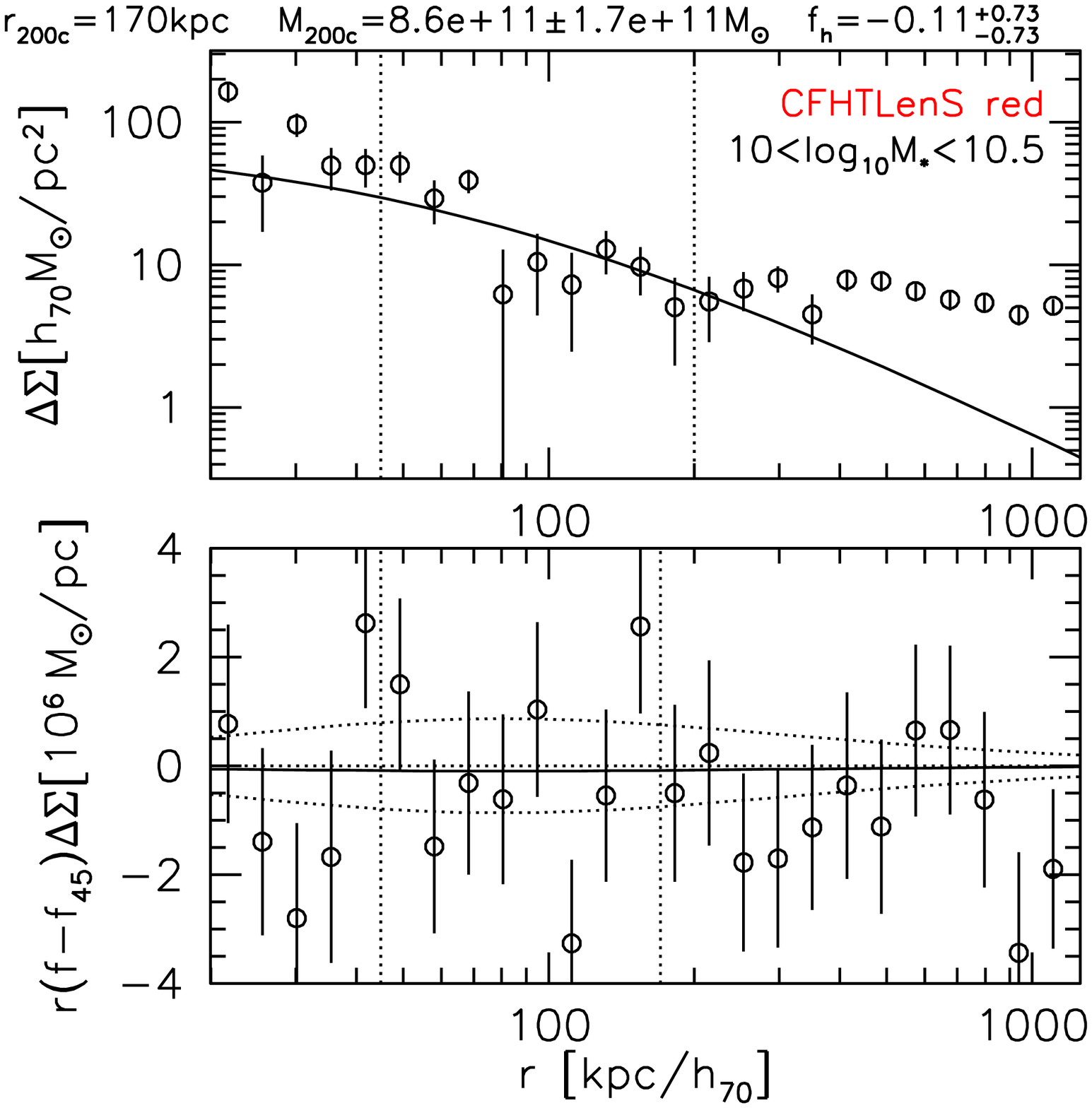}
  \includegraphics[width=5.8cm]{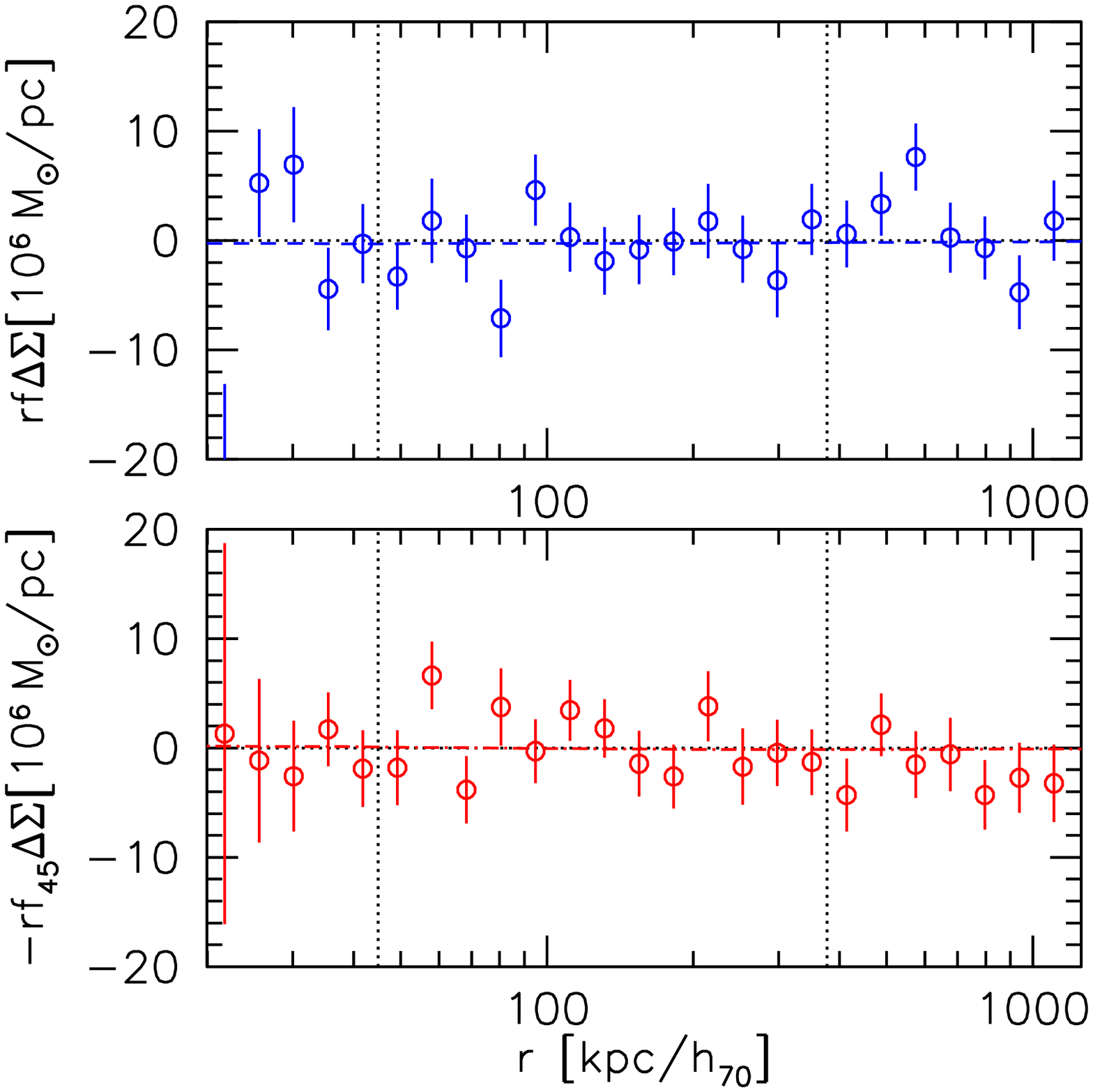}
  \includegraphics[width=5.8cm]{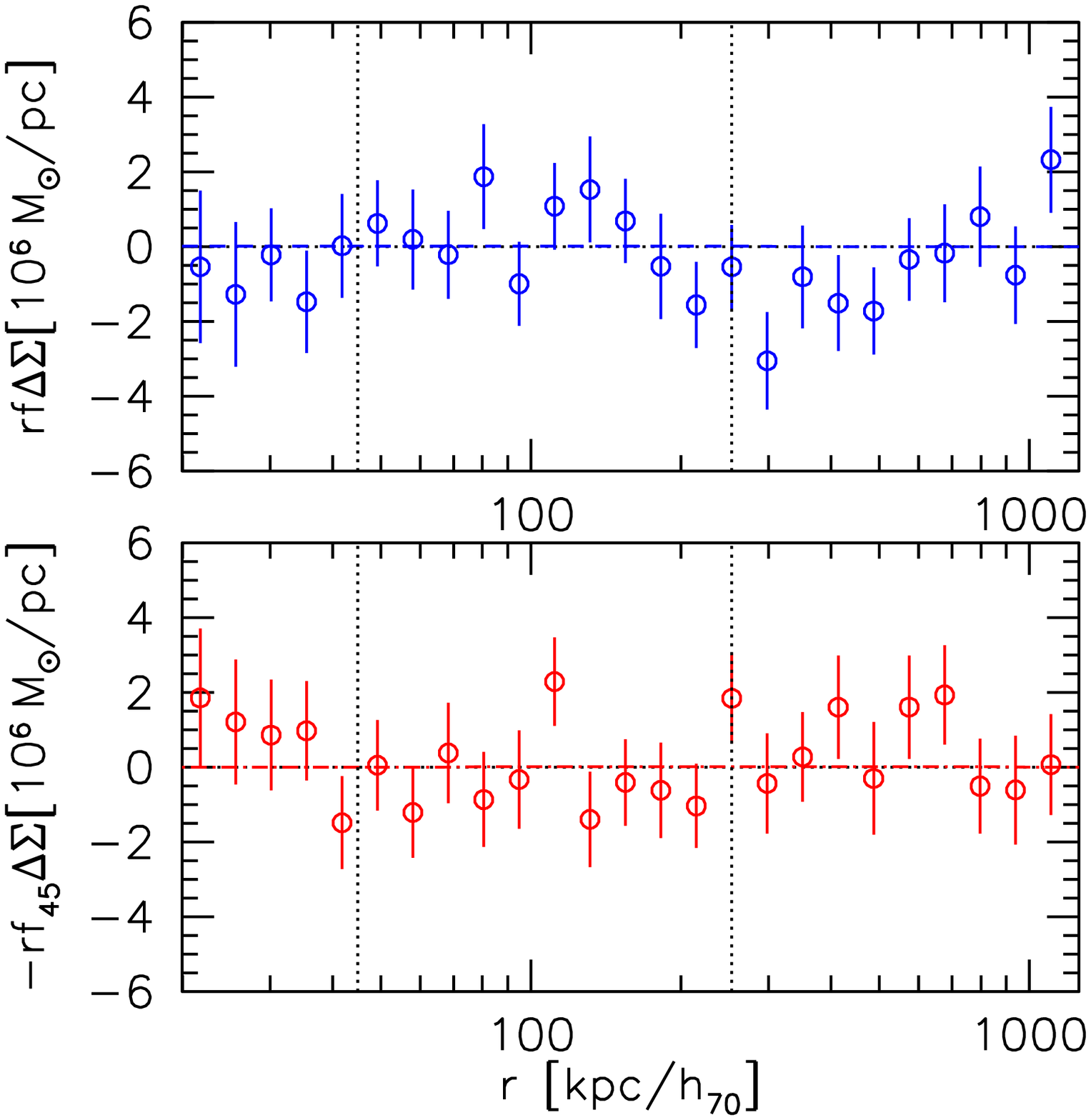}
  \includegraphics[width=5.8cm]{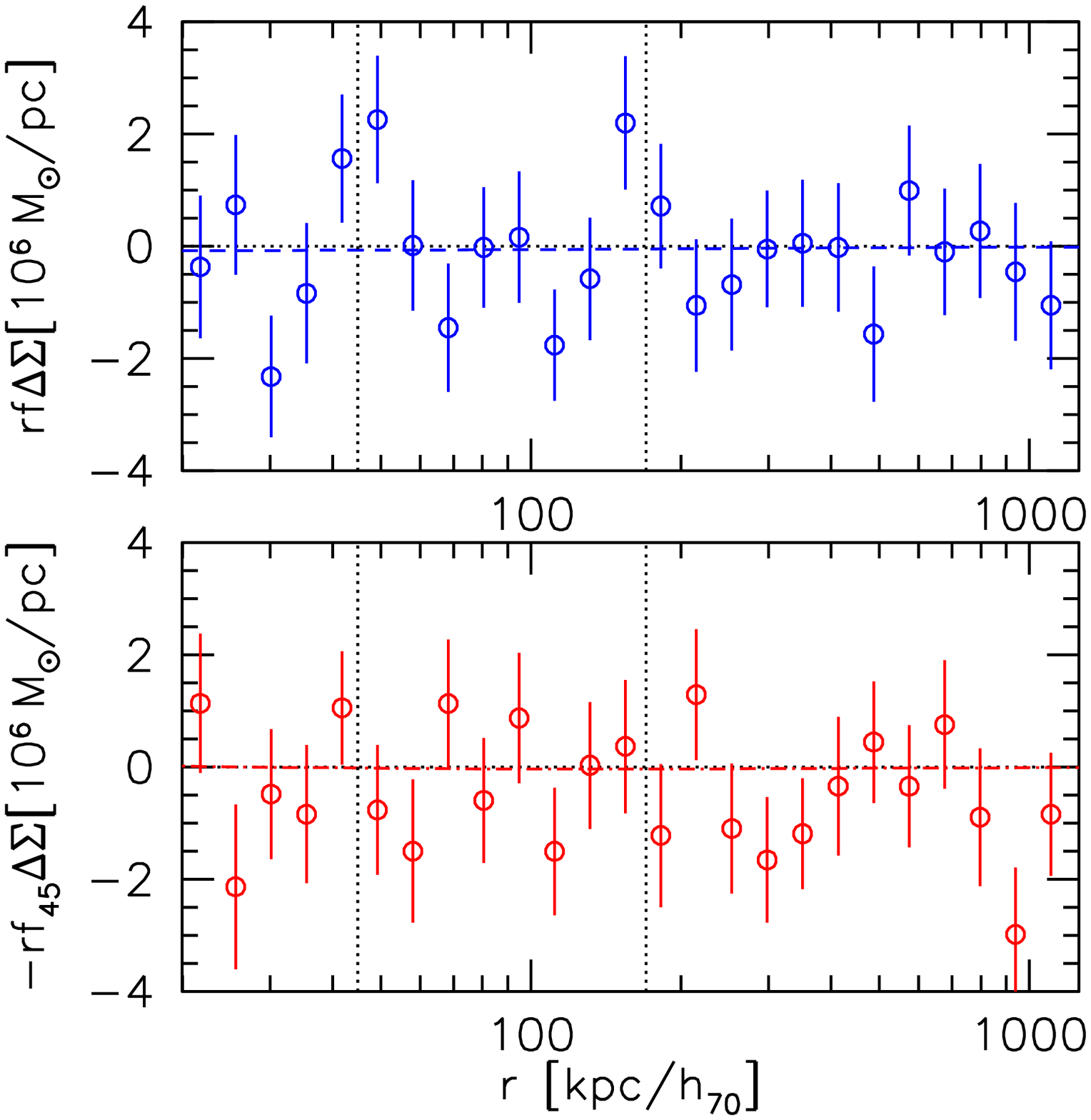}
    \caption{Measured isotropic ({\it top} row of panels) and anisotropic 
(rows two to four, note the varying $y$-axis scale)
shear signal around red lenses in 
 the CFHTLenS fields passing the systematics tests for cosmic shear
 as function of radial distance $r$. The anisotropic signal has been
scaled by $r$ for better readability,
 where rows two, three, and four show the signal components
  \mbox{$(f-f_{45})\Delta\Sigma$}, \mbox{$f\Delta\Sigma$}, and
  \mbox{$-f_{45}\Delta\Sigma$}, respectively.
From {\it left} to {\it right}, we show the stellar mass bins \mbox{$\log_{10} M_*>11$},
\mbox{$10.5<\log_{10} M_*<11$}, and \mbox{$10<\log_{10} M_*<10.5$},  combining all lens redshift slices.
 For the isotropic signal the curve shows the best-fitting NFW shear profile
 constrained within
 \mbox{$45\thinspace\mathrm{kpc}/h_{70}<r<200\thinspace\mathrm{kpc}/h_{70}$}. 
For \mbox{$(f-f_{45})\Delta\Sigma$} 
the  curves show 
 models corresponding to the  best-fit isotropic model and the
  best-fitting $f_\mathrm{h}$ (solid curves), as well as
  \mbox{$f_\mathrm{h}\in \{+1,0,-1\}$}
(dotted curves)
 for comparison.
For  \mbox{$f\Delta\Sigma$} and
  \mbox{$-f_{45}\Delta\Sigma$}, the dashed curves show model predictions for
  the best-fitting $f_\mathrm{h}$.
The best-fitting  $f_\mathrm{h}$ has been constrained from
\mbox{$(f-f_{45})\Delta\Sigma$} and \mbox{$\Delta\Sigma$}
within
 \mbox{$45\thinspace\mathrm{kpc}/h_{70}<r<r_{200\mathrm{c}}$} (indicated by
 the vertical dotted lines), with
 $r_{200\mathrm{c}}$ estimated from the fit to the isotropic signal. 
  }
   \label{fi:shearfitsred1}
    \end{figure*}

  \begin{figure*}
   \centering
  \includegraphics[width=5.8cm]{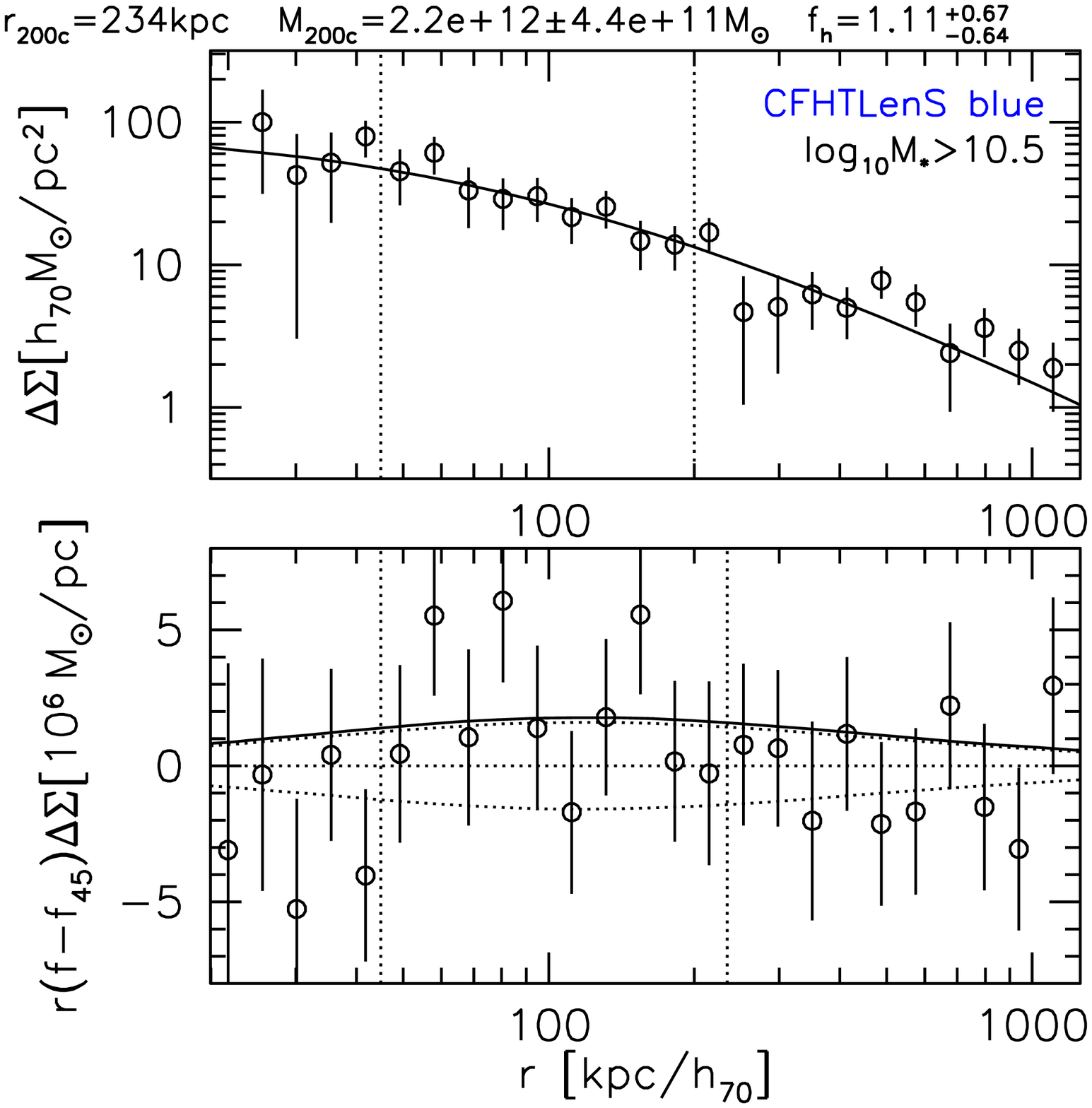}
  \includegraphics[width=5.8cm]{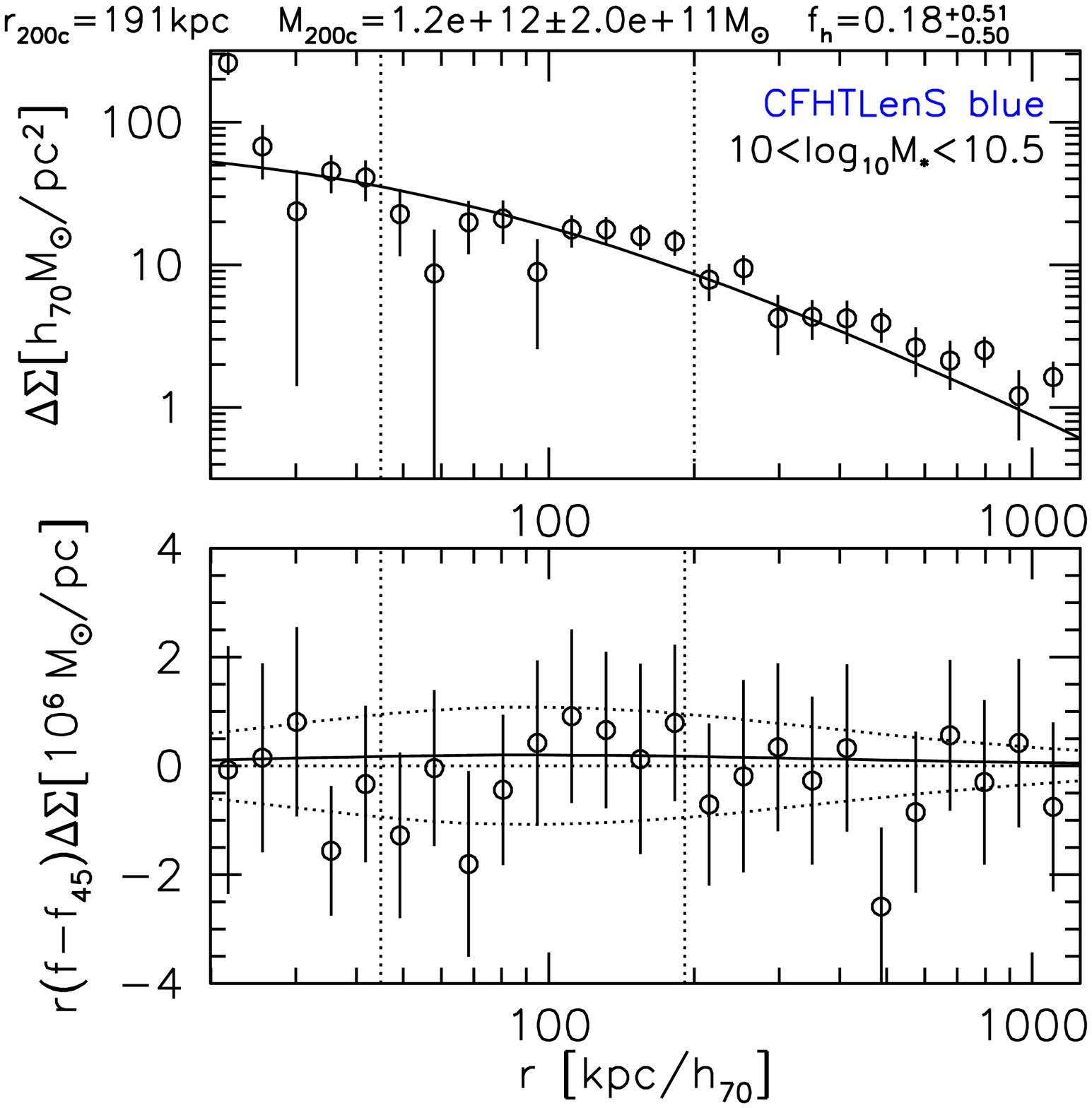}
  \includegraphics[width=5.8cm]{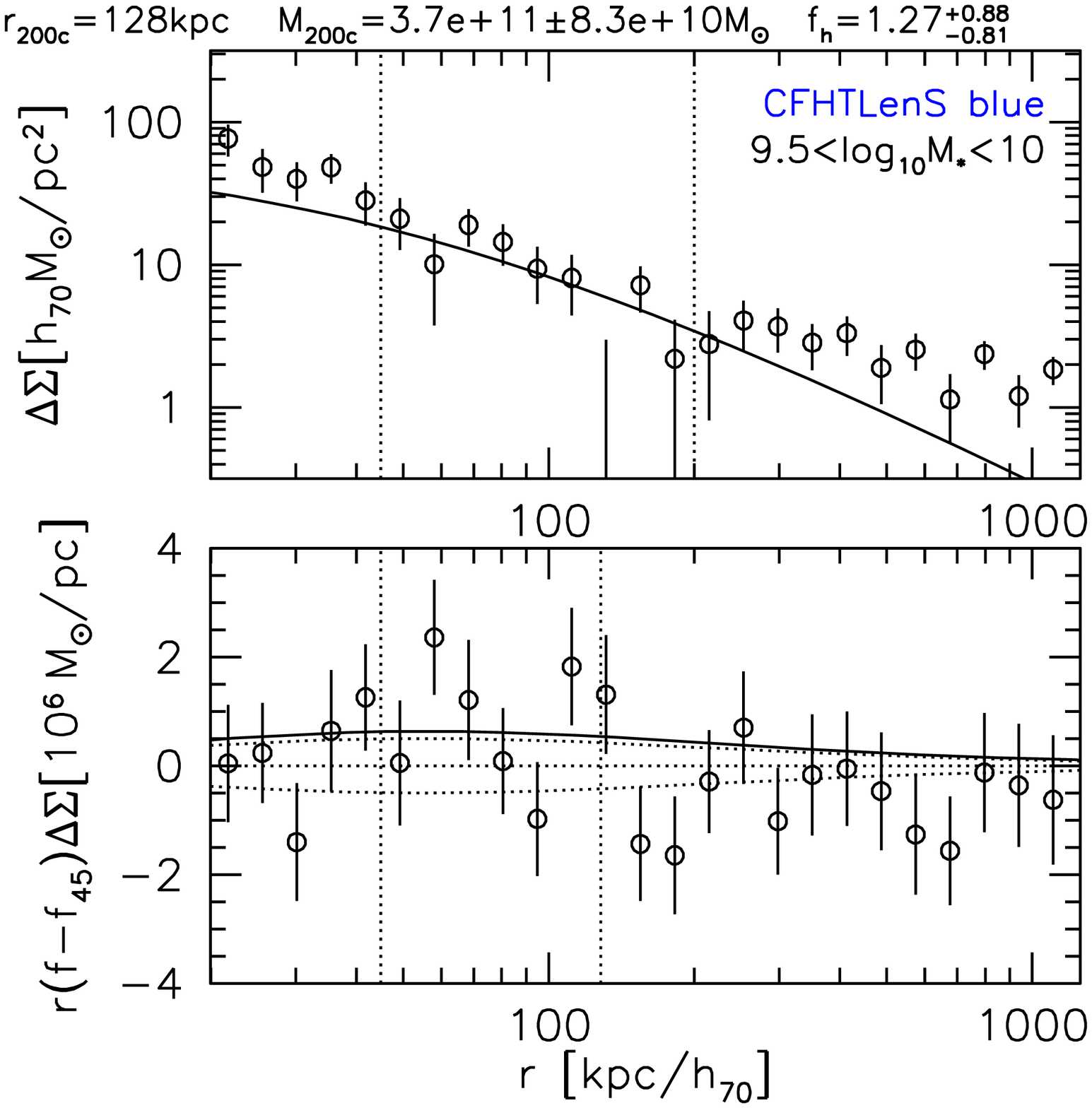}
  \includegraphics[width=5.8cm]{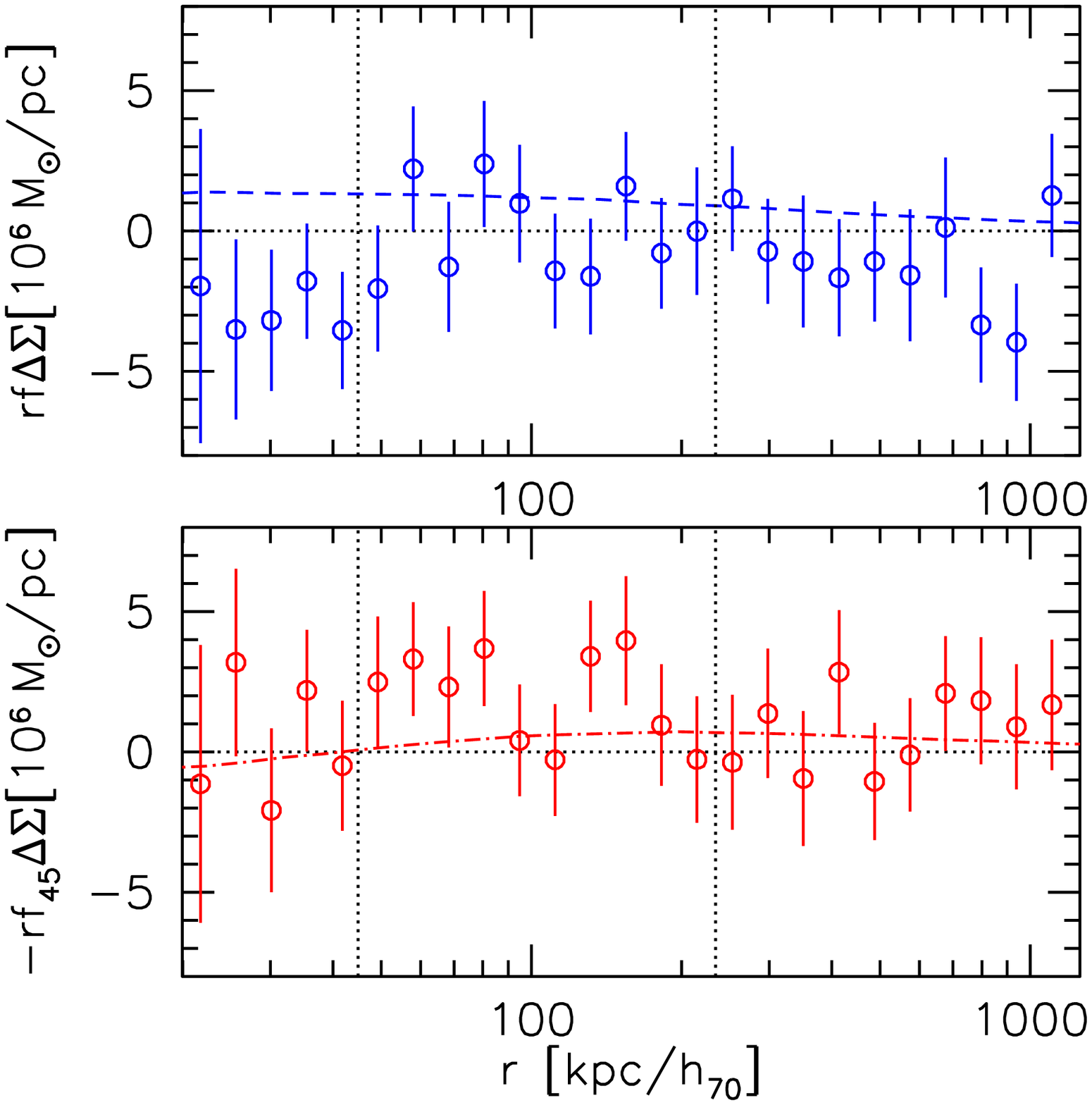}
  \includegraphics[width=5.8cm]{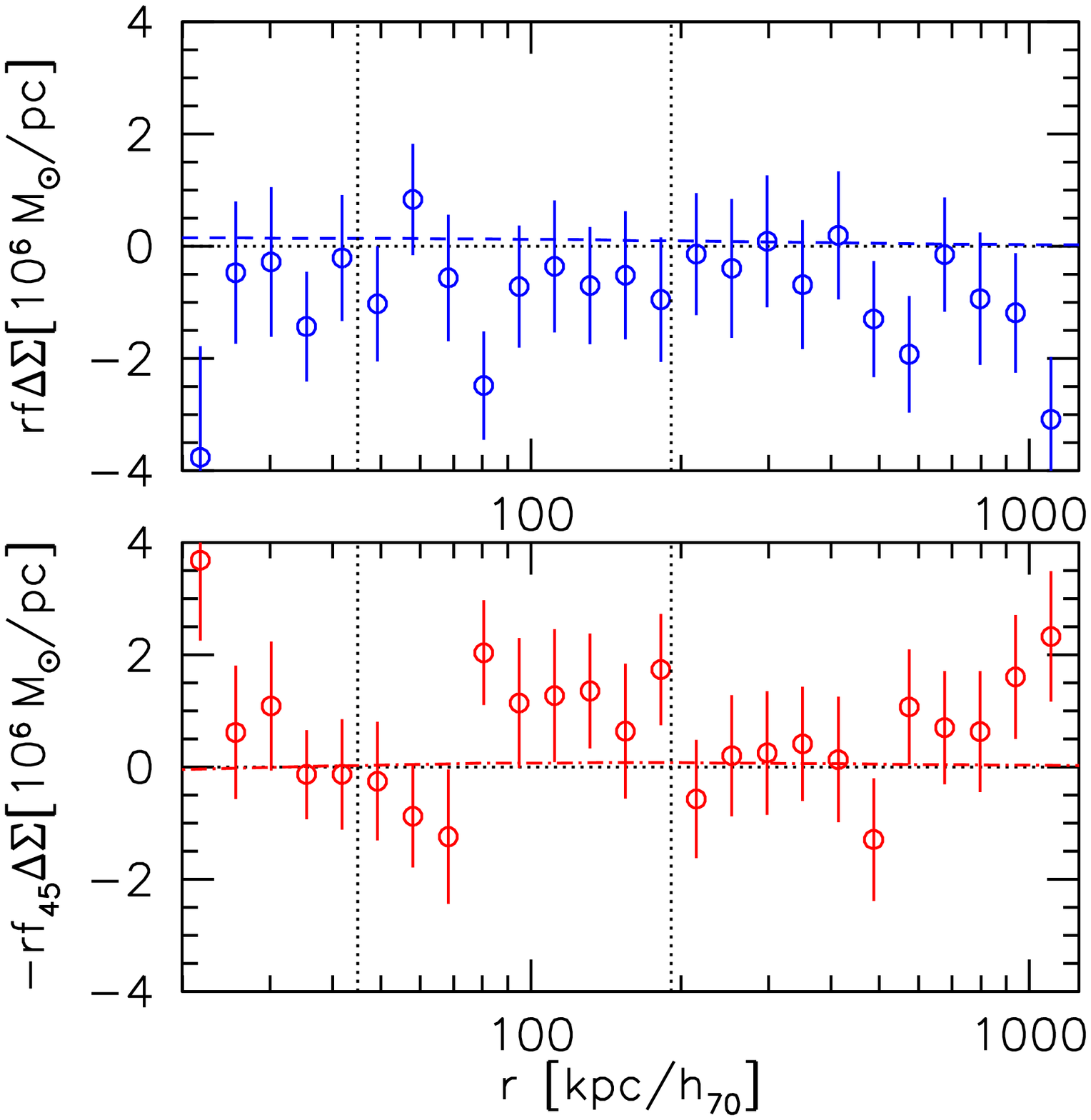}
  \includegraphics[width=5.8cm]{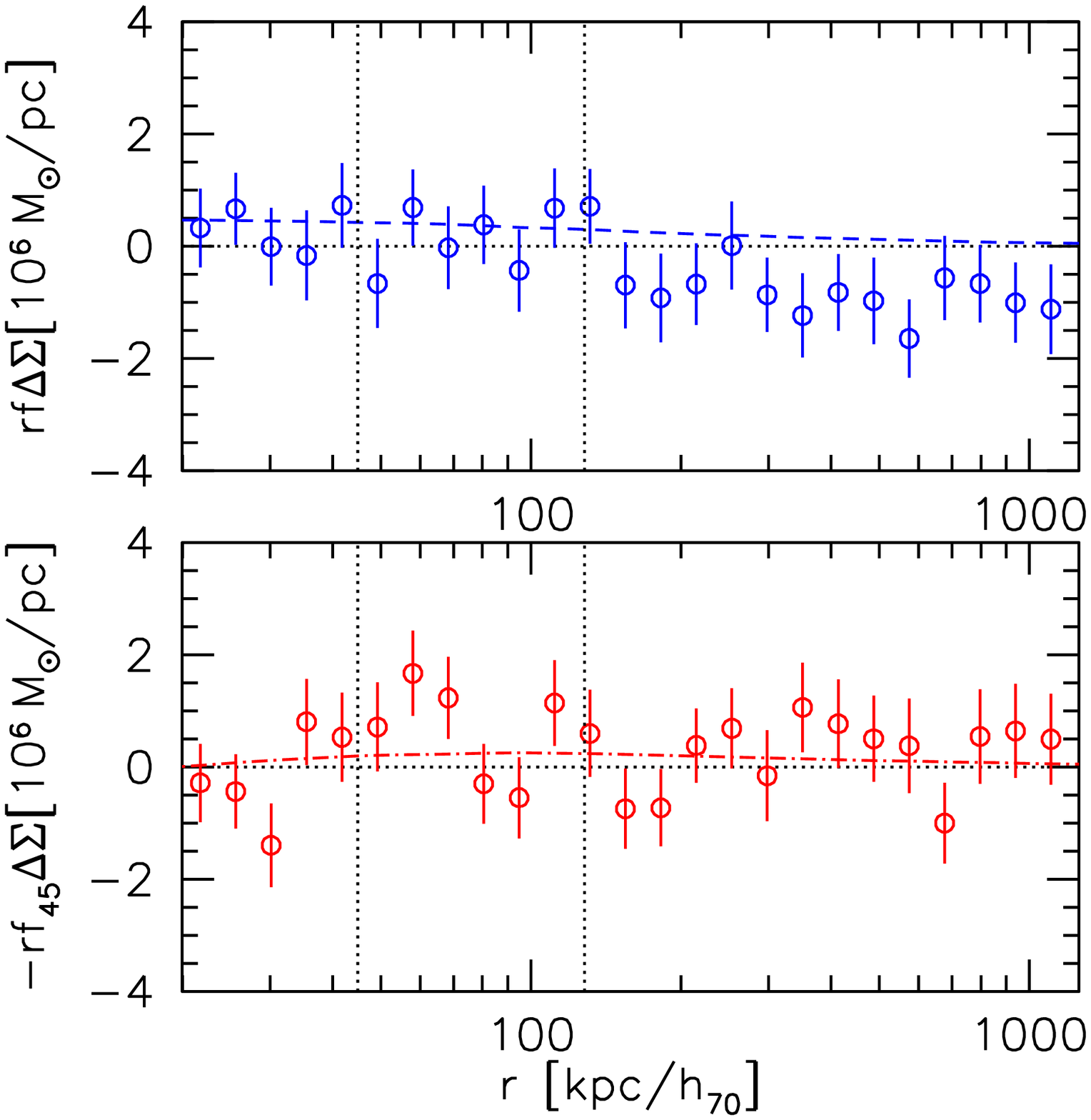}
   \caption{Measured isotropic ({\it top} panels) and anisotropic 
(rows two to four, note the varying $y$-axis scale)
shear signal around blue lenses in 
 the CFHTLenS fields passing the systematics tests for cosmic shear
as function of radial distance
$r$.
From {\it left} to {\it right}, we show the stellar mass bins
\mbox{$\log_{10} M_*>10.5$}, 
\mbox{$10<\log_{10} M_*<10.5$}, and
\mbox{$9.5<\log_{10} M_*<10$}, combining all lens redshift slices.
For further details see the caption of Fig.\thinspace\ref{fi:shearfitsred1}. 
  }
   \label{fi:shearfitsblue1}
    \end{figure*}

\subsection{Results}
\subsubsection{Measured signal and fit range}
\label{se:signal_errors_fitrange}
For each set of lens galaxies we measure the isotropic and anisotropic shear signal in 25 logarithmic 
bins of transverse physical separation between 20{\thinspace}kpc/$h_{70}$  and 1.2{\thinspace}Mpc/$h_{70}$. We compute the signal separately for the lenses in each \mbox{$\sim 1 \mathrm{deg}^2$} CFHTLS field, where we however use larger cut-outs from the mosaic catalogue  of source galaxies to ensure a good coverage for the lenses close to the edge of a CFHTLS field.
We then combine the signal from all fields according to their weight sums in each radial bin. Similarly, we compute a combined signal from all lens redshift slices for each field and the full survey.
Figs.\thinspace\ref{fi:shearfitsred1} and \ref{fi:shearfitsblue1}
show the signal combining all redshift slices for the different stellar mass
bins for red and blue lenses, respectively.
Error-bars are computed by bootstrapping the CFHTLS fields contributing to the
combined survey signal.
In this study we are primarily interested in the relative asymmetry in the shear field. For detailed investigations of the isotropic galaxy-galaxy lensing signal  in CFHTLenS 
please see \citet{vuh14,hgc15}, and \citet{caw15}.

Before we can obtain constraints on the halo ellipticity we have to choose the radial range which is included in the fit. 
We are interested in the shape of the dark matter-dominated matter halo surrounding the lens.
Thus, we exclude small scales which are
expected to be
 affected by the
baryonic component of the lens. 
This is visible from the isotropic galaxy-galaxy lensing signal
 (see the top panels of Figs.\thinspace\ref{fi:shearfitsred1} and
\ref{fi:shearfitsblue1}),  showing a strong excess signal at small scales
compared to the NFW fit constrained from larger scales 
 \citep[see also][who include a baryonic
component in the fit to the isotropic signal]{vuh14,hgc15}.

\begin{figure}
   \centering
  \includegraphics[width=8cm]{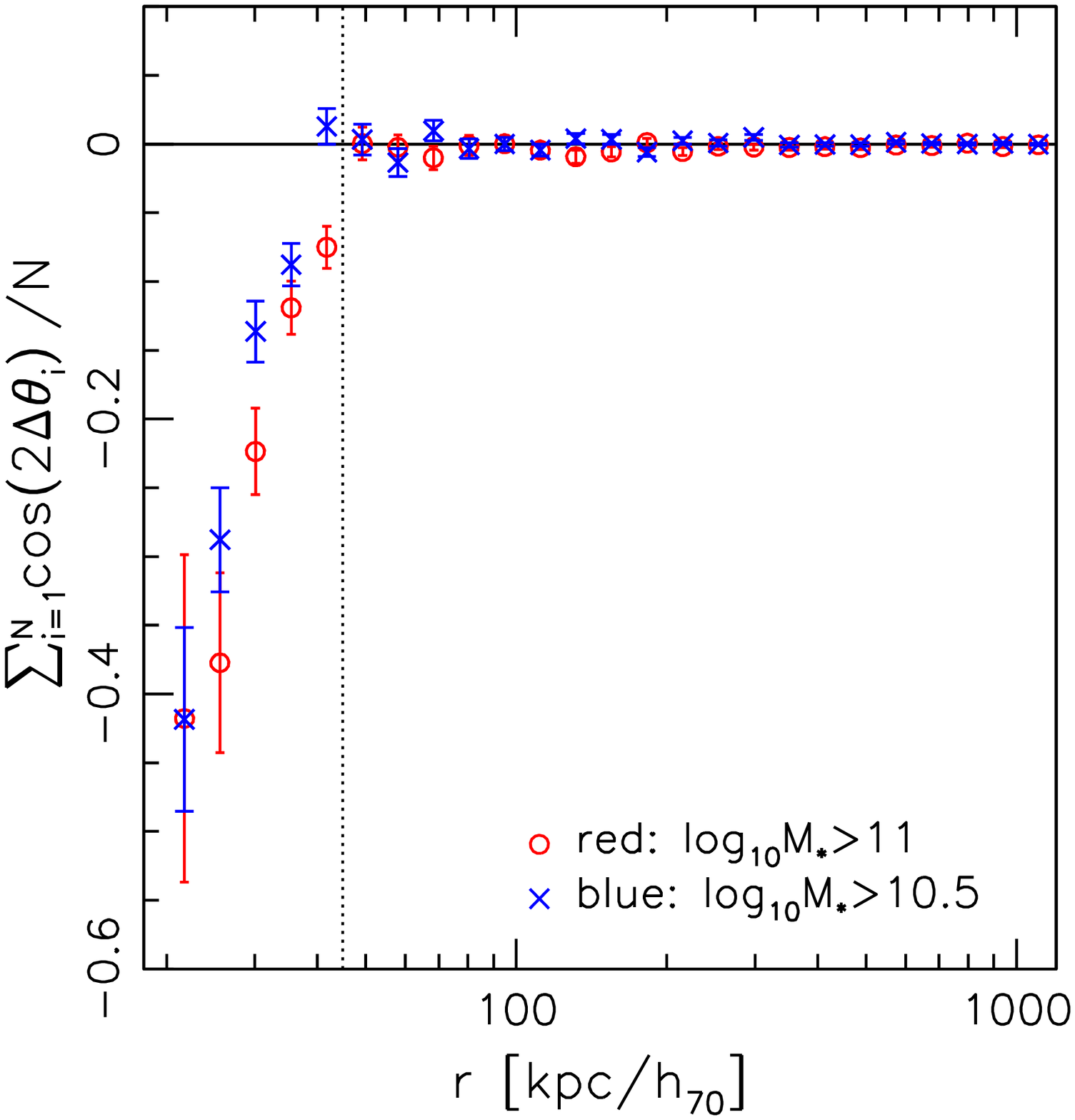}
   \caption{Measurement of the relative azimuthal variation in the source
     density  in CFHTLenS
     with respect to the lens major axis as
     function of radial distance. The open red circles (blue crosses) show 
  \mbox{$\sum_{i=1}^N\cos{(2\Delta\theta_i)}/N$} for the red (blue) lenses in
the correspondingly highest stellar mass bin combining all lens redshift slices.
The decrement at \mbox{$r<45$\thinspace{kpc}/$h_{70}$} indicates a higher source
density in the direction of the lens minor axis compared to the major
axis. To ensure that this cannot influence our analysis we only include
scales \mbox{$r>45$\thinspace{kpc}/$h_{70}$} in the halo shape analysis.
  }
   \label{fi:ncos}
    \end{figure}

In addition, the signal at very small scales might be   
 influenced by systematic effects in the source detection and
shape
measurement process originating from the presence of the nearby bright foreground lens galaxies. 
 As a first check for this we studied 
 the
relative azimuthal variation in the source density.
Fig.\thinspace\ref{fi:ncos} shows the radial dependence of
\mbox{$\sum_{i=1}^N\cos{(2\Delta\theta_i)}/N$} for all 
 source-lens pairs
 in
the highest stellar mass bins for both the blue and red lens samples,
combining all redshift slices.
The decrement at \mbox{$r<45$\thinspace{kpc}/$h_{70}$} shows that we have a higher
source density in the direction of the lens minor axis than the lens major
axis. This suggests that either the object detection and deblending, or the
shape measurement is more effective in the direction where the lens light
has a smaller impact, which is not surprising.
To ensure that this cannot affect our measurements we therefore only include
radial scales \mbox{$r>45$\thinspace{kpc}/$h_{70}$} for all lens bins, which 
also
 matches
our goal to remove the scales 
where baryons are important.
In Appendix \ref{app:testlf}
we present an additional test for shape measurements close to bright lens galaxies using simulated galaxy images.
It shows that any resulting spurious signal should be small compared to the statistical uncertainties from CFHTLenS within our fit range.

We also have to select an outer radius for the radial range which we
include in the halo shape analysis. Here we decide to include the signal out
to \mbox{$r_{200\mathrm{c}}$}. At significantly larger radii the isotropic
galaxy-galaxy lensing signal shows an excess signal for the lower stellar
mass bins  (see the middle and right panels in the top row of Fig.\thinspace\ref{fi:shearfitsred1}, showing excess signal at \mbox{$r\gtrsim 1.6 r_\mathrm{200c}$}). 
Interpreted in the halo model \citep[see e.g.][]{vuh14}, this
excess signal is a combination of the satellite term caused by the central galaxy (for lenses
which are satellites) and the two-halo term from neighbouring haloes.
Hence, at these scales the halo of the lensing galaxy can no longer be
regarded as isolated, which would require a much more complicated modelling scheme.
 We note that the galaxy bins with higher stellar mass show only a weak excess signal at large radii, presumably due to a low satellite fraction. So while we could in principle extend the fit range for these galaxies slightly, we decided to use $r_\mathrm{200c}$ as upper limit for all bins to keep the analysis more homogeneous.

Our approach requires that we first obtain an estimate for
\mbox{$r_{200\mathrm{c}}$}. For this we fit the isotropic signal in the fixed radial range
\mbox{$45\thinspace\mathrm{kpc}/h_{70}<r<200\thinspace\mathrm{kpc}/h_{70}$}.
\replya{Here we use the central redshift of our lenses \mbox{$z_\mathrm{l}=0.4$}
when computing 
the isotropic NFW model for the combined analysis of all redshift slices\footnote{\replya{The  central lens redshift agrees well with the effective lens redshift. The latter is computed according to the weight sums of the thin redshift slices for a fixed radial bin, yielding effective lens redshifts of 0.38 to 0.41 for the different galaxy types and stellar mass bins. We verified that the exact choice of the model $z_\mathrm{l}$ has minimal impact on our $f_\mathrm{h}$ constraints.}}.
}

For comparison, we also plot the components of the anisotropic signal
\mbox{$f\Delta\Sigma$} and \mbox{$-f_{45}\Delta\Sigma$}  in the bottom
panels of Figs.\thinspace\ref{fi:shearfitsred1} and \ref{fi:shearfitsblue1}.
We note that the blue galaxies shown in
Fig.\thinspace\ref{fi:shearfitsblue1}
show a tendency for \mbox{$f\Delta\Sigma<0$} and
\mbox{$-f_{45}\Delta\Sigma>0$}, especially towards larger radii, which is consistent with
the expected trend for cosmic shear. We will discuss this further in Sect.\thinspace\ref{se:discussion}.  

\replya{From the bootstrapping analysis we find that off-diagonal terms in the correlation matrix are
  small, justifying our analysis approach
 (see Sect.\thinspace\ref{se:estimate_fh}).
 Within the fit range of the} 
\replya{isotropic signal
 the average of the off-diagonal elements is consistent
with zero at the \mbox{$\sim 1-2\sigma$} level for all  lens bins with
\mbox{$|\langle\mathrm{cor}_{i,j}\rangle_{i>j}|\lesssim 2\%$}.}

\begin{table*}  
\caption{Weak lensing results using the CFHTLenS
  data.
\label{tab:results_shapes}}
\centering
\begin{tabular}{cccccccccc}
\hline
\hline
Colour & Stellar mass & \mbox{$r_{200\mathrm{c}}$}& \mbox{$M_{200\mathrm{c}}$}& \multicolumn{2}{c}{pass fields}& \multicolumn{2}{c}{all fields}\\
& $[M_\odot]$&  $[\mathrm{kpc}/h_{70}]$&  $[10^{11}M_\odot/h_{70}]$&
\mbox{$f_\mathrm{h}$}& \mbox{$\chi^2/\mathrm{d.o.f.}$}&
\mbox{$f_\mathrm{h}$}& \mbox{$\chi^2/\mathrm{d.o.f.}$} \\
\hline
Red & $ 10<\log_{10}M_*<10.5 $ &  170 & $ 8.6\pm1.7 $ & $ -0.11^{+0.73}_{-0.73} $ & $  9.6 / 7 $        &  $ -0.31^{+0.59}_{-0.60} $  & $  7.8 / 7 $ \\
Red & $ 10.5<\log_{10}M_*<11 $ &  253 & $ 28.2\pm2.9 $ & $ 0.01^{+0.35}_{-0.35} $ & $  9.5 / 10 $       &  $ -0.20^{+0.29}_{-0.29} $  & $  8.8 / 10 $ \\
Red & $ \log_{10}M_*>11 $ &  377 & $ 93.3\pm10.3 $ & $ -0.09^{+0.38}_{-0.38} $ & $  11.3 / 12 $         &  $ -0.01^{+0.33}_{-0.33} $  & $  14.0 / 12 $ \\
Red & $ \log_{10}M_*>10 $  &  &  & $ -0.04_{-0.25}^{+0.25} $ & $  30.5/31 $     &  $ -0.17_{-0.21}^{+0.21} $  & $  30.6/31 $ \\
\hline
Blue & $ 9.5<\log_{10}M_*<10 $ &  128 & $ 3.7\pm0.8 $ & $ 1.27^{+0.88}_{-0.81} $ & $  7.5 / 5 $         &  $ 0.89^{+0.75}_{-0.70} $  & $  11.2 / 6 $ \\
Blue & $ 10<\log_{10}M_*<10.5 $ &  191 & $ 12.1\pm2.0 $ & $ 0.18^{+0.51}_{-0.50} $ & $  2.8 / 8 $       &  $ 0.35^{+0.49}_{-0.49} $  & $  3.8 / 8 $ \\
Blue & $ \log_{10}M_*>10.5 $ &  234 & $ 22.2\pm4.4 $ & $ 1.11^{+0.67}_{-0.64} $ & $  7.3 / 9 $  &  $ 0.51^{+0.52}_{-0.51} $  & $  9.9 / 9 $ \\
Blue & $ \log_{10}M_*>9.5 $  &  &  & $ 0.69_{-0.36}^{+0.37} $ & $  20.5/24 $    &  $ 0.56_{-0.33}^{+0.34} $  & $  24.8/25 $ \\
\hline
\end{tabular}
\vspace{0.1cm}

{\flushleft
Note. --- Summary of the results from the analysis of the CFHTLenS data:
{\it Column 1:} Split between red and blue galaxies. 
{\it Column 2:} Stellar mass range. 
{\it Columns 3 and 4:} Radius  \mbox{$r_{200\mathrm{c}}$} and mass \mbox{$M_{200\mathrm{c}}$} as estimated from the
isotropic component of the shear signal from the  fields passing the
systematics tests (``pass fields''). 
{\it Columns 5 and 6:} Aligned ellipticity ratio \mbox{$f_\mathrm{h}$} 
and reduced $\chi^2$ for the \mbox{$f_\mathrm{h}$} fit from the analysis of
the ``pass fields''.
{\it Columns 7 and 8:} Aligned ellipticity ratio \mbox{$f_\mathrm{h}$} 
and reduced $\chi^2$ for the \mbox{$f_\mathrm{h}$} fit from the analysis of
all fields.
\\
}
\end{table*}

\subsubsection{Constraints on $f_\mathrm{h}$}
\label{se:cfhtles_constraints_fh}
The results of the fits to the 
 CFHTLenS data are presented in Table \ref{tab:results_shapes} and
we show the  model fits to the data in Figs.\thinspace\ref{fi:shearfitsred1} and
\ref{fi:shearfitsblue1}.
For each lens colour and stellar mass bin we fit the combined shear signal
from all redshift slices to ensure that the isotropic shear profile, which
determines $r_{200\mathrm{c}}$, is measured with high
significance\footnote{For comparison we repeated the
  measurement where we initially analyse
each redshift slice separately and combine the constraints when estimating
$f_\mathrm{h}$.
This led to nearly identical, and within the statistical
uncertainty fully consistent results.}.

For none of the individual lens bins do we detect an $f_\mathrm{h}$ significantly
different from zero.
We also compute joint constraints from the \mbox{$(f-f_{45})\Delta\Sigma$} and $\Delta\Sigma$ profiles of all stellar mass
bins as explained in Sect.\thinspace\ref{se:estimate_fh}, yielding 
\mbox{$f_\mathrm{h}=-0.04\pm 0.25$}
for the red lenses and 
\mbox{$f_\mathrm{h}=0.69_{-0.36}^{+0.37}$}
for the
blue lenses when restricting the analysis to the 129 fields passing the
systematics tests described in \citet[][``pass fields'']{hwm12}.
For comparison,  we estimate
\mbox{$f_\mathrm{h}=-0.17\pm 0.21$}
for the red  lenses and
\mbox{$f_\mathrm{h}=0.56_{-0.33}^{+0.34}$} 
for the blue lenses
when including all 171 CFHTLenS fields.
 In  Table \ref{tab:results_shapes} we also list  reduced $\chi^2$
  values 
which suggest that the models fit the data reasonably well in
  both cases when considering all lens bins together,
but we note slightly lower \mbox{$\chi^2/\mathrm{d.o.f.}$} for the blue
lenses when using the
``pass fields'' only.

\replya{As a consistency check for the possible impact of satellite galaxies
in the lens sample (see Sect.\thinspace\ref{sec:cfhtlens_lenssample}) we
also compute joint constraints for the red lenses now excluding the}
\replya{stellar mass
bin with the highest expected satellite fraction (\mbox{$10<\log_{10}
  M_*<10.5$}). In this case we obtain almost unchanged results  \mbox{$f_\mathrm{h}=-0.02\pm 0.26$} for the ``pass fields''
and
\mbox{$f_\mathrm{h}=-0.14\pm 0.22$} for all fields, suggesting
that the remaining satellites have a negligible impact on the  joint constraints.}

\section{Analysis of simulated data based on the Millennium simulation}
\label{sec:millennium}

To better understand the halo shape signal that we should
expect from non-idealised haloes and in the presence of misalignments between galaxies and their 
matter haloes, we analyse a simulated data set based on ray-tracing
 through the Millennium Simulation \citep{swj05} by \citet{hhw09}.
This also allows us to test the correction for systematic shear (see Sect.\thinspace\ref{su:method})
in the presence of a realistic cosmic shear field.
We refer the reader to \citet{hhw09} regarding the details of the
ray-tracing. Here we only summarise some of the main
characteristics relevant for our analysis.

\subsection{Mock shear catalogues}
\label{se:millen_shearcat}

The simulated catalogues comprise 64 light cones, each with an
area of \mbox{$4\times 4$}{\thinspace}deg$^2$, which we
subdivide into patches of 1{\thinspace}deg$^2$ to facilitate a bootstrap
analysis similar to the CFHTLenS fields. 
We use all galaxies at \mbox{$0.65<z<2.15$}
 as source sample (\mbox{$z_\mathrm{median}=1.187$}), providing a high source
density of 49.6/arcmin$^2$.
We wish to obtain high \mbox{$S/N$} estimates of the simulated halo shape
signal and therefore add only small shape noise with \mbox{$\sigma_e=0.03$},
which was chosen to be of the same order as the noise introduced by cosmic
shear.
\replya{This low level of shape noise leads to stronger relative noise
  contributions from cosmic shear compared to the CFHTLenS analysis.
This
  increases the} 
\replya{noise correlations between different radial bins.
While they can be substantial for the isotropic signal $\Delta \Sigma$  for some of the lens
bins with \mbox{$5\%
  < |\langle\mathrm{cor}_{i,j}\rangle_{i>j}|< 45\%$}, 
they are generally small for the anisotropic signal $(f-f_{45})\Delta \Sigma$
with \mbox{$|\langle\mathrm{cor}_{i,j}\rangle_{i>j}|\lesssim 4\%$}.
Noise in the latter dominates the uncertainties in the  $f_\mathrm{h}$
constraints.
We therefore expect that the net impact of the noise correlations on our
constraints is negligible.
As a consistency check we also repeat the analysis of the simulated data with more
realistic shape noise (\mbox{$\sigma_e=0.25$}).
Here we find consistent results, but the increased statistical uncertainty
makes it impossible to detect the signal for some of the weakly aligned lens models.}

\subsection{Mock lens galaxies}
\label{se:sims_mr_lenses}
For the foreground lenses we make use of galaxy shapes computed by
\citet[][hereafter \citetalias{jsb13}]{jsb13} and \citet[][hereafter \citetalias{jsh13}]{jsh13}.
Following \citet{bef07}, haloes and their member candidate particles
are first identified by a friends-of-friends algorithm \citep{def85}. In a second step, merger-tree data is used to remove particles belonging to substructures that are only
temporarily in the halo vicinity. Halo shapes are then estimated via the full quadrupole tensor of the halo's matter distribution that remains after removal of these transients. 
This information is complemented with semi-analytic galaxy evolution models
\citep{bbm06} and a classification of galaxy morphologies into early and
late types via the
bulge-to-total ratio of the rest-frame $K$-band luminosity \citep{pef09}.
They also separate galaxies into centrals (galaxies in the most massive substructure of a
halo) and satellites. 
Here we only use foreground galaxies classified as centrals, because
\citetalias{jsb13} and \citetalias{jsh13} had to assign simplistic galaxy shapes to the satellites
given that their dark matter haloes were poorly resolved.
Also, we only keep those foreground galaxies with a sufficient particle
number for robust shape estimation, see \citetalias{jsb13}.
For the centrals \citetalias{jsb13} and \citetalias{jsh13} assign shapes adopting the scheme of
\citet{hwh06}.
Here,  late type disc-dominated galaxies are aligned such that their spin  vector is parallel to the angular momentum vector of
their host dark matter halo.
These shapes are then projected onto the plane of the sky.
For early type galaxies it is assumed that the shapes of the galaxies follow
the shapes of their dark matter haloes. 
For this, \citetalias{jsb13} project the ellipsoid given by the eigenvectors
and eigenvalues of the inertia tensor for each halo on to the plane  of the sky and
use the resulting ellipse as the shape of the galaxy.

We additionally consider galaxy ellipticities that are misaligned with
respect to their host dark matter halo as detailed in \citetalias{jsh13}.
For early type galaxies the misalignment angles were drawn from a
Gaussian distribution with a scatter of 35$^\circ$ as estimated by
\citet{ojl09} from the distribution of satellites around luminous red galaxies (LRGs) in SDSS.
For late type galaxies \citetalias{jsh13} employ a misalignment distribution
based on a  fitting function that \citet{bet12} determined using a compilation
of simulated haloes with baryons and galaxy formation physics
\citep{bef10,dmf11,ctd09,oef05}.

Similarly to the analysis of the CFHTLenS data, we select lens galaxies in
the redshift range \mbox{$0.2<z_\mathrm{l}<0.6$}  split into two redshift
slices (which we analyse separately because of better $S/N$ compared to CFHTLenS) and stellar mass bins, see Table \ref{tab:results_shapes_millennium} for details.

\begin{table*}  
\caption{Weak lensing results using the mock data based on the Millennium Simulation, both for the analysis with and without foreground cosmic shear (c.s.) applied.
\label{tab:results_shapes_millennium}}
\begin{center}
\tabcolsep=0.16cm
\begin{tabular}{ccccccccccc}
\hline
\hline
Type & Stellar mass & $z_\mathrm{l,min}$ & $z_\mathrm{l,max}$ & 
$\sigma_e$& 
\mbox{$r_{200\mathrm{c}}$}& \mbox{$M_{200\mathrm{c}}$}&
\multicolumn{2}{c}{\mbox{$f_\mathrm{h}$} (aligned)}&
\multicolumn{2}{c}{\mbox{$f_\mathrm{h}$} (misaligned)}\\
&  & &  & & $[\mathrm{kpc}/h_{73}]$&  $[10^{11}M_\odot/h_{73}]$ & with c.s. & without c.s. & with c.s. & without c.s.\\
\hline
Early & $ 9.5<\log_{10}M_*<10 $ & 0.2 & 0.4 & 0.19 & 168 & $ 7.84 \pm 0.03 $ & $ 0.504^{+0.027}_{-0.025} $     &  $ 0.511^{+0.026}_{-0.026} $  &  $ 0.203^{+0.027}_{-0.026} $  &  $ 0.219^{+0.028}_{-0.026} $ \\
Early & $ 10<\log_{10}M_*<10.5 $ & 0.2 & 0.4 & 0.19 & 226 & $ 18.97 \pm 0.06 $ & $ 0.635^{+0.015}_{-0.015} $   &  $ 0.644^{+0.015}_{-0.015} $  &  $ 0.306^{+0.015}_{-0.014} $  &  $ 0.298^{+0.016}_{-0.014} $ \\
Early & $ 10.5<\log_{10}M_*<11 $ & 0.2 & 0.4 & 0.19 & 375 & $ 86.19 \pm 0.25 $ & $ 0.759^{+0.009}_{-0.009} $   &  $ 0.772^{+0.010}_{-0.008} $  &  $ 0.373^{+0.010}_{-0.008} $  &  $ 0.376^{+0.010}_{-0.008} $ \\
Early & $ 11<\log_{10}M_*<11.5 $ & 0.2 & 0.4 & 0.21 & 563 & $ 291.14 \pm 1.69 $ & $ 0.846^{+0.012}_{-0.011} $  &  $ 0.863^{+0.011}_{-0.011} $  &  $ 0.402^{+0.011}_{-0.010} $  &  $ 0.407^{+0.012}_{-0.010} $ \\
Early & $ 9.5<\log_{10}M_*<11.5 $ & 0.2 & 0.4 & 0.19  &  &  & $ 0.715_{-0.006}^{+0.007} $       &  $ 0.726_{-0.006}^{+0.007} $  &  $ 0.344_{-0.006}^{+0.007} $  &  $ 0.346_{-0.006}^{+0.007} $ \\
\hline
Early & $ 9.5<\log_{10}M_*<10 $ & 0.4 & 0.6 & 0.20 & 156 & $ 7.56 \pm 0.03 $ & $ 0.407^{+0.021}_{-0.021} $     &  $ 0.463^{+0.022}_{-0.020} $  &  $ 0.144^{+0.022}_{-0.021} $  &  $ 0.209^{+0.022}_{-0.021} $ \\
Early & $ 10<\log_{10}M_*<10.5 $ & 0.4 & 0.6 & 0.20 & 213 & $ 19.15 \pm 0.05 $ & $ 0.501^{+0.013}_{-0.012} $   &  $ 0.535^{+0.013}_{-0.012} $  &  $ 0.232^{+0.013}_{-0.012} $  &  $ 0.259^{+0.012}_{-0.012} $ \\
Early & $ 10.5<\log_{10}M_*<11 $ & 0.4 & 0.6 & 0.21 & 351 & $ 86.34 \pm 0.24 $ & $ 0.660^{+0.009}_{-0.008} $   &  $ 0.679^{+0.009}_{-0.008} $  &  $ 0.299^{+0.009}_{-0.007} $  &  $ 0.332^{+0.009}_{-0.007} $ \\
Early & $ 11<\log_{10}M_*<11.5 $ & 0.4 & 0.6 & 0.22 & 502 & $ 251.61 \pm 1.37 $ & $ 0.805^{+0.012}_{-0.010} $  &  $ 0.827^{+0.011}_{-0.010} $  &  $ 0.359^{+0.011}_{-0.010} $  &  $ 0.379^{+0.011}_{-0.010} $ \\
Early & $ 9.5<\log_{10}M_*<11.5 $ & 0.4 & 0.6 & 0.21  &  &  & $ 0.607_{-0.006}^{+0.006} $       &  $ 0.636_{-0.006}^{+0.006} $  &  $ 0.272_{-0.005}^{+0.007} $  &  $ 0.306_{-0.005}^{+0.006} $ \\
\hline
Early & $ 9.5<\log_{10}M_*<11 $ & 0.2 & 0.6 & 0.20  &  &  & $ 0.616_{-0.005}^{+0.006} $ &  $ 0.637_{-0.004}^{+0.006} $  &  $ 0.285_{-0.004}^{+0.006} $  &  $ 0.307_{-0.005}^{+0.005} $ \\
Early & $ 9.5<\log_{10}M_*<11.5 $ & 0.2 & 0.6 & 0.20 & &  & $ 0.657_{-0.004}^{+0.005} $       &  $ 0.678_{-0.004}^{+0.005} $  &  $ 0.304_{-0.004}^{+0.005} $  &  $ 0.324_{-0.004}^{+0.004} $ \\
\hline
Late & $ 9.5<\log_{10}M_*<10 $ & 0.2 & 0.4 & 0.32  & 131 & $ 3.68 \pm 0.01 $ & $ 0.073^{+0.016}_{-0.016} $      &  $ 0.068^{+0.017}_{-0.015} $  &  $ 0.034^{+0.017}_{-0.016} $  &  $ 0.025^{+0.017}_{-0.016} $ \\
Late & $ 10<\log_{10}M_*<10.5 $ & 0.2 & 0.4 & 0.32 & 171 & $ 8.17 \pm 0.03 $ & $ 0.123^{+0.014}_{-0.013} $     &  $ 0.133^{+0.015}_{-0.013} $  &  $ 0.038^{+0.014}_{-0.014} $  &  $ 0.064^{+0.014}_{-0.014} $ \\
Late & $ 10.5<\log_{10}M_*<11 $ & 0.2 & 0.4 & 0.32 & 245 & $ 24.02 \pm 0.10 $ & $ 0.129^{+0.012}_{-0.011} $    &  $ 0.132^{+0.012}_{-0.011} $  &  $ 0.031^{+0.012}_{-0.012} $  &  $ 0.059^{+0.012}_{-0.012} $ \\
Late & $ 9.5<\log_{10}M_*<11 $ & 0.2 & 0.4 & 0.32 &  &  & $ 0.108_{-0.007}^{+0.009} $  &  $ 0.111_{-0.008}^{+0.009} $  &  $ 0.034_{-0.008}^{+0.009} $  &  $ 0.049_{-0.008}^{+0.009} $ \\
\hline
Late & $ 9.5<\log_{10}M_*<10 $ & 0.4 & 0.6 & 0.32 & 124 & $ 3.82 \pm 0.01 $ & $ 0.041^{+0.014}_{-0.012} $      &  $ 0.068^{+0.013}_{-0.013} $  &  $ 0.007^{+0.014}_{-0.012} $  &  $ 0.013^{+0.014}_{-0.013} $ \\
Late & $ 10<\log_{10}M_*<10.5 $ & 0.4 & 0.6 &0.32 & 163 & $ 8.68 \pm 0.03 $ & $ 0.101^{+0.012}_{-0.010} $     &  $ 0.122^{+0.012}_{-0.011} $  &  $ 0.022^{+0.012}_{-0.011} $  &  $ 0.031^{+0.012}_{-0.011} $ \\
Late & $ 10.5<\log_{10}M_*<11 $ & 0.4 & 0.6 &0.32 & 230 & $ 24.38 \pm 0.08 $ & $ 0.115^{+0.010}_{-0.010} $    &  $ 0.128^{+0.010}_{-0.010} $  &  $ 0.029^{+0.011}_{-0.010} $  &  $ 0.037^{+0.011}_{-0.009} $ \\
Late & $ 9.5<\log_{10}M_*<11 $ & 0.4 & 0.6 &0.32 &  &  & $ 0.085_{-0.007}^{+0.007} $  &  $ 0.104_{-0.006}^{+0.008} $  &  $ 0.019_{-0.006}^{+0.007} $  &  $ 0.027_{-0.007}^{+0.007} $ \\
\hline
Late & $ 9.5<\log_{10}M_*<11 $ & 0.2 & 0.6 &0.32 &  &  & $ 0.095_{-0.005}^{+0.005} $  &  $ 0.107_{-0.004}^{+0.006} $  &  $ 0.025_{-0.004}^{+0.006} $  &  $ 0.036_{-0.005}^{+0.006} $ \\
\hline

\end{tabular}
\end{center}
\vspace{0.1cm}
{\flushleft
Note. --- Summary of the results from the analysis of the Millennium Simulation.
{\it Column 1:} Split between early and late types.
{\it Column 2:} Stellar mass range.
{\it Column 3:} Minimum lens redshift. 
{\it Column 4:}  Maximum lens redshift. 
{\it Column 5:} Ellipticity dispersion of the selected lenses with \mbox{$0.05<|e_\mathrm{g}|<0.95$} combining both components. 
{\it Columns 6 and 7:} Radius  \mbox{$r_{200\mathrm{c}}$} and mass \mbox{$M_{200\mathrm{c}}$} as estimated from the
isotropic component of the shear signal for the aligned lens models with cosmic shear (very small differences for the
other models).
{\it Columns 8 and 9:} Aligned ellipticity ratio \mbox{$f_\mathrm{h}$} for
aligned lens models (models \texttt{Est} and \texttt{Sa1} from \citetalias{jsh13})
with and without  foreground cosmic shear applied, respectively.
{\it Columns 10 and 11:} Aligned ellipticity ratio \mbox{$f_\mathrm{h}$} for
misaligned lens models (models \texttt{Ema} and \texttt{Sma} from \citetalias{jsh13})
with and without  foreground cosmic shear applied, respectively.\\
}
\end{table*}

\subsection{Results}

We summarise  the main results of the analysis based on the Millennium Simulation
in Table \ref{tab:results_shapes_millennium},
 listing estimated halo masses, as well as \mbox{$f_\mathrm{h}$} constraints for both aligned and misaligned galaxy shape models, each with and without foreground cosmic shear applied.
For this analysis we use modified $\Lambda$CDM cosmological parameters for the
evaluation of angular diameter distances, matching the input to the
Millennium Simulation, with \mbox{$\Omega_\mathrm{m}=0.25$},
\mbox{$\Omega_\Lambda=0.75$}, \mbox{$H_0=73 h_{73}$ km/s/Mpc}.

\subsubsection{Halo masses}
\label{se:mr_halomasses}
In the CFHTLenS analysis we use stellar mass as proxy for halo mass.
For the analysis of the Millennium Simulation we again split the lenses into stellar mass bins.
However, these stellar mass estimates are based on semi-analytic galaxy evolution models \citep{bbm06} which have some
uncertainty \citep[e.g.][]{kbc09,shs12}. 
For a direct comparison of the results from CFHTLenS and the simulation it is more relevant that the included bins roughly span the same range in halo mass.
Here we note that the halo mass estimates inferred from the isotropic shear signal agree reasonably well for the late type (blue) lenses, whereas the early type (red) lenses yield higher halo masses in a given stellar mass bin in the simulation.
To roughly match the halo mass range when computing joint
constraints on $f_\mathrm{h}$ from the simulation, we therefore  remove the highest stellar mass bin (\mbox{$11<\log_{10} M_*<11.5$}) and instead include one additional, lower stellar mass bin (\mbox{$9.5<\log_{10} M_*<10$}). 
Given the only moderate dependence of $f_\mathrm{h}$ on mass (Table \ref{tab:results_shapes_millennium}), this approximate matching is fully sufficient for our goal to provide model predictions.

\subsubsection{Shear profile plots}
\begin{figure*}
   \centering
  \includegraphics[width=5.8cm]{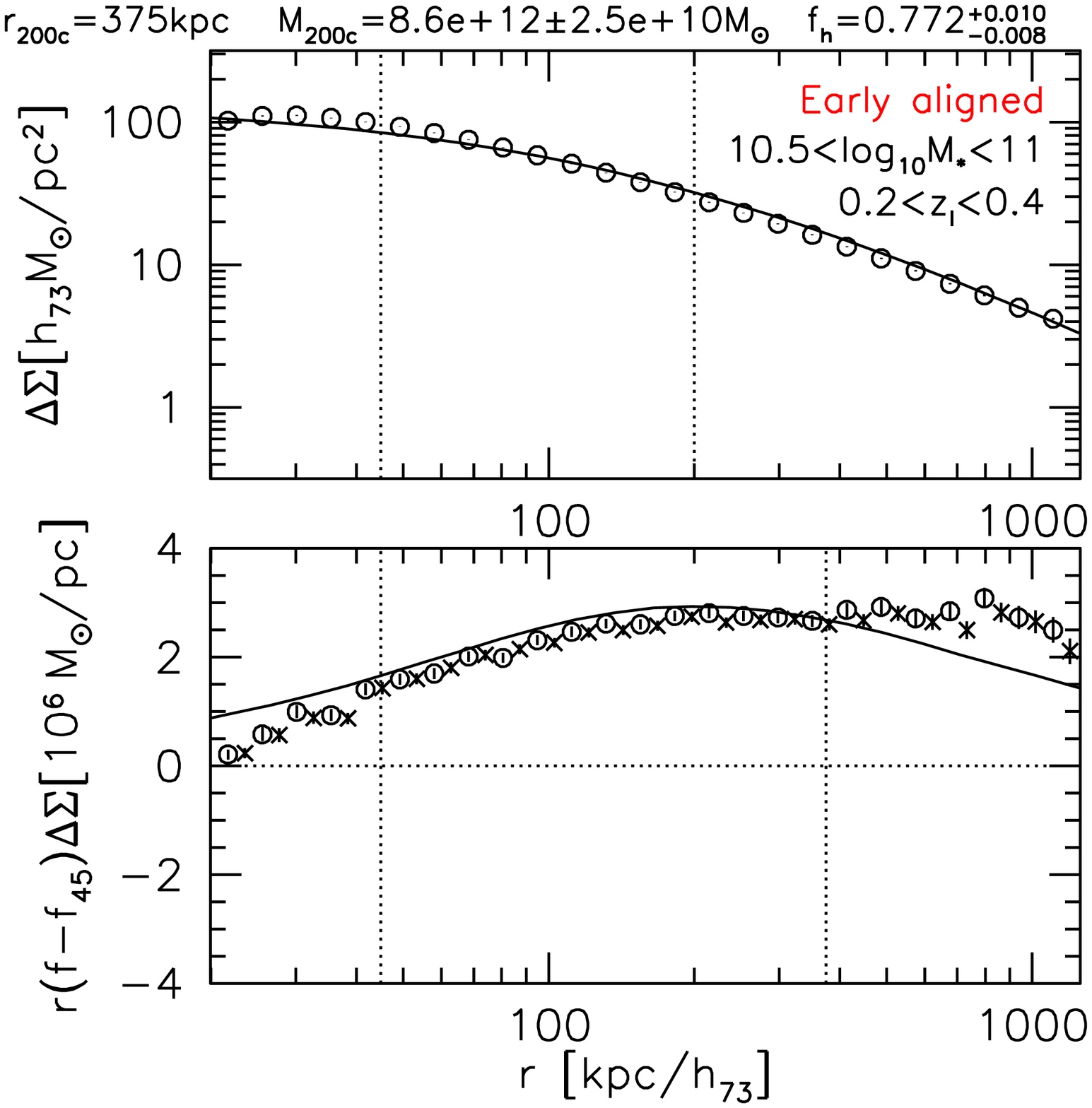}
  \includegraphics[width=5.8cm]{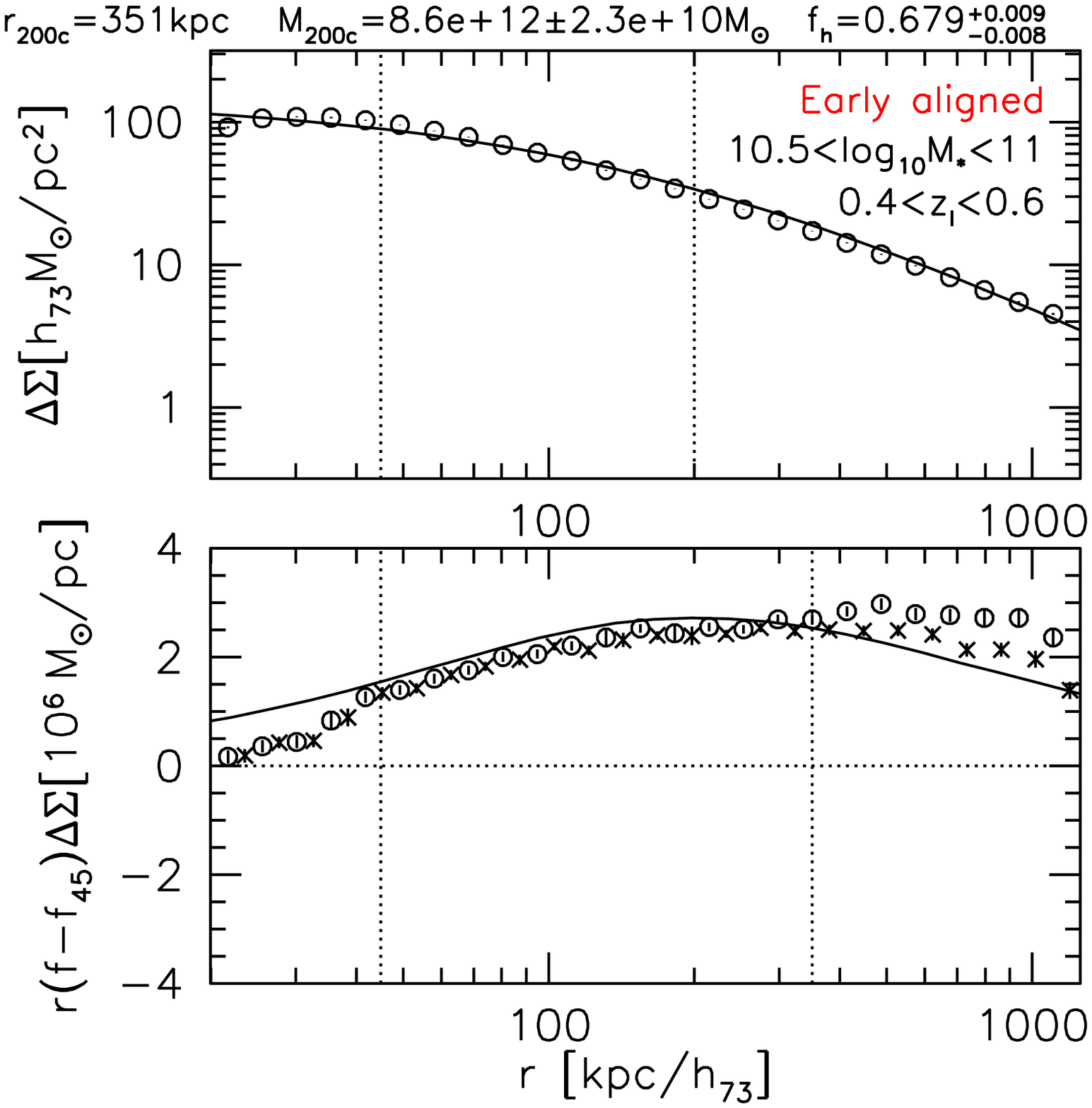}
  \includegraphics[width=5.8cm]{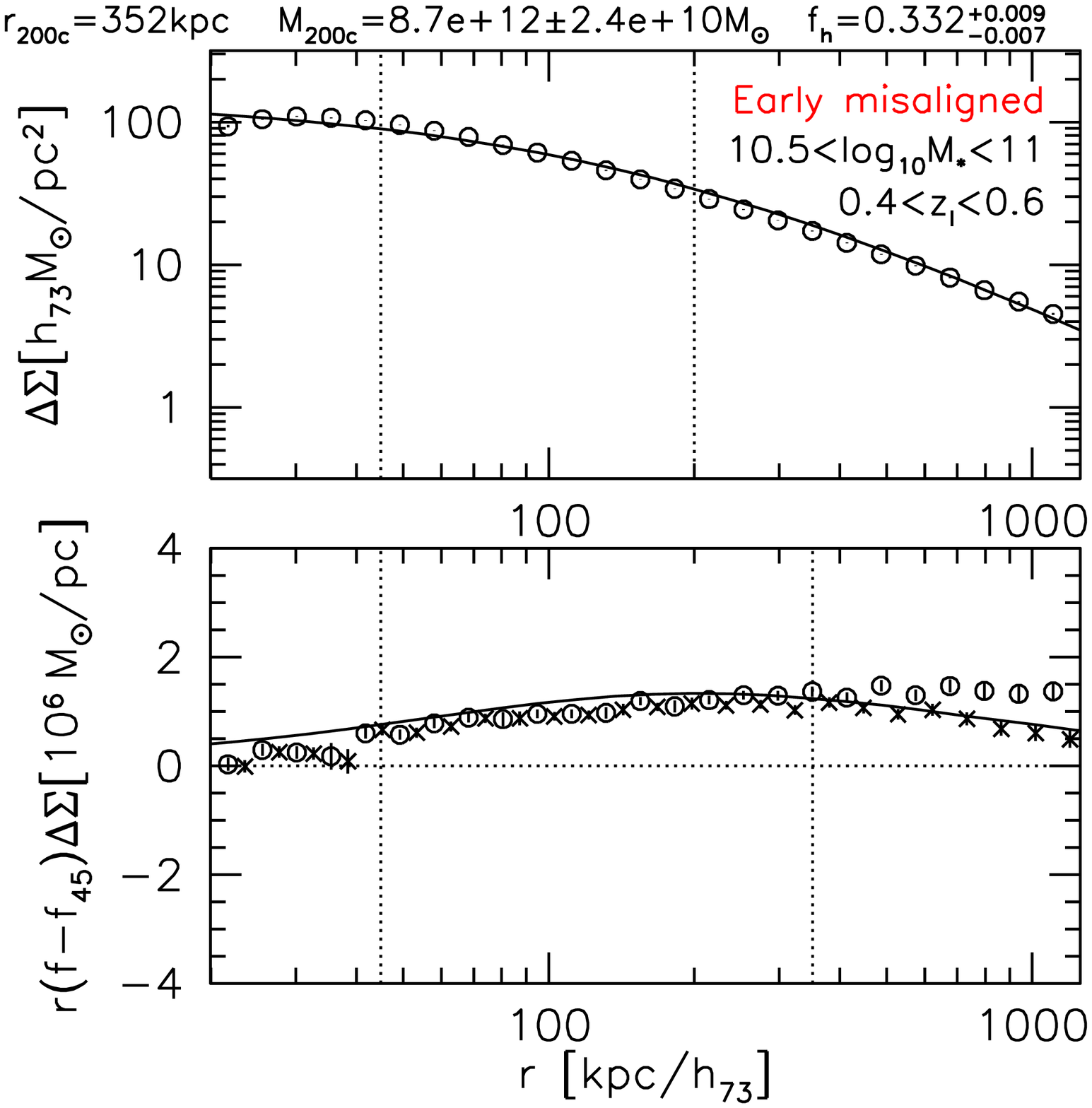}
  \includegraphics[width=5.8cm]{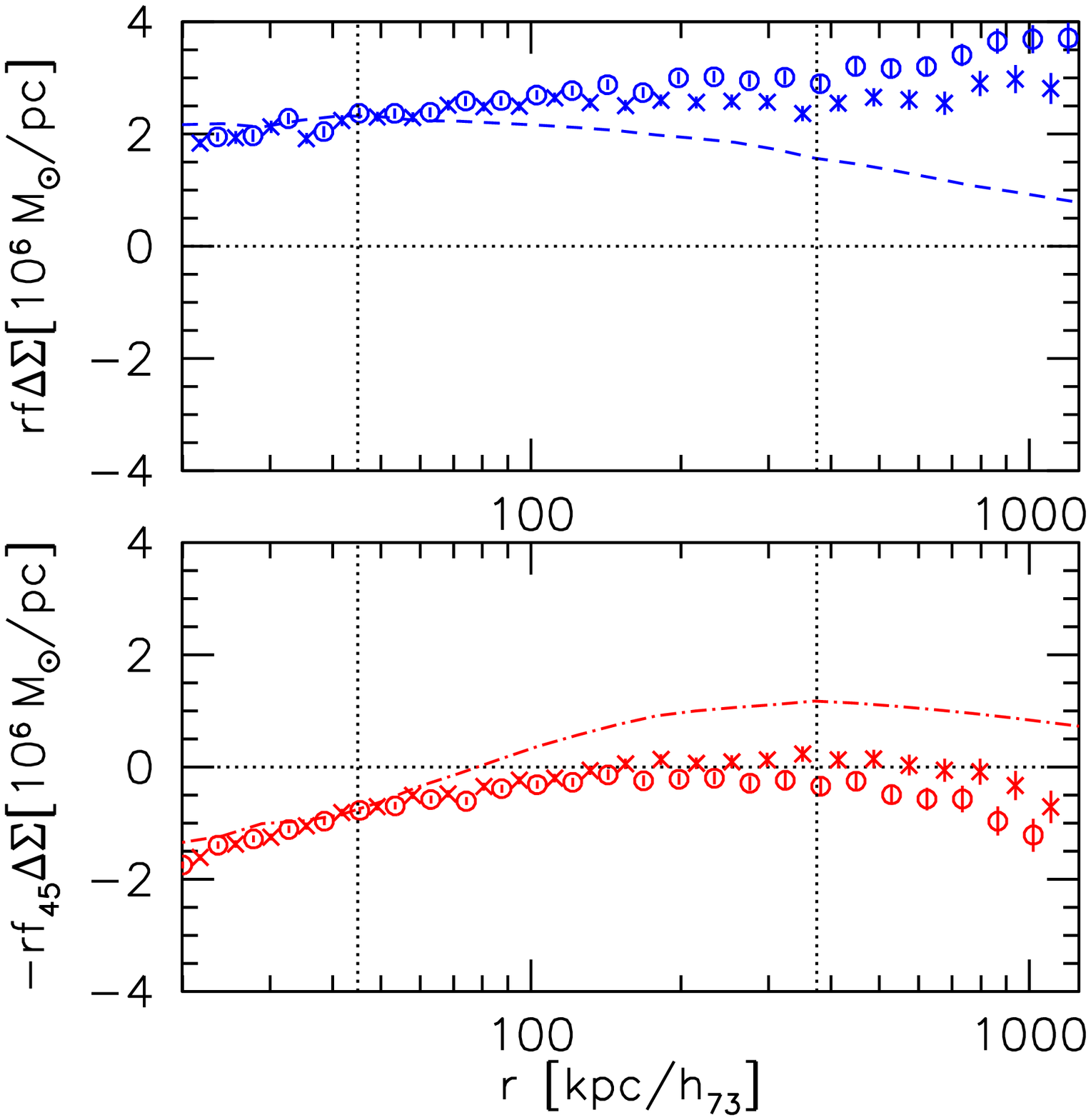}
  \includegraphics[width=5.8cm]{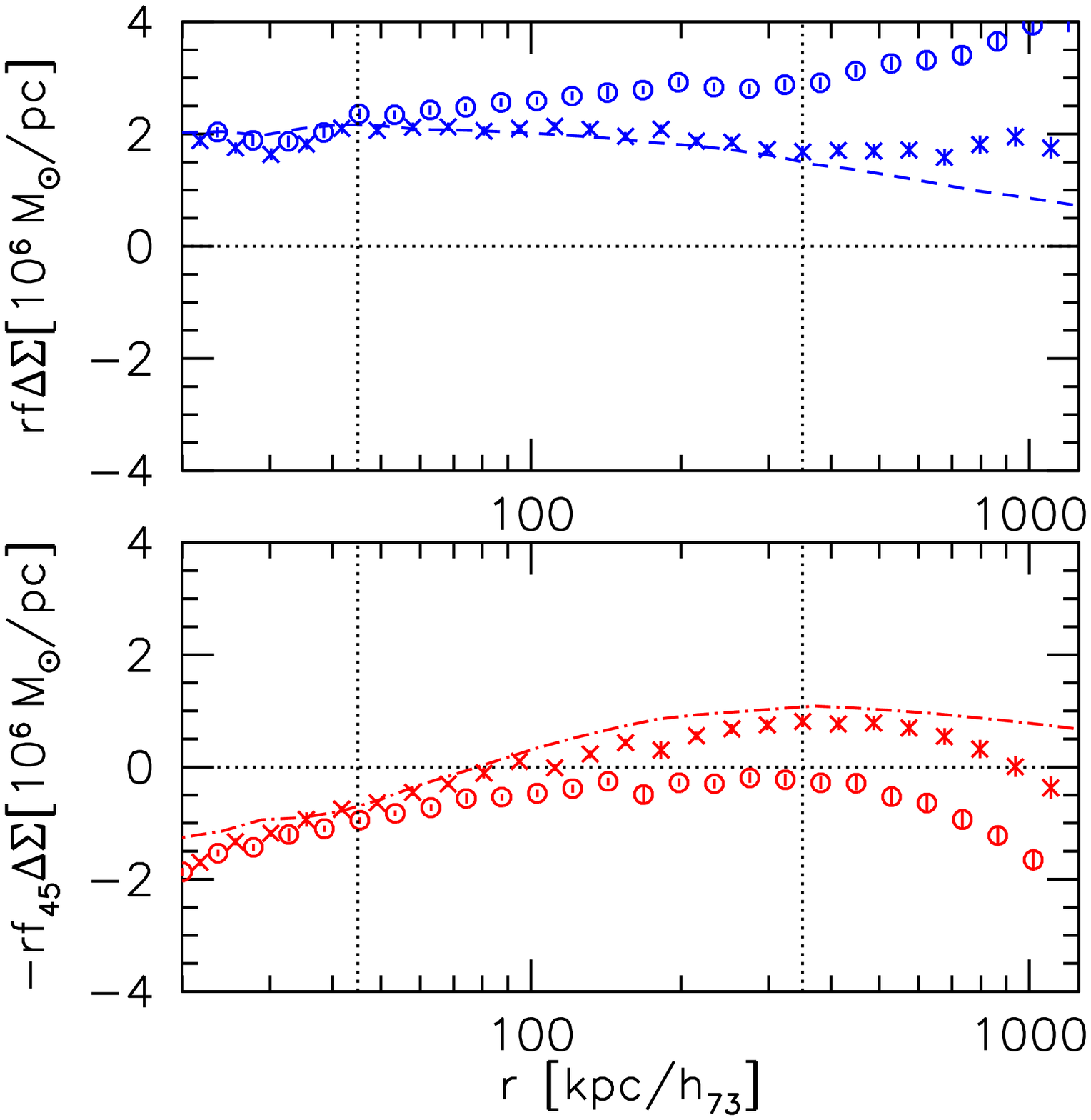}
  \includegraphics[width=5.8cm]{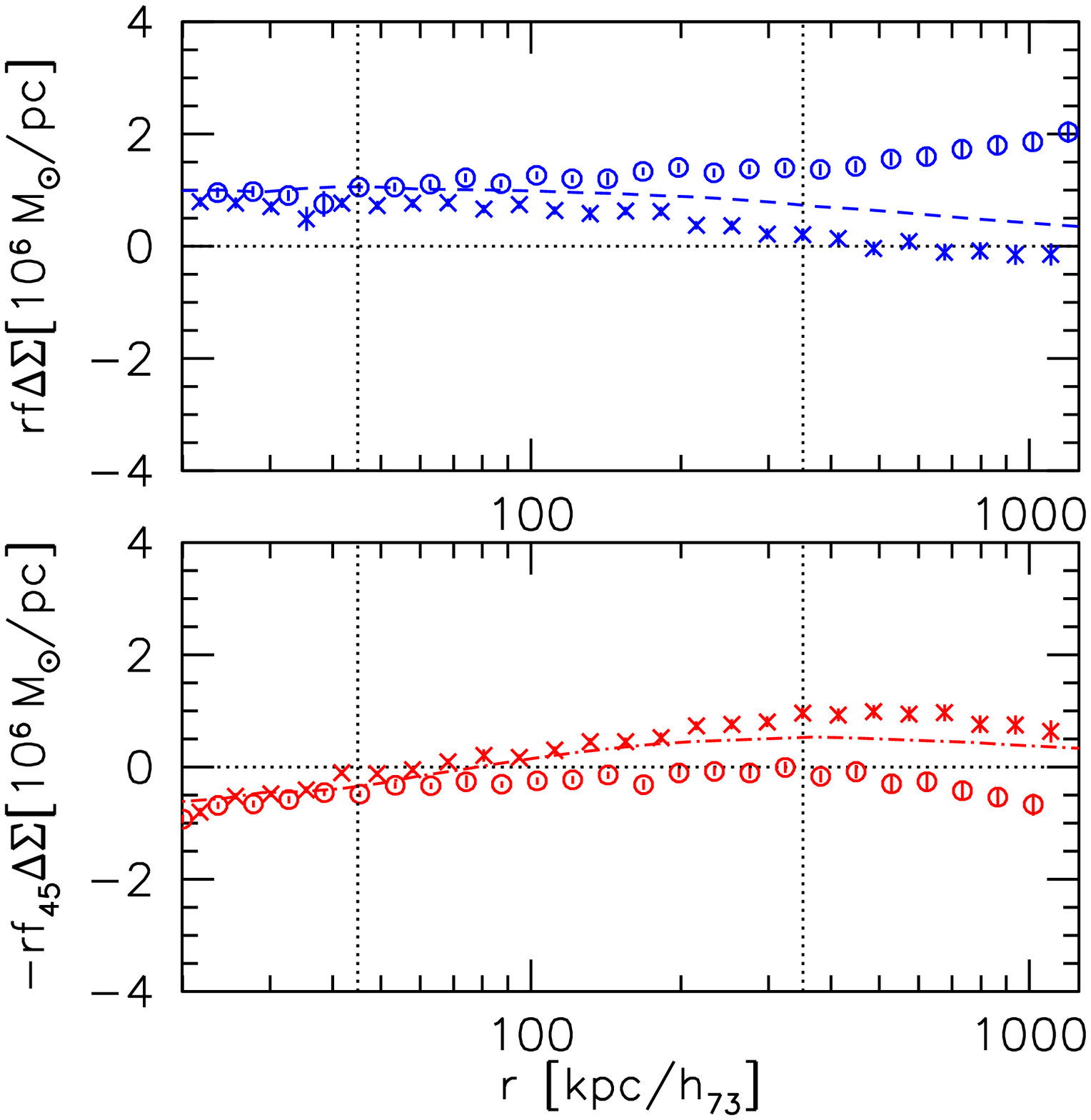}
   \caption{Analysis of the Millennium
Simulation 
showing early type foreground lens
galaxies 
in stellar mass bin \mbox{$10.5<\log_{10} M_*<11$}:
From {\it top} to {\it bottom} we show the isotropic signal $\Delta\Sigma$
and the anisotropic signal components \mbox{$(f-f_{45})\Delta\Sigma$},
$f\Delta\Sigma$, and $-f_{45}\Delta\Sigma$, respectively. 
For the
anisotropic signal components the crosses (circles) correspond to the cases with
(without) foreground cosmic shear applied, where the crosses are displayed
with an offset of half a bin for better readability (note the 
scaling by $r$ for better readability).
The {\it left} column
corresponds to  aligned shape models and
lens
redshift slice  \mbox{$0.2<z_\mathrm{l}<0.4$}.
The {\it middle} column also shows  aligned shape models
but higher lens redshifts  
 \mbox{$0.4<z_\mathrm{l}<0.6$}, illustrating the effect of an increasing cosmic shear contribution.
The {\it right} column 
corresponds to the same lens galaxies as the  {\it middle} column, but now
with misaligned shape models, illustrating the resulting suppression of the
anisotropic signal.
The curves indicate the model predictions for the
best-fitting $f_\mathrm{h}$  and best-fitting isotropic model.
$f_\mathrm{h}$ is determined from the ratio of
\mbox{$(f-f_{45})\Delta\Sigma$} and \mbox{$\Delta\Sigma$} for the case without cosmic shear, fitted within the
range indicated by the vertical dotted lines.
  }
   \label{fi:shearfitsearly_ms1}
    \end{figure*}

\begin{figure*}
   \centering
  \includegraphics[width=5.8cm]{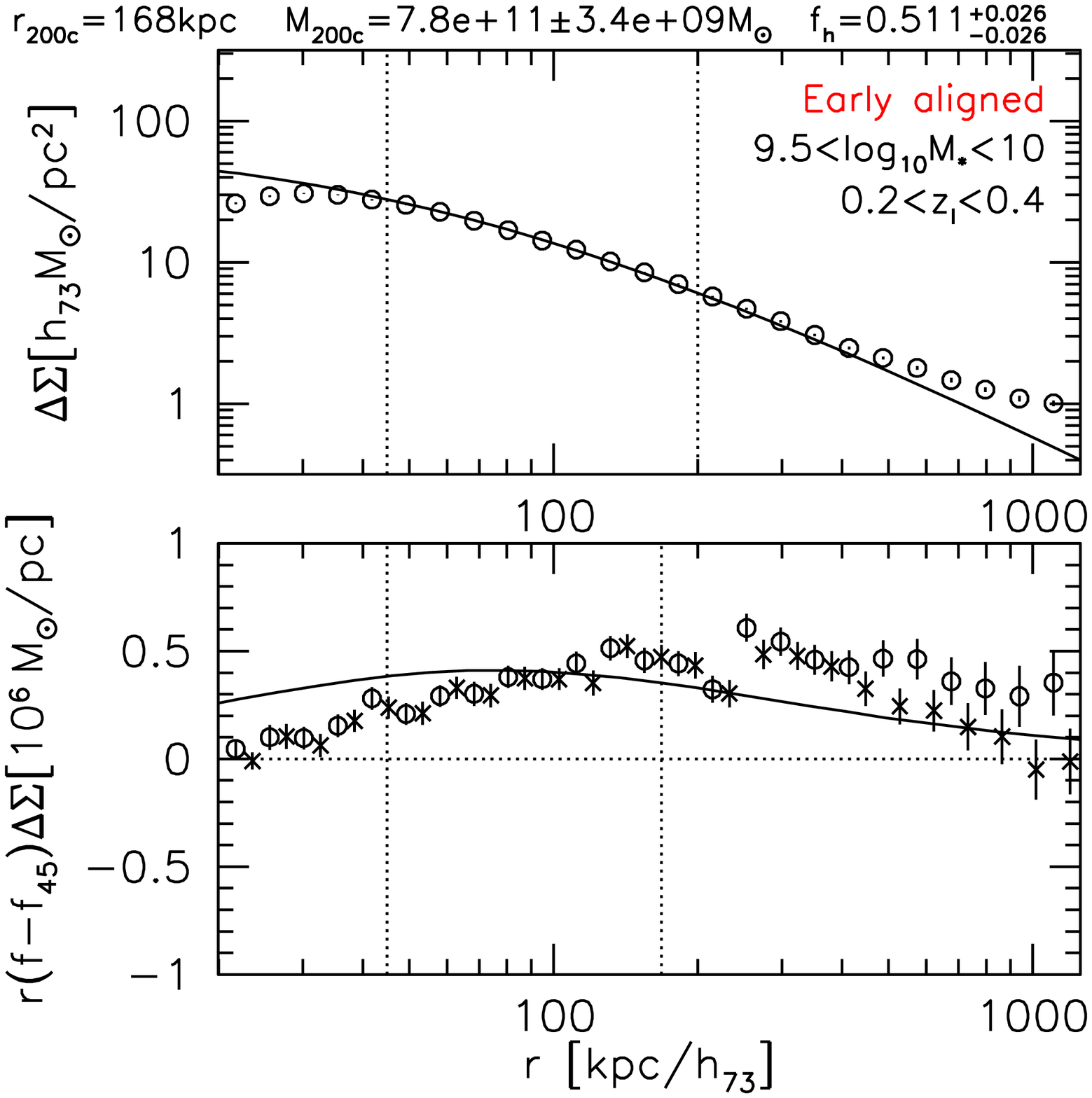}
 \includegraphics[width=5.8cm]{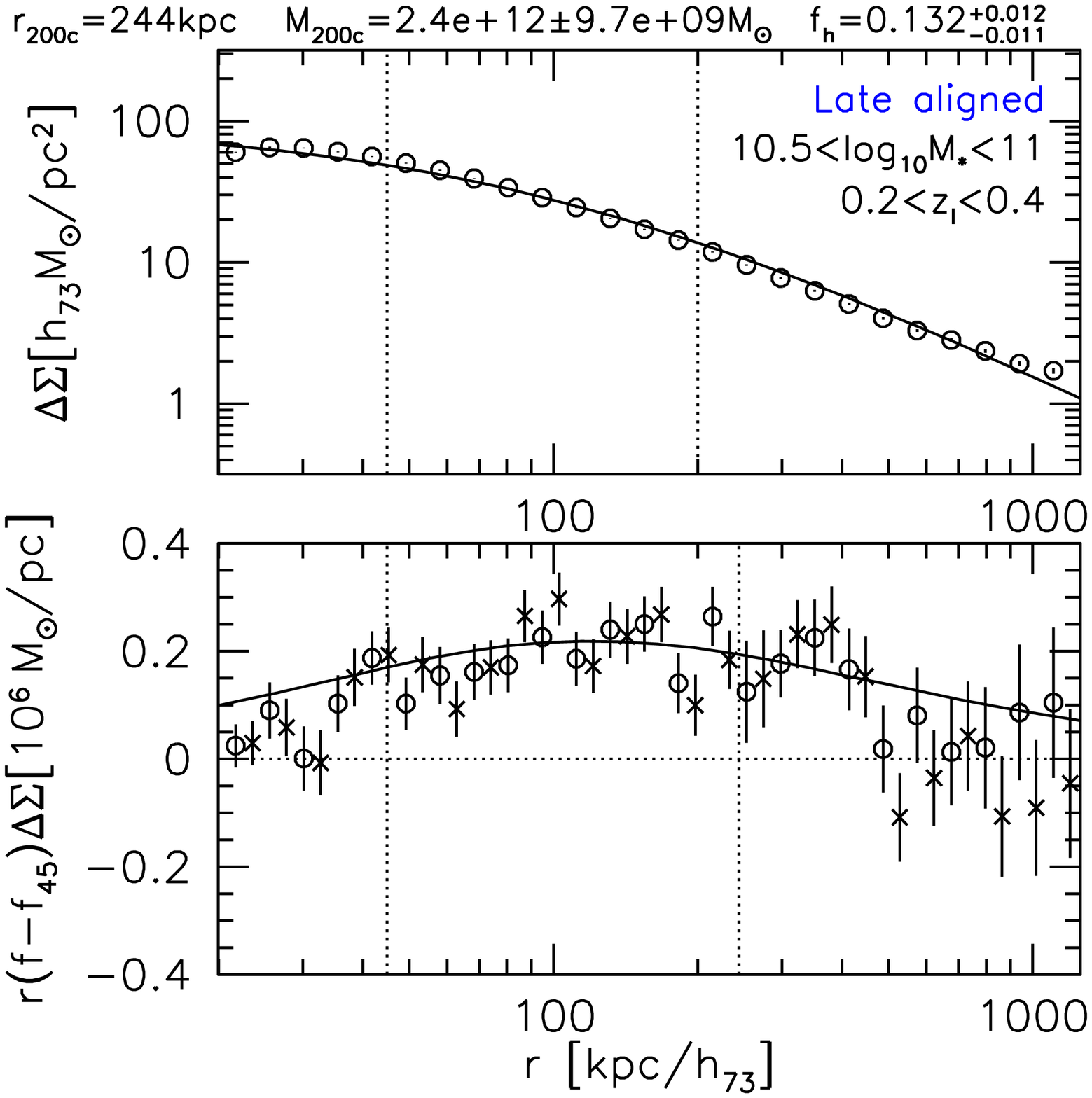}
 \includegraphics[width=5.8cm]{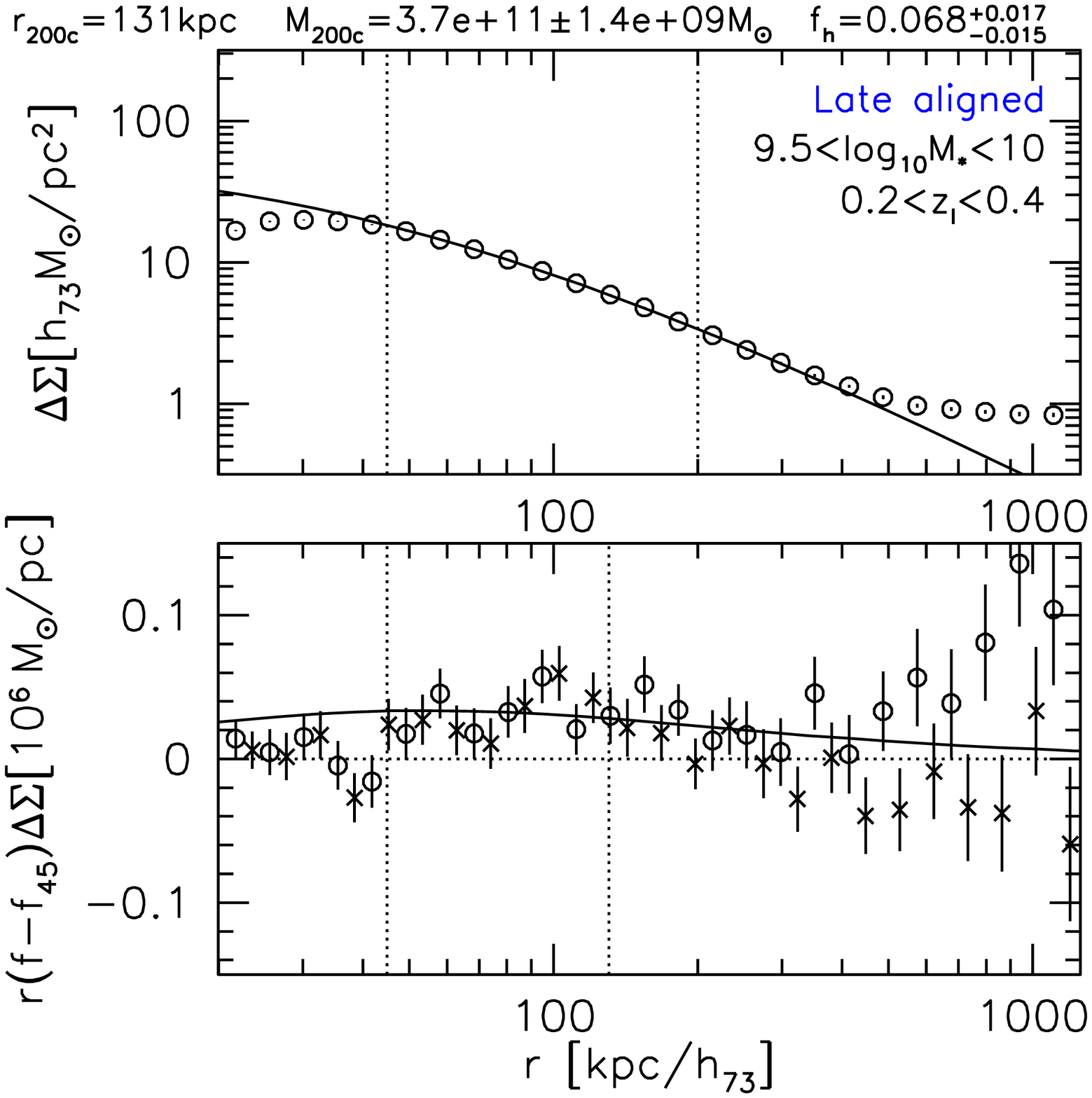}
 \includegraphics[width=5.8cm]{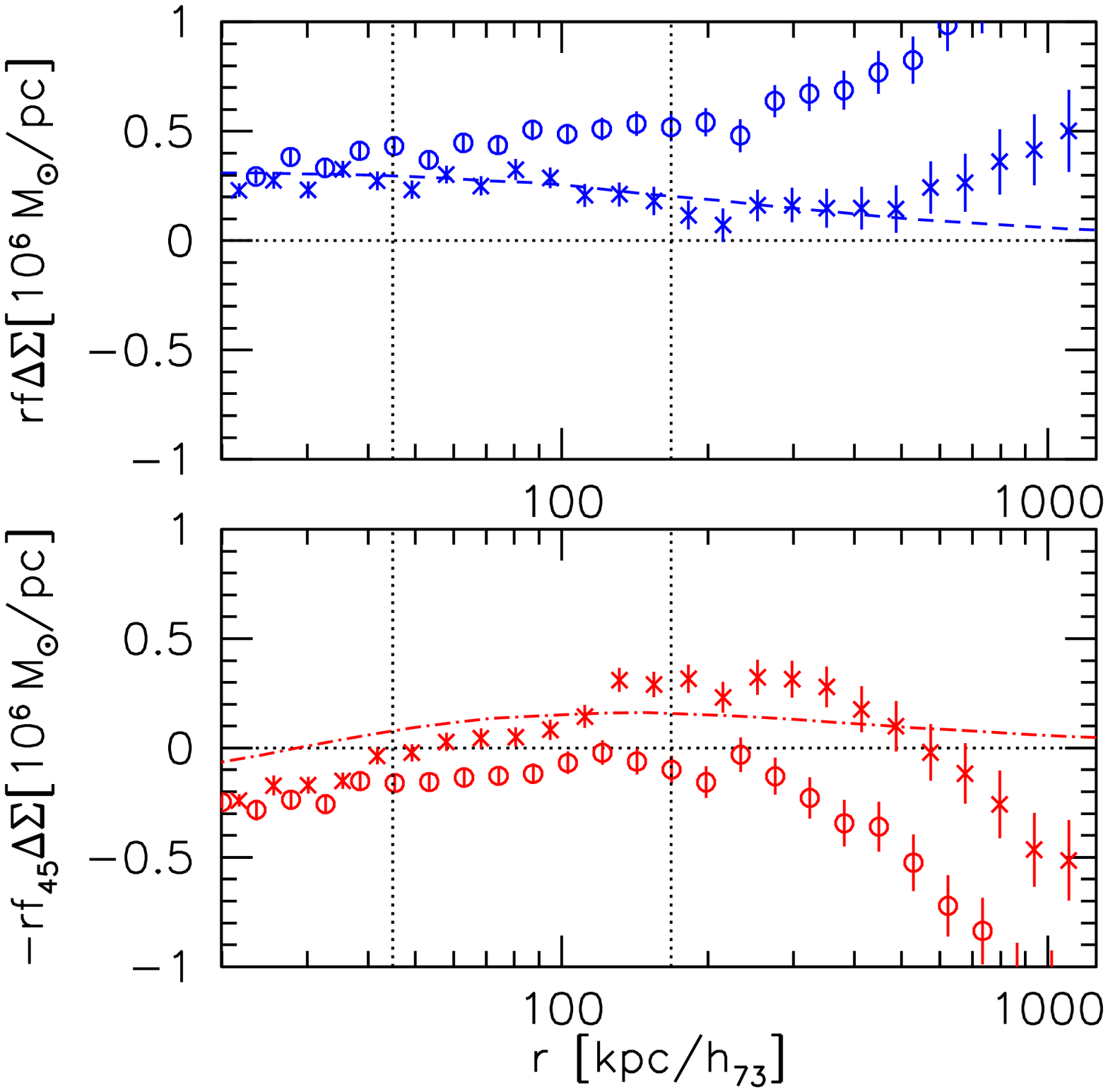}
  \includegraphics[width=5.8cm]{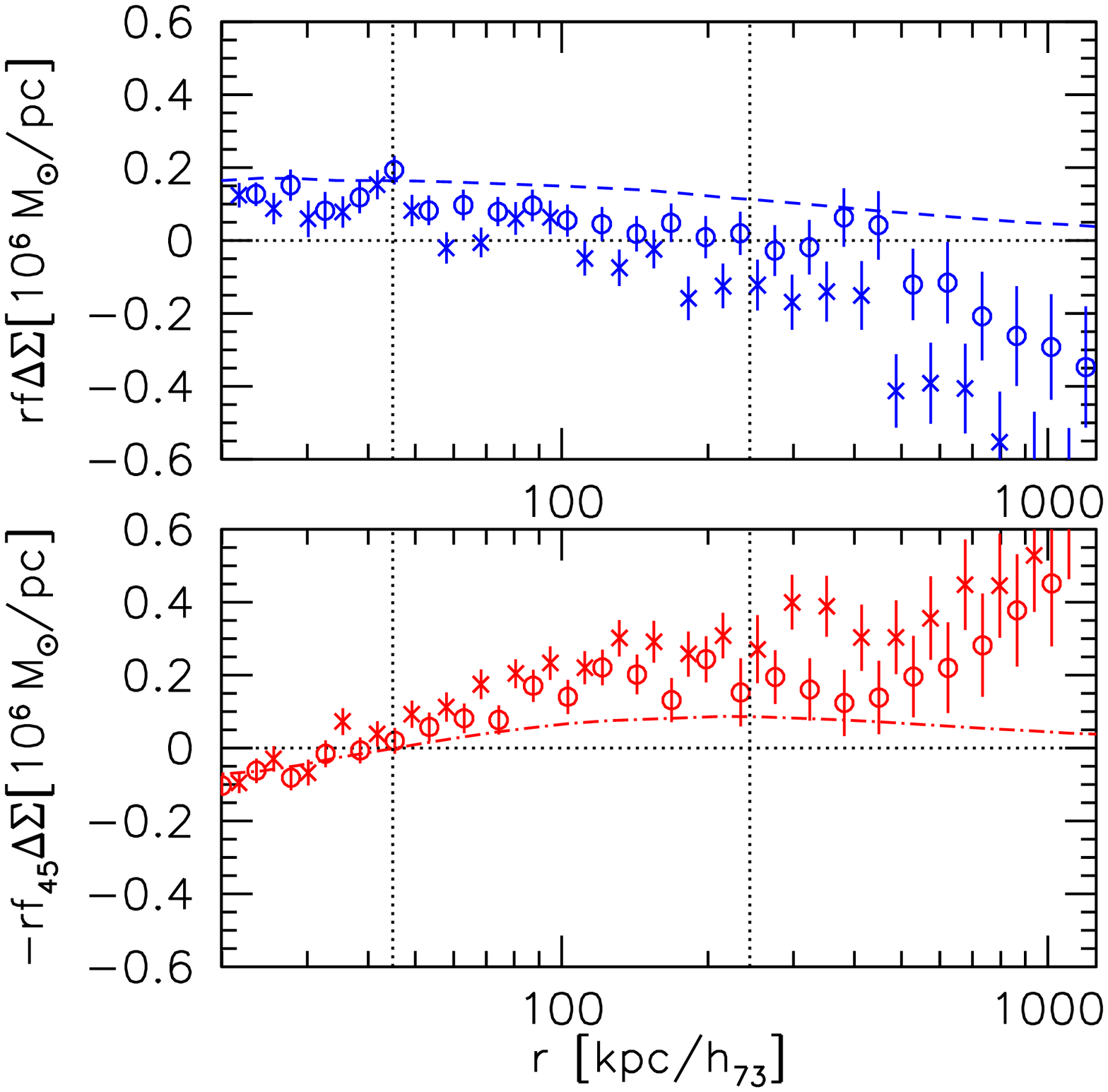}
  \includegraphics[width=5.8cm]{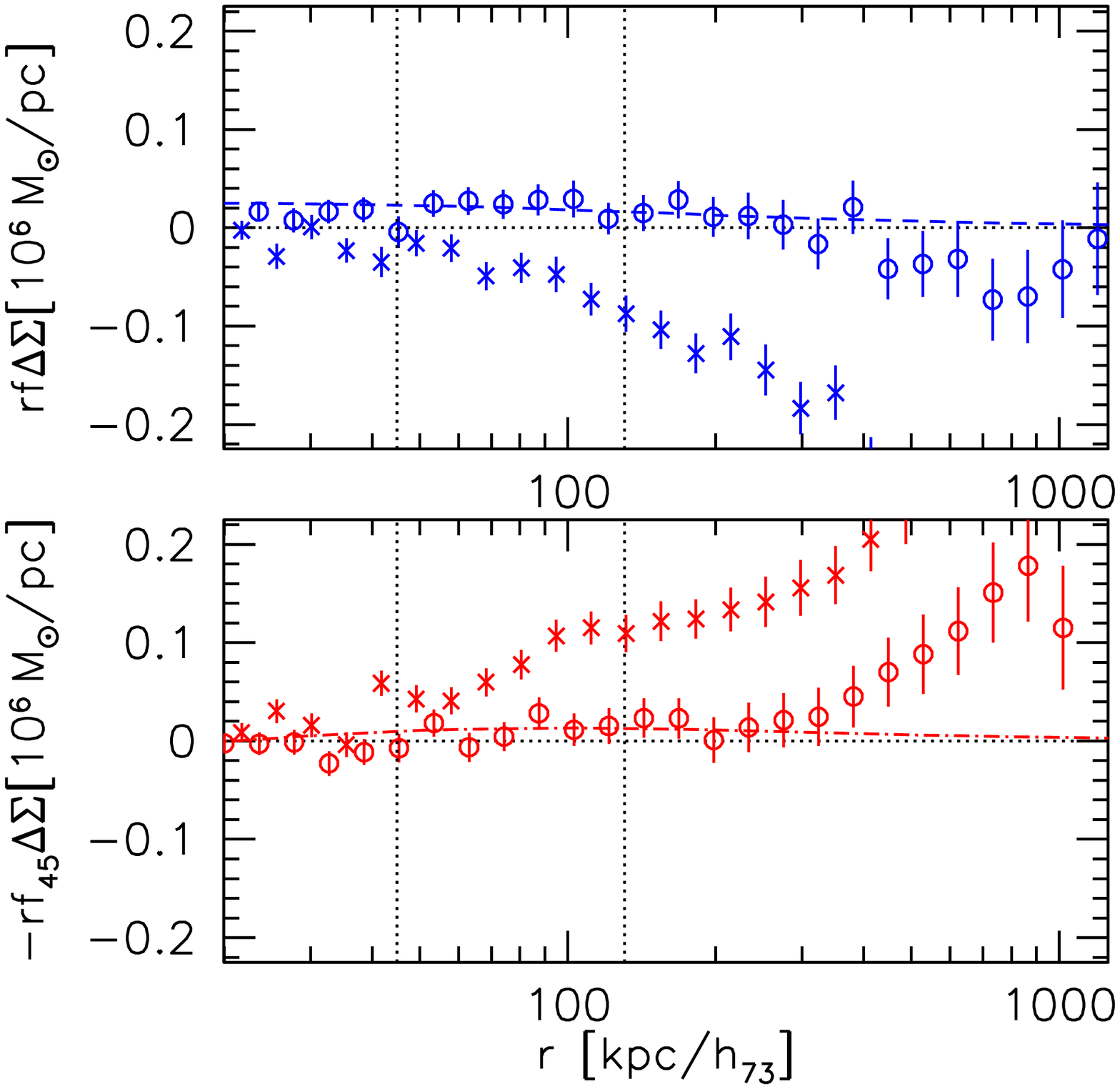}
   \caption{Analysis of the Millennium
Simulation showing lenses at \mbox{$0.2<z_\mathrm{l}<0.4$} with aligned shape models:
The {\it left} column
shows early type galaxies with stellar mass \mbox{$9.5<\log_{10} M_*<10$}.
 The {\it middle} column corresponds to late type galaxies in stellar mass bin
\mbox{$10.5<\log_{10} M_*<11$}.
The {\it right} column also shows the signal of late type galaxies, but with lower stellar masses  \mbox{$9.5<\log_{10} M_*<10$}.
For further details see the caption of Fig.\thinspace\ref{fi:shearfitsearly_ms1}. 
  }
   \label{fi:shearfitslate_ms1}
    \end{figure*}

We plot the isotropic and anisotropic shear profiles for an illustrative subset 
of the combinations of lens bins and shape models in  Figs.\thinspace\ref{fi:shearfitsearly_ms1} 
and \ref{fi:shearfitslate_ms1}. 
From top to bottom the panels show the isotropic profile $\Delta\Sigma(r)$
and the corresponding anisotropic profiles $(f-f_{45})\Delta\Sigma(r)$,
$f\Delta\Sigma(r)$, and $-f_{45}\Delta\Sigma(r)$, both 
 with (crosses) and without (circles)
foreground cosmic shear applied to
the lens ellipticities\footnote{The foreground cosmic shear is always applied to the sources, but it only has a net effect if it is applied to both lenses and sources.}.

 Fig.\thinspace\ref{fi:shearfitsearly_ms1} shows early type lenses with high
 stellar mass \mbox{$10.5<\log_{10} M_*<11$}.
Here, the left and middle columns show the signal of lenses with aligned
shape models at lower (\mbox{$0.2<z_\mathrm{l}<0.4$}) and higher
(\mbox{$0.4<z_\mathrm{l}<0.6$}) redshift, respectively,  illustrating the
increasing cosmic shear contribution.
The right column shows the signal of the same lenses as the middle column,
but now with misaligned shape models.

Fig.\thinspace\ref{fi:shearfitslate_ms1} shows 
the signal of other sets of lens galaxies at  \mbox{$0.2<z_\mathrm{l}<0.4$}
with aligned shape models, namely early types at lower stellar mass
(\mbox{$9.5<\log_{10} M_*<10$}) in the left column, as well as late types
with high (\mbox{$10.5<\log_{10} M_*<11$}) and low (\mbox{$9.5<\log_{10}
  M_*<10$}) stellar mass in the middle and right columns, respectively.

\subsubsection{Deviations in the isotropic shear profile}

We note that the isotropic $\Delta\Sigma$ profile falls below the NFW model 
at very small
scales, 
 especially for the lower mass haloes.
This is caused by the force softening used in the Millennium Simulation and smoothing applied in the ray-tracing \citep{hhw09,ghh13}.
This is not problematic for our analysis given
that we only include scales \mbox{$r>45\thinspace\mathrm{kpc}/h_{73}$} in the
fit. 
Also, the highest stellar mass bins for early types show a marginally steeper $\Delta\Sigma$ profile
than the best-fitting NFW model
 (see Fig.\thinspace\ref{fi:shearfitsearly_ms1}).
 This could be caused by our use of the  \citet[][]{dsk08}
 mass-concentration relation, which was derived using N-body simulations 
with a significantly different input cosmology than the one used in the  Millennium
Simulation (\mbox{$\sigma_8=0.796$} versus \mbox{$\sigma_8=0.9$}),  see e.g. \citet{lna14} regarding the cosmology dependence of the mass-concentration relation.
Since we estimate $f_\mathrm{h}$   directly from the ratio of the measured
  \mbox{$(f-f_{45})\Delta\Sigma$} and \mbox{$\Delta\Sigma$},
the impact of these deviations is negligible for our analysis,
 especially compared to the statistical uncertainties from CFHTLenS. However, as a
  result the model curve for    \mbox{$(f-f_{45})\Delta\Sigma$}, which is
  computed from the model for  \mbox{$\Delta\Sigma$} and the best-fitting $f_\mathrm{h}$, is slightly
  biased high compared to the data in Fig.\thinspace\ref{fi:shearfitsearly_ms1}.

\subsubsection{Influence of cosmic shear}
\label{se:ms_cs}
In the 
two bottom 
rows of panels in
Figs.\thinspace\ref{fi:shearfitsearly_ms1} and \ref{fi:shearfitslate_ms1}
 we 
show the anisotropic components $f\Delta\Sigma$ and $-f_{45}\Delta\Sigma$,
which are decreased and increased, respectively, due to the cosmic shear
contribution.
Here the relative effect is stronger towards larger radii,
higher redshifts, lower halo masses, and lower \mbox{$f_\mathrm{h}$}. 
In contrast, \mbox{$(f-f_{45})\Delta\Sigma$} is only weakly affected by the cosmic shear 
contribution within the fitted range
(\mbox{$45\thinspace\mathrm{kpc}/h_{70}<r<r_{200\mathrm{c}}$}) as visible in
the second row of panels, but we note that it
is decreased at larger radii compared to the case without cosmic shear. 
This is a
result of the non-vanishing $\xi_-$ at these scales (see Sect.\thinspace\ref{su:method}).
We stress that our approach using the \citetalias{mhb06} formalism and restricting the fit
range below $r_{200\mathrm{c}}$ efficiently suppresses the impact of cosmic
shear. As visible in Table \ref{tab:results_shapes_millennium} the remaining
net effect is \mbox{$\Delta f_\mathrm{h}\lesssim 0.02$} averaged over our redshift range, which is  an order
of magnitude smaller than current observational uncertainties.

\subsubsection{Indications of large-scale structure shape-shear correlations}
\label{se:shape_shear}
We note that the anisotropic shear signal shows deviations
from the simple elliptical  NFW model even in the case that no foreground
cosmic shear is applied to the lens ellipticities, as best visible for
the early type galaxies in 
Figs.\thinspace\ref{fi:shearfitsearly_ms1} and \ref{fi:shearfitslate_ms1}.
For the early type galaxies the measured signal is 
increased for $f\Delta\Sigma$
and
decreased for $-f_{45}\Delta\Sigma$ compared to the model that assumes an isolated single elliptical NFW halo. This trend is the
strongest at large radii (for the scaling by $r$ shown in the plot), but
especially for the lower mass haloes it is visible down to small radii 
(left column of Fig.\thinspace\ref{fi:shearfitslate_ms1}).
This  has the opposite effect compared to the foreground cosmic
shear which aligns the lens and the source.
In fact, both effects actually cancel fairly well 
for the middle column in
Fig.\thinspace\ref{fi:shearfitsearly_ms1}. 
Accordingly, this means that the ellipticities of the foreground lenses and the
background sources become anti-aligned because of this effect. 
This is precisely the signature of shape-shear correlations, which are one
of the intrinsic alignment contaminants to cosmic shear \citep[e.g.][]{his04,jma11,hgh13}, and which were studied using the same simulations in \citetalias{jsh13}.
They are caused by alignments of the foreground galaxies with their
surrounding large-scale structure, which lenses the background source
galaxies.
Note that the halo ellipticity signal we want to extract contributes
to shape-shear correlations itself, but an additional contribution comes
from the surrounding large-scale structure.
 We suspect that the additional signal we are detecting here in the context of halo shape measurements is caused by this large-scale structure contribution.

Note that this effect appears to be weaker for the late type
lenses,
in agreement with observational constraints  \citep{mbb11}, but also with the \citetalias{jsh13} intrinsic alignments analysis of these simulations.
For late type galaxies the strongest deviations from the single NFW models are visible for the highest stellar mass bin in the middle column of
Fig.\thinspace\ref{fi:shearfitslate_ms1}. Here they appear to have the same sign as
cosmic shear (alignment of lenses and sources), but in absolute terms the
effect is weaker and the constraints are more noisy than for the early types.

The observed profiles for \mbox{$f\Delta\Sigma$},
\mbox{$-f_{45}\Delta\Sigma$}, and \mbox{$(f-f_{45})\Delta\Sigma$}
experience a similar relative suppression when misaligned lenses are studied
compared to aligned lenses, if foreground cosmic shear is not applied (compare
the middle and the right column of Fig.\thinspace\ref{fi:shearfitsearly_ms1}).
This is expected given that random misalignment of lens galaxies 
suppresses both the halo shape signature and the additional signal from
shape-shear correlations  caused by  alignments of the lens galaxies with their
large-scale environment.
In contrast, the application of foreground cosmic shear introduces
additional additive signal which is the same for aligned and  misaligned
lens models.

\begin{figure}
   \centering
   \includegraphics[width=8cm]{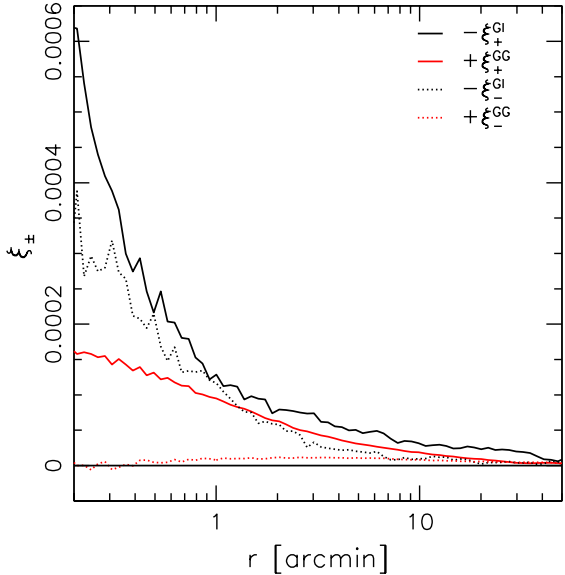}
   \caption{Shear-shear and shape-shear correlation functions
     $\xi_\pm^\mathrm{GG}$ and $-\xi_\pm^\mathrm{GI}$ for the   aligned early
     type galaxies in the Millennium Simulation
at  \mbox{$0.4<z_\mathrm{l}<0.6$} with \mbox{$10.5<\log_{10} M_*<11$}. For
these lenses $r_\mathrm{200c}$ equals 0\farcm98.
}
   \label{fi:xi}
    \end{figure}

To further test our hypothesis that the extra signal is caused
  by shape-shear correlations
  from the large-scale  environment, we further investigate the signal of
  the aligned early type galaxies
at  \mbox{$0.4<z_\mathrm{l}<0.6$} with \mbox{$10.5<\log_{10} M_*<11$}.
For these galaxies the plots in the  middle column of
Fig.\thinspace\ref{fi:shearfitsearly_ms1} suggest that the contributions
from cosmic shear and shape-shear correlations cancel approximately.
For these lenses and their surrounding background sources we compare
the two-point correlation functions  
\begin{equation}
\xi_\pm^\mathrm{GI}(r)=\langle e_\mathrm{l,t}^\mathrm{int} {\gamma}_\mathrm{s,t}\pm  e_\mathrm{l,\times}^\mathrm{int}   {\gamma}_\mathrm{s,\times}\rangle (r)\,
\end{equation}
between the (not-lensed) intrinsic lens ellipticities $e_\mathrm{l}^\mathrm{int}$ and the background shears (shape-shear signal), versus the corresponding cosmic shear correlation functions 
\begin{equation}
\xi_\pm^\mathrm{GG}(r)=\langle {\gamma}_\mathrm{l,t} {\gamma}_\mathrm{s,t}\pm  {\gamma}_\mathrm{l,\times}   {\gamma}_\mathrm{s,\times}\rangle (r)\, .
\end{equation}
As shown in Fig.\thinspace\ref{fi:xi}, $\xi_+^\mathrm{GI}$ and
$\xi_+^\mathrm{GG}$
are of roughly comparable amplitude (note the opposite sign) over a wide range in
scale ($r\gtrsim 0\farcm6$).
This is consistent with the interpretation that
the influence of cosmic shear and the influence of the shape-shear correlations  on the halo shape measurement cancel approximately. 
Note however that $\xi_\pm^\mathrm{GI}$ mixes contributions from both the
halo shape signal (primarily at small $r$)  and the shape-shear correlations from
the large-scale environment (affecting also larger $r$).
Therefore we do not attempt to interpret Fig.\thinspace\ref{fi:xi} more quantitatively.

As demonstrated in  Sect.\thinspace\ref{se:ms_cs},
halo shape estimates based on
\mbox{$(f-f_{45})\Delta\Sigma$} are insensitive to extra alignment from
cosmic shear if a limited radial range is included in  the fit as done in
our analysis.
Given 
that the radial dependence of the measured 
\mbox{$(f-f_{45})\Delta\Sigma$} signal is reasonably well described by the model prediction within the fit range for most of our lens bins,
we suspect that this formalism also provides an approximate correction for the extra alignment caused by the 
surrounding large-scale structure.
However, it is clear that this model of a
single, isolated, elliptical NFW halo is only a very crude approximation of
reality, and that some deviations due to the surrounding large-scale
structure are still expected. 
 For example, note the significant deviations from the best-fit model also within the fit range in the left column of Fig.\thinspace\ref{fi:shearfitslate_ms1}.
Nevertheless, as we perform a direct, consistent comparison of our CFHTLenS results to
the constraints from the simulation, which ideally includes the same physical
effects, the relative conclusions should not be
affected substantially.

\subsubsection{Constraints on $f_\mathrm{h}$ for early type galaxies}

\label{se:ms_halo_ell_early}

We summarise the estimates for the aligned ellipticity ratio $f_\mathrm{h}$
for the different lens samples 
in Table
\ref{tab:results_shapes_millennium}.
 For early types we generally measure larger  $f_\mathrm{h}$ for more
massive haloes.
For example, the most massive low-$z$ ellipticals yield \mbox{$
  f_\mathrm{h}=0.863\pm0.011 $}  for the aligned lens galaxy
models and no cosmic shear. 
For lower mass haloes this value decreases, and we find
\mbox{$f_\mathrm{h}=0.616_{-0.005}^{+0.006}$}
when applying cosmic shear and combining both redshift bins and the stellar mass range
\mbox{$9.5<\log_{10} M_*<11$}, to best mimic our red galaxy sample in
CFHTLenS.

This result is somewhat surprising:
Given that the projected lens shapes of early types are
defined via the projected quadrupole tensor of the mass distribution in the
simulation, we would naively expect to measure \mbox{$f_\mathrm{h}\simeq 1$}
in all cases.
We suggest three possible explanations for this, and we presume that
all three may contribute in practice at different levels:
First, the projected halo shapes are approximated via the  quadrupole tensor
of the mass distribution. 
However, the  projected mass
distributions will not  have exactly elliptical shapes, and therefore
deviate from the projected light distribution.
Second,  the haloes are embedded in a surrounding large-scale structure.
Components of that large-scale structure that have a random orientation with
respect to the orientation of the halo will add noise in the form of an on average
spherical mass distribution, reducing  \mbox{$f_\mathrm{h}$}.
Also the transient sub-haloes, which have been
removed (see Sect.\thinspace\ref{se:sims_mr_lenses}), contribute to this.
Finally, the surrounding large-scale structure  may have a component that is
aligned with respect to the halo orientation, causing the extra signal
discussed in Sect.\thinspace\ref{se:shape_shear}. While we expect that our
analysis using $(f-f_{45})\Delta\Sigma$ partially separates this component from
the  signal of the halo, there may be some residual effect impacting the
\mbox{$f_\mathrm{h}$} constraints. 
A hint for this is visible in the left column of
Fig.\thinspace\ref{fi:shearfitslate_ms1},
where the model fits the measured $(f-f_{45})\Delta\Sigma$ relatively poorly.
We note that the second and third effect
are likely more pronounced for less massive haloes for which the surrounding
large-scale structure has a larger relative impact, in agreement with the
 trend observed.

For the misaligned lens models $f_\mathrm{h}$  generally decreases as
expected, and we find \mbox{$f_\mathrm{h}=0.285_{-0.004}^{+0.006}$} for
the combined lens sample.

\subsubsection{Constraints on $f_\mathrm{h}$ for late type galaxies}

For  the late type galaxies the recovered values of  $f_\mathrm{h}$ are generally small,
 with \mbox{$f_\mathrm{h}=0.095\pm 0.005$} in the case of
aligned models 
and
\mbox{$f_\mathrm{h}=0.025_{-0.004}^{+0.006}$} for misaligned models, combining all lens samples and applying cosmic shear.
 This indicates that even if the angular momentum vector of each dark matter
halo was perfectly aligned with the spin axis of its disc
galaxy, the halo shape signal would be largely washed out
due to misalignment between the mass distribution and the angular
momentum vector of the halo. Once galaxy misalignments are included, the net signal
is so small that it should remain undetected even with much larger
surveys than CFHTLenS.
Despite this generally low signal there is a trend that
$f_\mathrm{h}$ increases in the simulated data for more massive haloes.

\subsubsection{Accounting for the differences in the ellipticity
  dispersions between CFHTLenS and the simulation}
\label{se:ellipticity_dispersion}
We can directly compare our constraints on $f_\mathrm{h}$ from
the Millennium Simulation to those from CFHTLenS if we assume that our procedure
to assign galaxy ellipticities to the haloes in the simulation is adequate.
However, if we compare the ellipticity dispersions $\sigma_e$ (including
both ellipticity components)
of the selected lenses between
CFHTLenS (Table \ref{tab:lenses}) and the simulation (Table
\ref{tab:results_shapes_millennium}),
we find that the average dispersions are somewhat higher in CFHTLenS, with
$\sigma_e^\mathrm{CFHTLenS}=0.29\,(0.35)$ 
for red (blue) lenses, compared to
$\sigma_e^\mathrm{simulation}=0.20\,(0.32)$ for the early (late) types in the Millennium
Simulation.
Hence, the simulation appears to deliver on average too round lenses, especially for early types.
As a first-order correction for this discrepancy one could  increase the
lens ellipticities in the simulation by a factor
\begin{equation}
\label{eq:scale}
s=\frac{\sigma_e^\mathrm{CFHTLenS}}{\sigma_e^\mathrm{simulation}}\simeq 1.46\, (1.10)
\end{equation}
if one assumes that the early (late) type galaxies in the simulation
resemble the red (blue) galaxies in  CFHTLenS.
We choose to not directly rescale the ellipticities, but to keep the analysis
comparable to earlier studies using these mock catalogues
\citep{jsh13,jsb13}.
From the definition of \mbox{$f_\mathrm{h}$} (see
Eq.\thinspace\ref{eq:fhfromeg}) it is directly evident that such a linear
increase of the lens ellipticities $e_\mathrm{g}$ will reduce  \mbox{$f_\mathrm{h}$} by the same
factor given that the shear signal (and hence $e_\mathrm{h}$) in the
simulation is unchanged.
Accordingly, when conducting a comparison in
Sect.\thinspace\ref{se:discussion} that assumes that the
red (blue) CFHTLenS sample
corresponds to the early (late) type sample in the simulation, 
 we will reduce the \mbox{$f_\mathrm{h}$} estimate from the simulation
by a factor $s$.

We also considered alternative
approaches to 
compute scale factors $s$ for the
approximate  matching of
the ellipticity distributions between CFHTLenS and the simulation, 
such as the ratios 
$\langle |e_\mathrm{g}| \rangle^\mathrm{CFHTLenS}/\langle |e_\mathrm{g}| \rangle^\mathrm{simulation}=1.39\,(1.09)$ 
or $\left(\langle |e_\mathrm{g}|^2 \rangle/\langle |e_\mathrm{g}| \rangle\right)^\mathrm{CFHTLenS}/\left(\langle |e_\mathrm{g}|^2 \rangle/\langle |e_\mathrm{g}| \rangle\right)^\mathrm{simulation}=1.52$ $ \,(1.11)$, where the latter is equivalent to matching the $\langle |e_\mathrm{g}| \rangle$ with an extra weighting by $|e_\mathrm{g}|$ as employed in 
Equations (\ref{eq:fdeltasigma_estimate}) and  (\ref{eq:fdeltasigma_estimate45}).
The resulting differences of $\pm 5\%$ ($\pm 1\%$) compared to Eq.\thinspace(\ref{eq:scale}) are small, especially compared to the statistical uncertainties from CFHTLenS.
Also note that we do not need to correct the measured lens ellipticity dispersions in CFHTLenS for measurement noise given the high signal-to-noise ratio of all lenses in our analysis$^7$.

\section{Discussion and conclusions}
\label{se:discussion}
Using CFHTLenS observational data we have obtained  weak lensing constraints on $f_\mathrm{h}$, the
ratio between the aligned projected ellipticity components of galaxy-scale matter haloes and
their galaxies.
In addition, we have used simulated data sets based on the  Millennium
Simulation
to test the methodology and obtain an estimate for the expected signal given
current
alignment models for galaxies and their dark matter haloes.

In this analysis we make use of the formalism introduced by \citetalias{mhb06} to correct
for additional alignments of lenses and sources, e.g.~due to residual shape
systematics or cosmic shear. 
We demonstrate that this method 
removes a simplistic, constant
alignment. 
In addition, we demonstrate that a realistic cosmic shear field, as present
in the Millennium Simulation, is well removed for the fit range we employ,
leaving residuals  \mbox{$\Delta f_\mathrm{h}\lesssim 0.02$} for our lens
redshift range (\mbox{$0.2<z_\mathrm{l}<0.6$}), which is an order
of magnitude smaller than current statistical uncertainties from CFHTLenS.
We demonstrate that a more simplistic estimator that only considers the
azimuthal variation in the tangential shear component
is highly biased (see e.g.~the  third panel in the right column 
of Fig.\thinspace\ref{fi:shearfitslate_ms1}) as previously pointed out by
\citet{bra10} and \citet{hob10}.

From the CFHTLenS data we estimate 
\mbox{$f_\mathrm{h}=-0.04\pm 0.25$}
for red lenses and 
\mbox{$f_\mathrm{h}=0.69_{-0.36}^{+0.37}$}
for 
blue lenses 
combining all stellar mass bins with \mbox{$\log_{10}M_*>10$}
(\mbox{$\log_{10}M_*>9.5$}) for the red (blue) lenses.
The colour separation is done via the photometric type from \textsc{BPZ}, which we
use as a proxy for the
separation into early and late type galaxies.
 As discussed below, these values are broadly consistent with
  theoretical models.

In our analysis of the Millennium Simulation we use different models for the
alignment of galaxies with their dark matter haloes for early and late type 
galaxies.
 For late type galaxies the analysis
assumes that the spin vectors of disc galaxies are
aligned with the angular momentum vectors of their dark matter haloes.
In the case of perfect alignment we find
\mbox{$f_\mathrm{h}=0.095\pm 0.005$} from the Millennium Simulation, which
reduces to
\mbox{$f_\mathrm{h}=0.025_{-0.004}^{+0.006}$}
 for models which assume the
misalignment distribution from \citet{bet12}.
This distribution is based on a compilation
of results from  simulations that include baryons and galaxy formation
physics \citep{oef05,ctd09,bef10,dmf11}.
If we rescale these results as discussed in
Sect.\thinspace\ref{se:ellipticity_dispersion}
to account for the differences in the lens ellipticity dispersions
in CFHTLenS versus the simulation, these values change to
\mbox{$f_\mathrm{h}=0.086\pm 0.005$} for the aligned models and
\mbox{$f_\mathrm{h}=0.023_{-0.004}^{+0.005}$} for the misaligned models.
Based on the simulation we do not expect to detect a significant signal for late type
galaxies in CFHTLenS.
 Our CFHTLenS constraint for blue galaxies 
is higher than the scaled prediction  for the aligned models by $1.7\sigma$, 
and
higher by $1.9\sigma$ compared to the scaled prediction for the misaligned models.

 For early type galaxies the analysis of the simulation assumes that
the lens ellipticities follow the ellipticity of the projected inertia
tensor of the halo mass distribution.
Hence, the projected galaxy shapes follow approximately the projected dark matter shapes.
In the case of no misalignment we 
estimate
\mbox{$f_\mathrm{h}=0.616_{-0.005}^{+0.006}$}
from the simulation when matching
the measured
halo mass range of the CFHTLenS constraints approximately. 
When applying a Gaussian misalignment distribution with an rms scatter of
35$^\circ$ \citep[as suggested by the distribution of satellites around LRGs
in SDSS, see][]{ojl09},
 we estimate a 
value of
\mbox{$f_\mathrm{h}=0.285_{-0.004}^{+0.006}$}.
If we rescale these results as discussed in
Sect.\thinspace\ref{se:ellipticity_dispersion}
to account for the differences in the lens ellipticity dispersions
in CFHTLenS versus the simulation, the values change to
\mbox{$f_\mathrm{h}=0.422_{-0.003}^{+0.004}$} for the aligned models and
\mbox{$f_\mathrm{h}=0.195_{-0.003}^{+0.004}$} for the misaligned models.
Assuming that the red CFHTLenS galaxies directly correspond to the simulated early
type galaxies would mean that our CFHTLenS constraints are lower than the
rescaled aligned prediction by $1.8\sigma$, and  lower by $0.9\sigma$ compared to
the rescaled misaligned prediction.
Hence, they are poorly described by the perfectly aligned model but fully 
consistent with the misaligned model.

It is interesting to compare our constraints to the results of recent
hydrodynamical simulations. \citet{tmm14} find a mean 3D misalignment angle
at \mbox{$z=0.3$} of 25.20$^\circ$ for a halo mass bin
\mbox{$10^{11.5}h^{-1}M_\odot<M<10^{13}h^{-1}M_\odot$} which most closely
matches the mass range of our red galaxies, but this increases to
33.47$^\circ$ for lower halo masses
\mbox{$10^{10}h^{-1}M_\odot<M<10^{11.5}h^{-1}M_\odot$}. The misalignment
distribution estimated by \citet{ojl09} and assumed in our simulated
analysis (rms scatter of
35$^\circ$) would correspond to a similar {\it mean} value of \mbox{$\sim 28^\circ$}
\citep{tmm14}.
Thus, our assumed misalignment model  approximately matches the
results of this  hydrodynamical simulation.
However, for the direct comparison between observations and simulations there are additional relevant effects to be
  considered: 
for example, baryons appear to make the inner halo more spherical
\citep{bkd13}. In addition, sub-haloes are found to be rounder than haloes
\citep{kdm07,tmm14}. 
 Accordingly, as our analysis of early type galaxies in the simulation uses central
haloes only, and given that it does not include baryons, we expect that our
simulation should somewhat over-predict $f_\mathrm{h}$.
 A further complication for the comparison arises from the observation that numerical simulations of galaxy formation suggest that dark matter halo shapes are misaligned at different radii \citep[e.g.][]{sfc12,wlk14}.

A number of previous studies have attempted to constrain halo ellipticity
with weak lensing observationally.
Two previous studies were able  to split the lens sample into red and
blue galaxies as done in our study: 
\citetalias{mhb06} find \mbox{$f_\mathrm{h}=0.60 \pm 0.38$}
for red and  \mbox{$f_\mathrm{h}=-1.4^{+1.7}_{-2.0}$} for blue lenses,
assuming an elliptical NFW mass profile and employing data from the SDSS.
Using data from the RCS2 and employing the same formalism as \citetalias{mhb06}, \citet{uhs12} find
\mbox{$f_\mathrm{h}=0.20^{+1.34}_{-1.31}$} for red lenses and \mbox{$f_\mathrm{h}=-2.17^{+1.97}_{-2.03}$}
for blue lenses when assuming an elliptical NFW mass profile and using
linear weighting with the lens ellipticity as done in our study.
While our error-bars appear to be substantially tighter than those of these
two studies,
we note that our analysis of simulated data suggests a sign error in the
numerical model prediction  computed by \citetalias{mhb06} for $f_{45}$, which
was also employed by \citet{uhs12}. This sign error has likely biased
  their derived constraints.
With the corrected sign the predicted signal for 
$(f-f_{45})\Delta\Sigma$ is 
higher at a given \mbox{$f_\mathrm{h}$} over the entire  radial fit
range and does not drop as quickly towards large $r$ (see
Fig.\thinspace\ref{fi:simnfw}). Accordingly, this correction leads to
significantly tighter
constraints on  \mbox{$f_\mathrm{h}$}.

\citet{hyg04} and \citet{phh07} use single-band data from RCS and early
CFHTLS observations, respectively, to constrain halo ellipticity without
subdivision into red and blue galaxies.
They do not correct for systematic alignment between lenses and sources as
introduced by \citetalias{mhb06}. \citet{hyg04} conduct a maximum likelihood
analysis assuming elliptical, truncated isothermal sphere (TIS) models from which they
find \mbox{$f_\mathrm{h}=0.77^{+0.18}_{-0.21}$}.  
\citet{phh07} compute the ratio of the shears measured in quadrants along
the lens  minor and major axes, for which they find a tentative signal \mbox{$0.76\pm 0.10$} when
averaged out to 70$^{\prime\prime}$. 
There are differences in the  anisotropic shear fields for
NFW and TIS profiles 
(see \citetalias{mhb06}),
but nonetheless the 
\mbox{$f_\mathrm{h}$} constraint from \citet{hyg04} may appear
somewhat high even if their lens selection based on magnitude (without colours) would provide a
perfect selection of early types, which is certainly not the case.
The reason for this is that their formalism does not account for
spurious alignment caused e.g. by
foreground cosmic shear,
which  leads to a lower measured value of \mbox{$f_\mathrm{h}$}.
Our analysis of the Millennium Simulation may provide a possible explanation
for the high value of \mbox{$f_\mathrm{h}$} measured by \citet{hyg04}:
As discussed in Sect.\thinspace\ref{se:shape_shear}, we detect an excess
signal in $f\Delta\Sigma$ for early type lenses, 
which approximately corresponds to the signal probed by \citet{hyg04} and
\citet{phh07}.
This excess signal has the opposite 
 sign than the signal caused by
cosmic shear (also
visible in $-f_{45}\Delta\Sigma$). We interpret this signal as the impact of shape-shear
intrinsic alignments, which are a major contaminant to cosmic shear
measurements \citep[e.g.][]{his04,jma11,hgh13}.
They are caused by an alignment of foreground galaxies with their
surrounding large-scale structure \citep[as e.g.~detected in the distribution of red galaxies, see][]{mhi06,ljf13,zyw13}, which also lenses the background
sources. While the halo ellipticity signal 
contributes to
small-scale shape-shear correlations itself \citep{bra07}, there appears to be an additional component
generated by the  large-scale structure the halo is embedded in. In
other words, the assumed model of an isolated elliptical NFW halo is too
simplistic. The net effect of this signal is a net
anti-alignment (orthogonal alignment) of the ellipticities of foreground lenses and background
sources.
We expect that the formalism introduced by \citetalias{mhb06}  and employed by
us 
also provides a partial correction for this contaminant,
but  conclude that the most robust halo shape results
can be obtained from
the direct, relative comparison between observations and
simulations that include this effect, as done in our study.

We note that our analysis of blue galaxies in CFHTLenS shows indications for
the influence of cosmic shear on the (uncorrected) anisotropic shear
components $f\Delta\Sigma$ (slightly negative signal in the third row of
panels in Fig.\thinspace\ref{fi:shearfitsblue1}) and   $-f_{45}\Delta\Sigma$ (slightly positive signal in the fourth row of
panels in Fig.\thinspace\ref{fi:shearfitsblue1}).
In contrast, we do not observe this trend for red galaxies (see
Fig.\thinspace\ref{fi:shearfitsred1}).
A likely explanation for this may be that the additional shape-shear signal
discussed above appears to roughly cancel the cosmic shear contribution for
our red galaxy sample.
In contrast, 
the magnitude-selected lens samples from  \citet{hyg04} and
\citet{phh07} will likely be dominated by galaxies at lower redshift.
Here, the shape-shear contribution will dominate given the smaller cosmic
shear signal.

As discussed in Sect.\thinspace\ref{se:ms_halo_ell_early},
 the
influence of neighbouring large-scale structure is likely one of the main 
reasons  why the analysis of the aligned early type galaxies in the
Millennium Simulation yields somewhat smaller $f_\mathrm{h}$ than the naive
expectation of \mbox{$f_\mathrm{h}\simeq 1$}, in particular for the lower
mass haloes. 

The early type lenses with the highest halo mass show the
largest $f_\mathrm{h}$ in our simulation (see Table \ref{tab:results_shapes_millennium}).
In addition, more massive haloes are less spherical in simulations
\citep[e.g.][]{bas05,dgt14}.
Furthermore, galaxies with more massive haloes are expected to be less
misaligned \citep{tmm14}.
From all three effects we expect that future studies with larger samples might have the best prospects for
detecting halo ellipticity with weak lensing for very massive early type galaxies.
Very recently, \citet{clj15} reported the detection
  of a significant halo shape signal for exactly such luminous red galaxies
  from SDSS,  employing a new estimator from \citet{acd15}.
In the future, it will be interesting to  test
  this estimator on  large
mock data-sets, such as the data provided by the Millennium Simulation.

Weak lensing studies constraining halo ellipticity are not only interesting
for  a better understanding of the link between galaxies and their
surrounding 
matter haloes,  but have also been discussed as a possible test for
theories of modified
gravity such as MOND \citep[Modified Newtonian Dynamics,][]{mil83}, TeVeS
\citep[Scalar--Tensor--Vector theory,][]{bek04}, and MOG/STVG \citep[Modified
Gravity and Scalar--Tensor--Vector Gravity theory,][]{mof06,mot09}, for which
lensing prescriptions have been developed \citep{mo01,bek04,ckt06,mot09}.
The equivalent signals to halo shapes are discussed for isolated galaxies in
\citet{mil01} and \citet{sek02}, predicting an isotropic shear signal towards large
radii. Accordingly, a significant detection of halo ellipticity from weak
lensing,
 which would be expected within $\Lambda$CDM for the analysis of early type galaxies in upcoming experiments, could 
 be interpreted as evidence against such theories of modified
gravity.
 However, our analysis of the simulated data has shown that already within $\Lambda$CDM neighbouring structures have a considerable influence, which needs to be taken into account.
Thus, better model predictions need to be developed also for theories of modified
gravity, that include both large-scale structure contributions and the influence of baryon physics.

The prospects for near-future improvements on halo shape constraints from
weak lensing are relatively good:
On the one hand, weak lensing surveys are improving rapidly in size.
Surveys which are already underway include the Dark Energy Survey
\citep[DES,][]{des05}, the Hyper Suprime-Cam  Subaru Strategic Program 
\citep[HSC-SSP,][]{miyazaki12}, the Kilo Degree Survey
\citep[KiDS,][]{dejong13}, as well as the
Panoramic Survey Telescope and Rapid Response System
\citep[PanSTARRS,][]{kaiser10},
and future programmes such as  LSST \citep{lsst09} and {\it Euclid}
\citep{laureijs11} will tighten parameter constraints even further.
On the other hand, the use of new estimators may also provide additional
insight:
  \citet{ssk12} and \citet{acd15} suggest that the position angle dependence of the galaxy-shear-shear
correlation function, which is one of the observables of
galaxy-galaxy-galaxy lensing \citep{scw05},
may  provide constraints on 
halo ellipticity independent of the orientation of the lens galaxy ellipticity. 
This measurement would be unaffected by misalignment, but is expected to have lower signal-to-noise.
In addition, the measurement of higher-order lensing (``flexion'') may
also provide additional sensitivity for  constraining halo ellipticity \citep{ers11,erb13}.

\section*{Acknowledgements}

We thank Rachel Mandelbaum and Peter Schneider for useful discussions 
and comments on this manuscript.
We also thank Rachel Mandelbaum 
for making the
tabulated predictions for $f_\mathrm{rel}$ and $f_\mathrm{rel,45}$ for an elliptical NFW profile 
available to us.
\replya{We thank the referee for his or her comments, which have helped to improve 
the manuscript significantly.}

This work is based on
observations obtained with MegaPrime/MegaCam, a joint project of the
Canada-France-Hawaii Telescope (CFHT) and CEA/Irfu, at CFHT, which is
operated by the National Research Council (NRC) of Canada, the
Institut National des Sciences de l'Univers (INSU) at the Centre
National de la Recherche Scientifique (CNRS) of France, and the
University of Hawaii. This research used the facilities of the
Canadian Astronomy Data Centre operated by the NRC of Canada with the
support of the Canadian Space Agency. We thank the CFHT staff, in
particular J.-C.~Cuillandre and E.~Magnier, for the observations, data
processing and continuous improvement of the instrument
calibration. We also thank TERAPIX for quality assessment, and
E.~Bertin for developing some of the software used in this
study. CFHTLenS data processing was made possible thanks to support
from the Natural Sciences and Engineering Research Council of Canada
(NSERC) and HPC specialist O.~Toader. 
The
early stages of the CFHTLenS project were made possible thanks to the
European Commissions Marie Curie Research Training
Network DUEL (MRTN-CT-2006-036133) and its support of CFHTLenS team members LF, HHi, and BR.
This analysis makes use of data derived from the Millennium
  Simulation, which was carried out as part of the programme of the Virgo Consortium on the Regatta supercomputer of the Computing Centre of the Max-Planck-Society in Garching.

TS acknowledges support from NSF through grant
AST-0444059-001, SAO through grant GO0-11147A, and NWO.
SH acknowledges support by the National Science Foundation (NSF) grant number AST-0807458-002, and by the DFG cluster of excellence \lq{}Origin and Structure of the Universe\rq{}.
HHo acknowledges support from NWO VIDI grant number 639.042.814 and ERC FP7
grant 279396.
PS receives support by the DFG through the project SI 1769/1-1.
HHi is supported by the DFG Emmy Noether grant Hi 1495/2-1.
PEB was supported by the Deutsche Forschungsgemeinschaft under
  the project SCHN 342/7–1 in the framework of the Priority Programme
  SPP-1177, and the Initiative and Networking Fund of the Helmholtz
  Association, contract HA-101 ('Physics at the Terascale').
LF acknowledges support from NSFC grant 11333001 \& Shanghai Research grant
13JC1404400 of STCSM.
BJ acknowledges support by an STFC Ernest Rutherford Fellowship, grant reference ST/J004421/1.

{\it Author Contributions:} All authors contributed to the development and writing of this paper. The authorship list reflects the lead authors of this paper (TS, SH, HHo, PS, EvU) followed by two alphabetical groups. 
The first alphabetical group includes key contributors to the science analysis and interpretation in this paper, the founding core team and those whose long-term significant effort produced the final CFHTLenS data product. The second group covers members of the CFHTLenS team and collaborators who made a significant contribution to the project and/or this paper. The CFHTLenS collaboration was co-led by CH and LVW.

\bibliographystyle{mn2e}
\bibliography{paper1d}

\begin{appendix}
\section{Tests with the ``clone'' image simulation}
\label{app:testlf}

  \begin{figure*}
   \centering
   \includegraphics[width=5.8cm]{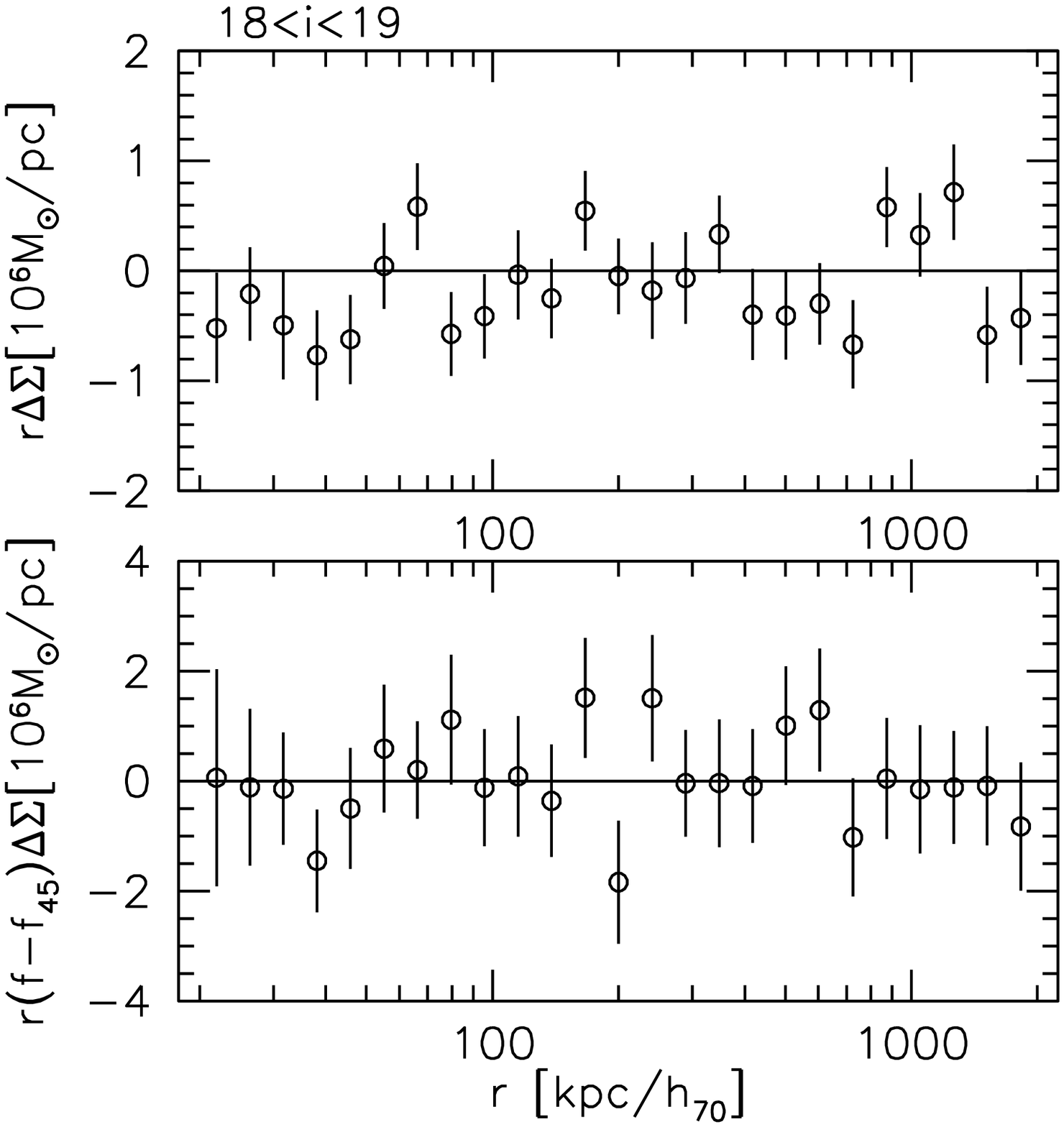}
   \includegraphics[width=5.8cm]{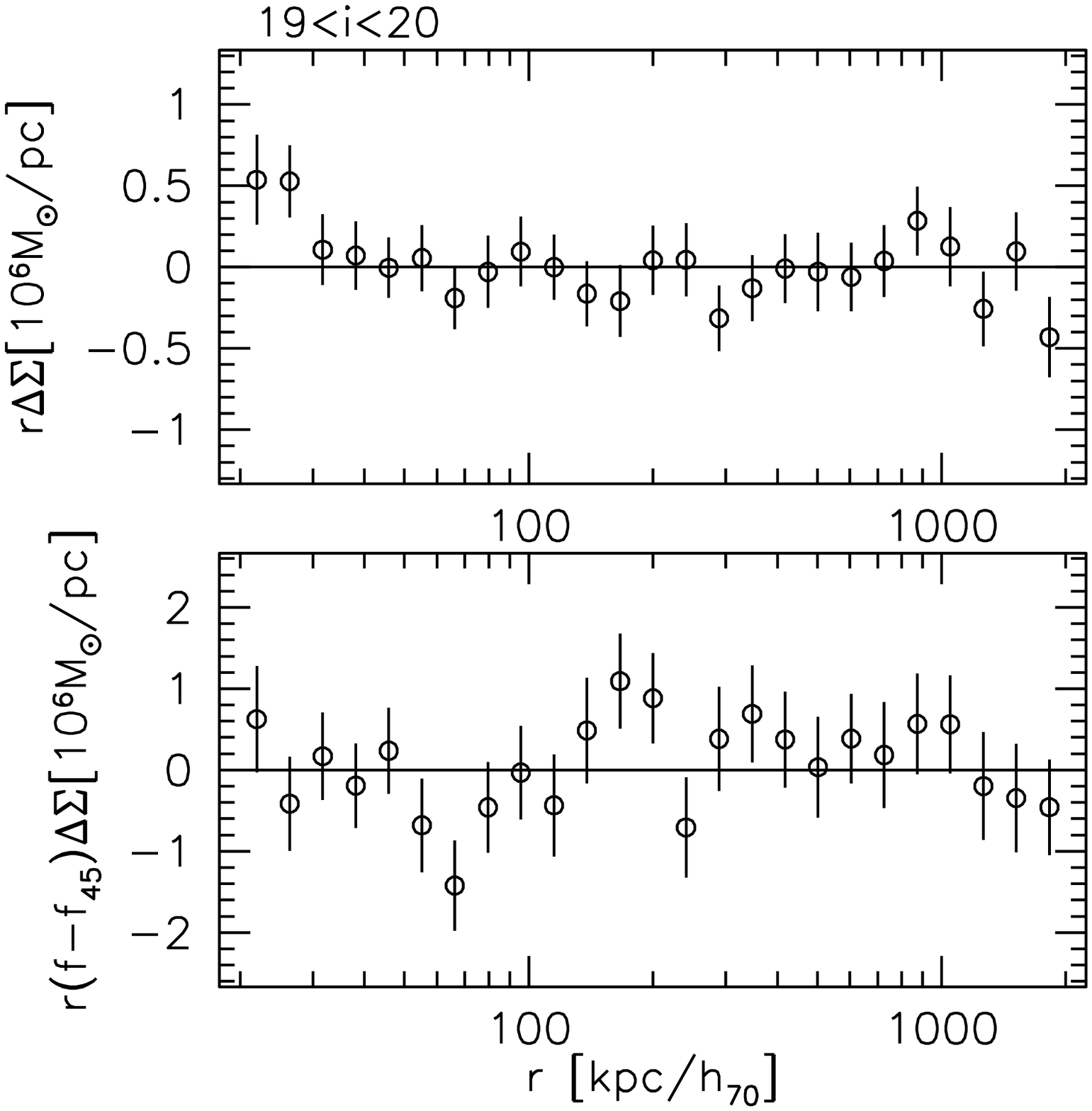}
   \includegraphics[width=5.8cm]{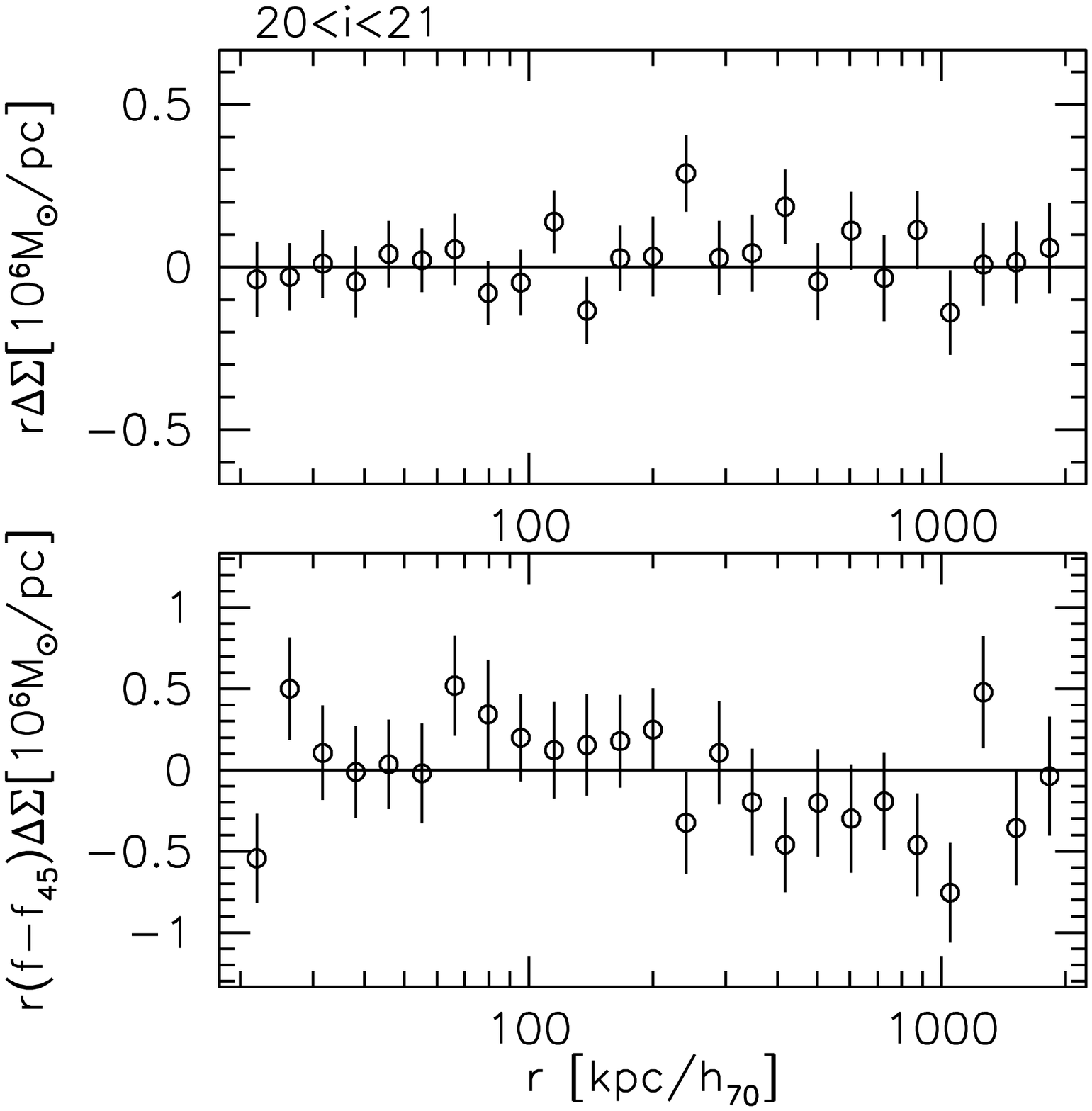}
   \caption{Test for a potential systematic contamination of the small-scale halo shape
   signal originating from \emph{lens}fit shape measurements in the
   presence of a nearby
     bright galaxy (lens): We employ \emph{lens}fit shape measurements of the
     ``Clone'' image simulation, which uses actual galaxy positions and
     magnitudes from CFHTLenS as input but no galaxy-galaxy lensing signal. 
To optimise the sensitivity of the test,
     all bright galaxies in the indicated $i$-band magnitude intervals
     with  \emph{lens}fit shape estimates in the CFHTLenS data were used as
     lens sample ({\it left}: \mbox{$18<i<19$}, {\it middle}:
     \mbox{$19<i<20$}, {\it right}: \mbox{$20<i<21$}),
 irrespectively of their assigned CFHTLenS redshift (we assume our
  central lens redshift \mbox{$z=0.4$} for the computation of angular
  diameter distances).
     The {\it top} ({\it bottom}) panels show the isotropic (anisotropic) shear signal, where both have been scaled by $r$ for better readability of the plot.  
}
   \label{fi:shearfitsclone}
    \end{figure*}

\citet[][]{mhk13a} and \citet{hwm12} present detailed tests of the \emph{lens}fit shape estimation
algorithm for cosmic shear measurements. 
As the required level of systematics control is 
less demanding for galaxy-galaxy lensing 
studies than for cosmic shear, we do not present general
shape measurement tests again in the current work.
However, what has not been tested before in detail, is the recovery of
galaxy shapes in the presence 
 of a nearby bright lens galaxy,
whose light might affect the shape measurement process for sources at small angular separations (e.g. smaller than the $9^{\prime\prime}$ postage stamp size)\footnote{For example, a circular source might appear slightly elliptical due to contamination by light from the nearby lens, where the net ellipticity would point towards the lens, thus generating spurious signal with \mbox{$\Delta\Sigma<0$}. This effect would likely be stronger along the direction of the lens major axis, and could thus also generate \mbox{$f\Delta\Sigma<0$}.}.
To test if this could be a concern for our study, we investigate
\emph{lens}fit shape measurements of the ``Clone'' image simulations
\citep{mhk13a}.
These simulations contain  simulated galaxy images with the positions and magnitudes from CFHTLenS  and cosmological shear from an N-body simulation \citep{hvw12} as input, but no galaxy-galaxy lensing signal. 
Hence, any detected galaxy-galaxy lensing signal would indicate a spurious effect introduced by the measurement process.

The result of this test is shown in Fig.\thinspace\ref{fi:shearfitsclone}. 
Here we split the ``lenses'' into magnitude bins and have subtracted the
intrinsic ellipticity and cosmological shear 
from the measured ellipticities
to maximise the \mbox{$S/N$} of the test and achieve statistical error-bars
which are significantly smaller than for the actual survey\footnote{In
  contrast to \citet[][]{mhk13a}, we do not make use of rotated galaxy pairs
  in the simulations, which are often used to partially cancel shape noise
  from the intrinsic source ellipticities. This combination is not useful
  for our test, as also the nearby ``lenses'' will be rotated, therefore
  leading to different lens-light contributions in the rotated source galaxy
  pairs.}.
 Given the lack of a clear signal at small scales, and the significantly smaller
uncertainties compared to the CFHTLenS constraints,
we conclude that the light of nearby bright lenses does
not appear to introduce significant spurious signal for our halo shape analysis.

\end{appendix}

\label{lastpage}

\end{document}